\documentclass[aps,preprint,superscriptaddress]{revtex4-1}

\usepackage{graphicx}
\usepackage{amsmath}
\usepackage{amssymb}
\usepackage{slashed}

\newcommand{\tr}{\rm tr \,}

\newcommand{\der}{\partial}
\newcommand{\CD}{\ooalign{\hfil$\cdot$\hfil\crcr$\der$}}

\begin{document}


\title{On chiral extrapolations of charmed meson masses \\and coupled-channel reaction dynamics}


\author{Xiao-Yu Guo}
\affiliation{GSI Helmholtzzentrum f\"ur Schwerionenforschung GmbH, \\Planckstra\ss e 1, 64291 Darmstadt, Germany}
\author{Yonggoo Heo}
\affiliation{GSI Helmholtzzentrum f\"ur Schwerionenforschung GmbH, \\Planckstra\ss e 1, 64291 Darmstadt, Germany}
\author{Matthias F.M. Lutz}
\affiliation{GSI Helmholtzzentrum f\"ur Schwerionenforschung GmbH, \\Planckstra\ss e 1, 64291 Darmstadt, Germany}
\affiliation{Technische Universit\"at Darmstadt, D-64289 Darmstadt, Germany}
\date{\today}

\begin{abstract}

We perform an analysis of QCD lattice data on charmed meson masses. The quark-mass dependence of the data set is used 
to gain information on the size of counter terms of the chiral Lagrangian formulated with open-charm states with $J^P= 0^-$ and 
$J^P =1^-$ quantum numbers. Of particular interest are those counter terms that are active in the exotic flavour sextet channel. 
A chiral expansion scheme where physical masses enter the extrapolation formulae is developed and applied to the 
lattice data set. Good convergence properties are demonstrated and an accurate reproduction of the lattice data based on ensembles 
of PACS-CS, MILC, ETMC and  HSC with pion and kaon masses smaller than 600 MeV is achieved.
It is argued that a unique set of low-energy parameters is obtainable only if additional information from  HSC
on some scattering observables is included in our global fits. The elastic and inelastic s-wave $\pi D$ and $\eta D$ scattering as considered by HSC 
is reproduced faithfully. Based on such low-energy parameters we predict 15 phase shifts and in-elasticities at physical quark masses 
but also for an additional HSC ensemble at smaller pion mass. In addition we find a clear signal for a member of the exotic flavour sextet states in the $\eta D$ channel, below the 
$\bar K D_s$ threshold. For the isospin violating strong decay width of the $D^*_{s0}(2317)$ we obtain the range $(104\,-\,116)$ keV.

\end{abstract}


\maketitle
\tableofcontents

\newpage


\section{Introduction}
\label{sec:1}

Systems with one heavy and one light quark play a particularly important role in the spectroscopy of QCD \cite{Yan:1992gz,Casalbuoni:1996pg,Lutz:2015ejy,Chen:2016spr}. 
Two distinct approximate symmetries characterize the spectrum of open-charm mesons. While in the limit of an infinitely heavy charm quark the
heavy quark spin-symmetry arises, the opposite limit with vanishing masses for the up, down and strange quark mass leads to the flavour SU(3) chiral symmetry. 
The approximate chiral symmetry of the up, down and strange quarks guides the construction of effective field theory approaches 
based on the chiral Lagrangian. 
There are two complementary approaches feasible. 
Either one may construct an effective chiral Lagrangian formulated in terms of heavy-quark multiplet fields \cite{Yan:1992gz,Casalbuoni:1996pg,Manohar:2000dt} or one may start with an effective chiral 
Lagrangian with fully relativistic fields, wherein the low-energy constants are correlated by constraints from the heavy-quark spin symmetry \cite{Lutz:2007sk,Lutz:2014jja}.
The former approach may be more economic in applications where the coupled-channel unitarity constraint is implemented by means of partial 
summation techniques \cite{Kolomeitsev:2003ac,Hofmann:2003je,Liu:2012zya,Altenbuchinger:2013vwa,Cleven:2014oka,Du:2016tgp}.

A striking prediction of the leading order chiral interaction of the Goldstone bosons with the D mesons with either $J^P = 0^-$
or $J^P =1^-$ is an attractive short-range force in the exotic flavour sextet channel \cite{Kolomeitsev:2003ac,Hofmann:2003je,Lutz:2007sk}. The strength of this 
interaction is somewhat reduced as compared to a corresponding force in the conventional flavour triplet channel that can be successfully used to describe the 
lowest scalar and axial-vector states in the open-charm meson spectrum \cite{Kolomeitsev:2003ac,Hofmann:2003je,Lutz:2007sk,Guo:2006fu,Guo:2006rp,Altenbuchinger:2013vwa,Cleven:2014oka,Du:2016tgp,Albaladejo:2016lbb}. Whether the chiral force in the flavour sextet sector leads to 
the formation of exotic open-charm meson states is an open issue. The possible existence of such an exotic flavour sextet multiplet of states depends on the precise 
form of chiral correction terms \cite{Hofmann:2003je,Lutz:2007sk}. 

In this work we wish to study the size of such chiral counter terms. First rough studies \cite{Hofmann:2003je,Lutz:2007sk} suffer from limited  empirical constraints.
Additional information from first QCD lattice simulation on a set of s-wave scattering lengths was used in a series of later works 
\cite{Liu:2012zya,Altenbuchinger:2013vwa,Cleven:2014oka,Du:2016tgp}. Results are obtained that in part show unnaturally large counter terms and/or illustrate some residual dependence on how to  set up the 
coupled-channel computation. Here we follow a different path and try to use the recent data set on the quark-mass dependence of the D meson 
ground-state masses \cite{Aoki:2008sm,Mohler:2011ke,Na:2012iu,Kalinowski:2015bwa,Cichy:2016bci,Cheung:2016bym,Moir:2016srx}. 
This dynamics is driven in part by the counter terms that also have significant impact on the open-charm coupled-channel systems as discussed above. One may hope to obtain 
results that are less model dependent in this case.

However, it is well known that chiral perturbation theory formulated with three light flavours does not always show a convincing convergence pattern 
\cite{Young:2002ib,Leinweber:2003dg,Beane:2004ks,Leinweber:2005xz,McGovern:2006fm,Djukanovic:2006xc,Schindler:2007dr,Hall:2010ai}. 
How is this for the case at hand? Only few studies are available in which this issue is addressed for open-charm meson systems.
In a recent work the authors presented a novel chiral extrapolation scheme for the quark-mass dependence of the baryon octet and decuplet states that is formulated in terms 
of physical masses \cite{Semke:2005sn,Semke:2011ez,Lutz:2014oxa,Lutz:2018a}. It is the purpose of our study to adapt this scheme to the open-charm sector of QCD and apply it to the available lattice data set. 
This requires in particular to consider the  D mesons with  $J^P = 0^-$ and $J^P =1^-$ quantum numbers on equal footing. For a given set of low-energy constants 
each set of the four D meson masses has to be determined numerically as a solution of a non-linear system.   

The work is organized as follows. In section II the part of the chiral Lagrangian that is relevant here is recalled. It follows a section where the one-loop contributions 
to the D meson masses are derived in a finite box. We do not consider discretization effects in our study. 
In section III and IV power counting in the presence of physical masses is discussed. The application to available lattice 
data sets is presented in sections V and VI. Lattice data taken on ensembles of PACS-CS, MILC, ETMC and HSC are considered.
In section VII we present our predictions for phase shifts and in-elasticities based on a parameter set obtained form the considered lattice data.  
In addition the fate of possible exotic states but also the isospin violating strong decay width of the $D^*_{s0}(2317)$ is discussed. 
With a summary and outlook the paper is closed.


\newpage

\section{The chiral Lagrangian with open-charm meson fields}
\label{sec:2}

We recall the chiral Lagrangian formulated in the presence of two anti-triplets of $D$ mesons with $J^P =0^-$ and  $J^P =1^-$ 
quantum numbers \cite{Yan:1992gz,Casalbuoni:1996pg}. In the relativistic version the Lagrangian was 
developed in \cite{Kolomeitsev:2003ac,Hofmann:2003je,Lutz:2007sk}.  
The kinetic terms read
\begin{eqnarray}
&& \mathcal{L}_{\mathrm{kin}}=(\CD_\mu D)(\CD^\mu \bar D)-  M^2\, D \, \bar D
-(\CD_\mu D^{\mu\alpha})(\CD^\nu \bar D_{\nu\alpha})+\frac{1}{2}\, (M + \Delta)^2\,D^{\mu\alpha} \,\bar D_{\mu\alpha}
\nonumber\\
&& \qquad -\, f^2\,{\tr }\big\{ U_\mu\,U^\mu \big\}+\frac{1}{2}\,f^2\,{\tr } \big\{\chi_+ \big\}\,,
\label{def-kin}
\end{eqnarray}
where
\begin{eqnarray}
&& U_\mu = {\textstyle \frac{1}{2}}\,e^{-i\,\frac{\Phi}{2\,f}} \left(
    \partial_\mu \,e^{i\,\frac{\Phi}{f}} \right) e^{-i\,\frac{\Phi}{2\,f}} \,, \qquad \qquad 
    \Gamma_\mu ={\textstyle \frac{1}{2}}\,e^{-i\,\frac{\Phi}{2\,f}} \,\partial_\mu  \,e^{+i\,\frac{\Phi}{2\,f}}
+{\textstyle \frac{1}{2}}\, e^{+i\,\frac{\Phi}{2\,f}} \,\partial_\mu \,e^{-i\,\frac{\Phi}{2\,f}}\,,
\nonumber\\
&& \chi_\pm = {\textstyle \frac{1}{2}} \left(
e^{+i\,\frac{\Phi}{2\,f}} \,\chi_0 \,e^{+i\,\frac{\Phi}{2\,f}}
\pm e^{-i\,\frac{\Phi}{2\,f}} \,\chi_0 \,e^{-i\,\frac{\Phi}{2\,f}}
\right) \,, \qquad \chi_0 =2\,B_0\, {\rm diag} (m_u,m_d,m_s) \,,
\nonumber\\
&& \CD_\mu \bar D = \partial_\mu \, \bar D + \Gamma_\mu\,\bar D \,, \qquad \qquad \qquad \quad \;\; 
\CD_\mu D = \partial_\mu \,D  - D\,\Gamma_\mu \,.
\label{def-chi}
\end{eqnarray}
Following \cite{Lutz:2007sk} we represent the $1^-$ field in terms of an antisymmetric tensor field $D_{\mu \nu}$.  The covariant 
derivative $\CD_\mu$ involves the chiral connection $\Gamma_\mu$, the quark masses enter via the symmetry breaking fields $\chi_\pm$ and
the octet of the Goldstone boson fields is encoded into the $3\times3$ matrix $\Phi$. The parameter $f$ is the chiral limit value of the 
pion-decay constant. Finally, given our particular renormalization scheme, the parameters $M$ and $M + \Delta$ give the masses of the $D$ and $D^*$ mesons in that limit 
with $m_u=m_d =m_s =0$.

We continue with first order interaction terms
\begin{eqnarray}
{\mathcal L}^{(1)} &=& 2\,g_P\,\Big\{D_{\mu \nu}\,U^\mu\,(\CD^\nu \bar D)
 - (\CD^\nu D )\,U^\mu\,\bar D_{\mu \nu} \Big\}
\nonumber\\
&-& \frac{i}{2}\,\tilde g_P\,\epsilon^{\mu \nu \alpha \beta}\,\Big\{
D_{\mu \nu}\,U_\alpha \,
(\CD^\tau \bar D_{\tau \beta} )
+ (\CD^\tau D_{\tau \beta})\,U_\alpha\,\bar D_{\mu \nu}) \Big\} \,,
\label{def-gP}
\end{eqnarray}
which upon an expansion in powers of the Goldstone boson fields provide the 3-point coupling constants of the Goldstone bosons to
the $D$ mesons. While the decay of the charged $D^*$-mesons \cite{Lutz:2007sk} implies
\begin{eqnarray}
|g_P| = 0.57 \pm 0.07 \,,
\label{value-gP}
\end{eqnarray}
the parameter $\tilde g_P$ in (\ref{def-gP}) can not be extracted from empirical data directly. The size of $\tilde g_P$
can be estimated using the heavy-quark symmetry of QCD \cite{Yan:1992gz,Casalbuoni:1996pg}. At leading order one expects
$\tilde g_P = g_P $.

Second order terms of the chiral Lagrangian were first studied in \cite{Hofmann:2003je,Lutz:2007sk}, where the focus was on counter 
terms relevant for s-wave scattering of Goldstone bosons with the $D$ mesons. A list of eight terms with dimension less 
parameters $c_i$ and $\tilde c_i$ was identified. This list was extended by further terms relevant for p-wave 
scattering in \cite{Guo:2008gp}. A complete collection of relevant terms is
\begin{eqnarray}
&&\mathcal{L}^{(2)}=-\big( 4\,c_0-2\,c_1\big)\, D \,\bar{D}  \,{\tr} \chi_+ -2\,c_1\,D \,\chi _+\,\bar{D}
\nonumber\\
&& \qquad \quad + \, 4\,\big(2\,c_2+c_3\big)\,D\bar{D}\,{\tr} \big(U_{\mu }\,U^{\mu \dagger }\big)- 4\,c_3\, D \,U_{\mu }\,U^{\mu \dagger }\,\bar{D}
\nonumber\\
&& \qquad \quad +\, \frac{1}{M^2}\,\big( 4\,c_4+2\,c_5\big)\, ({\CD_\mu } D) ({\CD_\nu }\bar{D}) \,{\tr} \big[ U^{\mu }, \,U^{\nu \dagger }\big]_+
-\frac{1}{M^2}\,2\,c_5\,({\CD_\mu } D) \big[ U^{\mu }, \,U^{\nu \dagger }\big]_+({\CD_\nu }\bar{D})
\nonumber\\
&& \qquad \quad +\,i\, c_6\,\epsilon ^{\mu \nu \rho \sigma }\Big(D \left[U_{\mu }, U_{\nu }^{\dagger }\right]_-\bar{D}_{\rho \sigma }-D_{\rho \sigma }\left[U_{\nu }^{\dagger }, U_{\mu }\right]_-\bar{D}\Big)
\nonumber\\
&& \qquad \quad +\,\big(2\,\tilde{c}_0-\tilde{c}_1\big)\,D^{\mu \nu }\,\bar{D}_{\mu \nu }\,{\tr}\chi _+
 +\tilde{c}_1\,D^{\mu \nu }\,\chi _+\,\bar{D}_{\mu \nu }
\nonumber\\
&&\qquad \quad - \big( 4\,\tilde{c}_2+2\,\tilde{c}_3\big)\,D^{\alpha \beta }\bar{D}_{\alpha \beta }\,{\tr}\big(U_{\mu }U^{\mu \dagger }\big)
+2\,\tilde{c}_3\,D^{\alpha \beta }\,U_{\mu }\,U^{\mu \dagger }\,\bar{D}_{\alpha \beta }
\nonumber\\
&& \qquad \quad -\,\frac{1}{(M +\Delta)^2}\,\big(2\,\tilde{c}_4+\tilde{c}_5\big)\, ({\CD_\mu }D^{\alpha \beta })\,({\CD_\nu }\bar{D}_{\alpha \beta } )\,{\tr} \big[U^{\mu }, \,U^{\nu \dagger }\big]_+
\nonumber\\
&&\qquad \quad + \, \frac{1}{(M + \Delta)^2}\, \tilde{c}_5 \,({\CD_\mu } D^{\alpha \beta })\,\big[U^{\mu },\, U^{\nu \dagger }\big]_+ ({\CD_\nu }\bar{D}_{\alpha \beta })
-4\,\tilde{c}_6\,D^{\mu \alpha }\,\big[U_{\mu },\, U^{\nu \dagger }\big]_-\bar{D}_{\nu \alpha } \,,
\end{eqnarray}
where the parameter $M$  and $M+ \Delta $ are the  $D$ and $D^*$ meson masses as evaluated at $m_u=m_d =m_s=0$. In  the limit of a very large charm-quark mass it follows 
$M \rightarrow \infty $  but $ \Delta \to 0 $. All parameters $c_i$ and $\tilde c_i$ are expected to scale linearly in the parameter $M_0$. As 
illustrated in \cite{Lutz:2007sk} it holds $\tilde c_i = c_i$  in the heavy-quark mass limit. 

A first estimate of some parameters can be found in \cite{Lutz:2007sk} based on large-$N_c$ arguments. Since at leading order 
in a $1/N_c$ expansion single-flavour trace interactions are dominant, the corresponding couplings should go to zero in the
$N_c\rightarrow \infty $ limit, suggesting
\begin{eqnarray}
&& c_0 \simeq \frac{c_1}{2} \,, \qquad \qquad c_2 \simeq -\frac{c_3}{2}\,, \qquad \qquad c_4 \simeq -\frac{c_5}{2}\,,
\nonumber\\
&& \tilde c_0 \simeq \frac{\tilde c_1}{2} \,, \qquad \qquad \tilde c_2 \simeq -\frac{\tilde c_3}{2} \,,\qquad \qquad 
\tilde c_4 \simeq -\frac{\tilde c_5}{2}\,.
\label{large-Nc}
\end{eqnarray}
In the combined heavy-quark and large-$N_c$ limit we are left with 4 free parameter only, $c_1,c_3, c_5, c_6$. 
For two of them  approximate ranges
\begin{eqnarray}
c_1  \simeq 0.44 - 0.47\,,\qquad \qquad \qquad 
c_3 + c_5 \simeq 1.0 -1.4  \,,
\label{estimate-cn}
\end{eqnarray}
were obtained previously in \cite{Lutz:2007sk}. While the parameter $c_1$ can be estimated from the $D$ meson masses, the parameter $c_3$ 
is constrained by the empirical $\pi D$ invariant mass spectrum \cite{Hofmann:2003je,Lutz:2007sk}. A complementary estimate was 
explored in \cite{Liu:2012zya}, where the parameter $c_3+ c_5$ was adjusted to first QCD lattice computations for s-wave scattering lengths of the 
Goldstone bosons with the $D$ mesons. It is remarkable that their range for $c_3 +c_5\simeq 1$ is quite consistent with the earlier estimates 
\cite{Hofmann:2003je,Lutz:2007sk} based on the empirical $\pi D$ invariant mass spectrum. The $c_3$  parameter 
is of crucial importance for the physics of two exotic sextets of $J^P =0^+$ and $J^P = 1^+$  resonances. Such multiplets are predicted 
by the leading order chiral Lagrangian (\ref{def-kin}), which entails in particular the Tomozawa-Weinberg coupled-channel 
interactions of the Goldstone bosons with the $D$ mesons  \cite{Kolomeitsev:2003ac}. The latter predicts weak attraction in the 
flavour sextet channel. If used as the driving term in a coupled-channel unitarization exotic signals appear.  
A reliable estimate of the correction terms proportional to $c_3$ and $\tilde c_3$ is important in order to arrive at a detailed 
picture of this exotic sector of QCD \cite{Hofmann:2003je,Lutz:2007sk}.

We close this section with a first construction of the symmetry breaking counter terms proportional to the product of two quark masses:
\begin{eqnarray}
&& \mathcal{L}^{(4)}=-{d_1}\,D \,\chi _+^2\,\bar{D}
-d_2\,D\, \chi _+\bar{D}\, {\tr } \left(\chi _+\right)-d_3\,D\,\bar{D}\, {\tr } \big(\chi _+^2\big)-d_4\, D\,\bar{D}\,\big(\tr \chi _+\big)^2
\nonumber\\
&& \qquad  + \,\frac{1}{2}\,\tilde{d}_1\,D^{\mu \nu }\,\chi _+^2\,\bar{D}_{\mu \nu }+\frac{1}{2}\,\tilde{d}_2\, D^{\mu \nu } \,\chi _+\bar{D}_{\mu \nu } \tr\big(\chi _+\big)
+\frac{1}{2}\,\tilde{d}_3\,D^{\mu \nu }\,\bar{D}_{\mu \nu } \tr \,\big(\chi _+^2\big)
\nonumber\\
&& \qquad  + \,\frac{1}{2}\,\tilde{d}_4\,D^{\mu \nu }\,\bar{D}_{\mu \nu }\big(\tr \chi _+\big)^2 \,.
\end{eqnarray}
Such terms are relevant in the chiral extrapolation of the $D$ meson masses. For the pseudo-scalar mesons we provide the 
tree-level contributions to the polarization $\Pi_H^{(2)}$  and $\Pi_H^{(4-\chi)}$ of the $D$ and $D_s$ mesons. We use a convention with
\begin{eqnarray}
&&M_{H \in \left[0^-\right]}^2 = M^2 + \Pi_H^{(2)} + \Pi_H^{(4-\chi)} + \cdots\,, \qquad  \; 
 M_{H \in \left[1^-\right]}^2 = (M + \Delta)^2 + \Pi_H^{(2)} + \Pi_H^{(4-\chi)} + \cdots\,,
\nonumber\\ \nonumber\\
&&\Pi^{(2)}_D   =  2 \, B_0 \big(4\,c_0-2\,c_1\big) \big(m_s+2\,m\big)+4 \, B_0 \,c_1 \,m \,,
\nonumber\\
&& \Pi^{(4-\chi)}_D = 4\, B_0^2\, \big(d_1+2\,d_2+2\,d_3+4\,d_4\big) m^2+4\, B_0^2 \,\big(d_3+d_4\big) m_s^2+ {4 \, B_0^2 }\big(d_2+4\,d_4\big) \,m \,m_s \,,
\nonumber\\
&& \Pi^{(2)}_{D_s}  =  2 \, B_0\, \big(4\,c_0-2\,c_1\big) \big(m_s+2\,m\big)+4 \, B_0 \,c_1 \,m_s \,,
\nonumber\\
&& \Pi^{(4-\chi)}_{D_s} = 4 \, B_0^2 \,\big(2\,d_3+4\,d_4\big) m^2+{4 \, B_0^2 }\,\big(d_1+d_2+d_3+d_4\big) \,m_s^2+{4 \, B_0^2 }\big(2\,d_2+4\,d_4\big)\, m\, m_s \,,
\label{res-tree-level}
\end{eqnarray}
where we consider the isospin limit with $m_u = m_d = m$. 
Analogous expressions hold for the vector mesons polarization $\Pi^{(2)}_{H \in \left[1^-\right]}$ and 
$\Pi^{(4-\chi)}_{H \in \left[1^-\right]}$ where the 
replacements  $c_i \to \tilde c_i$ and $d_i \to \tilde d_i$ are 
to be applied to (\ref{res-tree-level}). 
With $\tilde c_i = c_i$ and $\tilde d_i = d_i$ and $\Delta \to 0$ the heavy-quark spin symmetry is recovered exactly. 

We need to mention a technical issue. The propagator $S^{\alpha \beta}_{\mu \nu}(p)$ 
of our $1^-$ fields involves four Lorentz indices, which are pairwise antisymmetric. Either interchanging 
$\alpha \leftrightarrow \beta$ or $\mu \leftrightarrow \nu$ generates a change in sign. A mass renormalization from a loop contribution 
arises from a particular projection $\Pi(p^2)$ of the polarization tensor $\Pi^{ \mu \nu}_{\alpha \beta}(p)$ with
\begin{eqnarray}
&& \Pi (p^2) =  
\frac{-1}{(d-1)\,p^2}\,\big( g_{\mu \alpha }\,p_{\nu }\,p_{\beta }-g_{\mu \beta }\,p_{\nu }\,p_{\alpha }
-g_{\nu \alpha }\,p_{\mu }\,p_{\beta }+g_{\nu \beta }\,p_{\mu }\,p_{\alpha }\big)\,\Pi^{ \mu \nu, \,\alpha \beta}(p)\,,
\end{eqnarray}
where $d$ is the space-time dimension. This is the part which is used in (\ref{res-tree-level}) and will be used also in the following.

\newpage

\section{One-loop mass corrections in a finite box}
\label{sec:2}

The chiral Lagrangian of section \ref{sec:1} is used to compute the D-meson masses at the one-loop level. In order to prepare 
for a comparison of QCD lattice data this computation is done in a finite box of volume $V$.  
A direct application of the relativistic 
chiral Lagrangian in the conventional $\overline{MS}$ scheme does lead to a plethora of power-counting violating contributions. 
There are various ways to arrive at results that are consistent with the expectations of power counting 
rules \cite{Ellis:1997kc,Becher:1999he,Gegelia:1999gf,Semke:2005sn}.

We follow here the $\chi$-MS approach developed previously for the chiral dynamics of baryons \cite{Lutz:1999yr,Lutz:2001yb,Semke:2005sn}, 
which is based on the Passarino-Veltman reduction scheme \cite{Passarino:1978jh}. Recently this scheme was generalized for computations 
in a finite box \cite{Lutz:2014oxa}. This implies that all finite box effects are exclusively determined by the volume dependence of a set of universal scalar 
loop functions as discussed and presented in \cite{Lutz:2014oxa}. Our results will be expressed in terms 
of Clebsch coefficients $G_{QR}^{(H)}$ and  $G_{H Q}^{(\chi) }$, $G_{H Q}^{(S)}$, $G_{H Q}^{(V)}$ and a set of generic loop functions. 
While the index $H$ or $R$ runs either over the triplet of pseudo-scalar or vector $D$ mesons the index $Q$ runs over the octet of 
Goldstone bosons (see Tab. \ref{Cf_Q3} and Tab. \ref{Cf_Q4}). In our case there will 
be two tadpole integrals $\bar I^{(0)}_Q$ and $\bar I^{(2)}_Q$ from the Goldstone bosons and the scalar bubble-loop 
integral $\bar I_{QR}$.  
In addition there may be tadpole contributions $\bar I^{(n)}_R$ involving an intermediate $D$ meson. In order to render the power counting 
manifest it suffices to supplement the Passarino-Veltman reduction scheme 
by a minimal and universal subtraction scheme \cite{Semke:2005sn}: 
\begin{itemize}
\item{} any tadpole integral  involving a heavy particle is dropped
\item{} the scalar bubble-loop integral requires a single subtraction.
\end{itemize}
The required loop functions have been used and detailed in a previous work \cite{Semke:2005sn,Lutz:2014oxa,Lutz:2014oxa} for finite box 
computations. For the readers' convenience we recall the loop functions 
in the infinite box limit \cite{Semke:2005sn,Lutz:2014oxa} with 
\allowdisplaybreaks[1]
\begin{eqnarray}
&& \bar I^{(0)}_Q =\bar I_Q =\frac{m_Q^2}{(4\,\pi)^2}\,
\log \left( \frac{m_Q^2}{\mu^2}\right) \,, \qquad \qquad \qquad \bar I^{(2)}_Q = \frac{1}{4}\,m_Q^2\,\bar I_Q \,,  
\nonumber\\
&& \bar I_{Q R}=\frac{1}{16\,\pi^2}
\left\{ \gamma^H_{R} - \left(\frac{1}{2} + \frac{m_Q^2-M_R^2}{2\,M_H^2}
\right)
\,\log \left( \frac{m_Q^2}{M_R^2}\right)
\right.
\nonumber\\
&& +\left.
\frac{p_{Q R}}{M_H}\,
\left( \log \left(1-\frac{M_H^2-2\,p_{Q R}\,M_H}{m_Q^2+M_R^2} \right)
-\log \left(1-\frac{M_H^2+2\,p_{Q R}\,M_H}{m_Q^2+M_R^2} \right)\right)
\right\}\;,
\nonumber\\
&& \qquad  {\rm with} \qquad \qquad p_{Q R}^2 =
\frac{M_H^2}{4}-\frac{M_R^2+m_Q^2}{2}+\frac{(M_R^2-m_Q^2)^2}{4\,M_H^2}  \,,
\label{def-master-loop}
\end{eqnarray}
where we note that in the infinite volume limit the two tadpole integrals $\bar I^{(0)}_Q$ and $\bar I^{(2)}_Q$  turn dependent 
and can no longer be discriminated in that case. The finite volume corrections for  $\bar I^{(0)}_Q, \bar I^{(2)}_Q$ and $\bar I_{Q R}$ are detailed in 
\cite{Lutz:2014oxa}.

\begin{table}[b]
\setlength{\tabcolsep}{4.5mm}
\renewcommand{\arraystretch}{1.}
\begin{tabular}{ll|ll}
\hline
 $G_{\pi  D^*}^{(D)}=2\,\sqrt{3}\,g_P$ & $G_{K D^*}^{\left(D_s\right)}=4\,g_P$ &   &   \\
 $G_{\eta  D^*}^{(D)}=\frac{2}{\sqrt{3}}\,g_P$ & $G_{\eta  D_s^*}^{\left(D_s\right)}=\frac{4}{\sqrt{3}}\,g_P$ &  &   \\
 $G_{\bar{K} D_s^*}^{(D)}=2\,\sqrt{2}\,g_P$ &   & &   \\ \hline 
 $G_{\pi  D}^{\left(D^*\right)}=2\,\sqrt{3}\,g_P$ & $G_{K D}^{\left(D_s^*\right)}=4\,g_P$ & $G_{\pi  D^*}^{\left(D^*\right)}=2\,\sqrt{3}\,\tilde{g}_P$ & $G_{K D^*}^{\left(D_s^*\right)}=4\,\tilde{g}_P$ \\
 $G_{\eta  D}^{\left(D^*\right)}=\frac{2}{\sqrt{3}}\,g_P$ & $G_{\eta  D_s}^{\left(D_s^*\right)}=\frac{4}{\sqrt{3}}\,g_P$ & $G_{\eta  D^*}^{\left(D^*\right)}=\frac{2}{\sqrt{3}}\,\tilde{g}_P$ & $G_{\eta  D_s^*}^{\left(D_s^*\right)}=\frac{4}{\sqrt{3}}\,\tilde{g}_P$ \\
 $G_{\bar{K} D_s}^{\left(D^*\right)}=2\,\sqrt{2}\,g_P$ &  & $G_{\bar{K} D_s^*}^{\left(D^*\right)}=2\,\sqrt{2}\,\tilde{g}_P$ & \\[4pt]  
 \hline
\end{tabular}
\caption{Coefficients $G_{QR}^{(H)}$}
\label{Cf_Q3}
\end{table}

We point at the presence of the additional subtraction term $\gamma^H_{R}= \gamma^H_{R}(M, \Delta)$  
with
\begin{eqnarray}
 \gamma^H_{R} = -  \lim_{m, m_s\to 0}\,\frac{M_R^2-M_H^2}{M_H^2}\,\log \left|\frac{M_R^2-M_H^2}{M_R^2}\right| \,,
 \label{def-gammaHR}
\end{eqnarray}
as suggested recently in \cite{Lutz:2014oxa} in the analogous case of a baryon self-energy computation. 
The subtraction term depends on the chiral limit values $M$ and $M+ \Delta$ of the $D$ and $D^*$ meson masses only.
It was not yet imposed in earlier 
computations \cite{Semke:2005sn,Semke:2011ez,Lutz:2014oxa}. As was discussed in \cite{Lutz:2014oxa} the request of such a term
comes from a study of the chiral regime with 
\begin{eqnarray}
m_Q \ll \Delta  \qquad \qquad {\rm with }\qquad Q \in\{\pi, K ,\eta \}\,.
\label{def-chiral-regime}
\end{eqnarray}
Within  a counting scheme with $m_Q \sim \Delta \sim Q$ there is no need 
for any additional subtractions beyond the ones enforced by the $\chi$-MS approach. However, to arrive at consistent results 
for $m_Q \ll \Delta$ this subtraction is instrumental. While for $\Delta \sim m_Q \sim Q$ and $\gamma^H_{R} =0$ the scalar bubble 
scales with $\bar I_{QR} \sim Q $ as expected from dimensional analysis, in the chiral regime with $m_Q \ll \Delta$ and $m_Q \sim Q$ 
one would expect $\bar I_{QR} \sim Q^2 \sim m_Q^2$. This expectation turns true only, 
for $\gamma^H_{R} \neq 0 $ as chosen in (\ref{def-gammaHR}).

We are now well prepared to collect all contributions to the D-meson self energies at the one-loop level.
Consider the bubble loop and tadpole contributions. The Passarino-Veltman reduction scheme in combination with the $\chi$-MS approach leads to the following 
expressions 
\allowdisplaybreaks[1]
\begin{eqnarray}
&& \Pi _{H\in \left[0^-\right]}^{\rm bubble }= \sum _{Q\in [8]} \sum _{R\in \left[1^-\right]} 
\Bigg(\frac{G_{QR}^{(H)}}{2\,f}\Bigg)^2 
\Bigg\{-\frac{1}{4}\,\Big(M_H^2-M_R^2+m_Q^2\Big) \,\bar I_Q  - M_H^2 \,p^2_{QR}\, 
\bar I_{Q R} \Bigg\} \,,
\label{result-loop-0_full}\\ 
&& \Pi _{H\in \left[0^-\right]}^{\rm tadpole }=  \frac{1}{4\,f^2}\sum _{Q\in [8]}\,\Big( G_{H Q}^{(\chi) }\,\bar I_Q-G_{H Q}^{(S)} \,m_Q^2\,\bar I_Q
- G_{H Q}^{(V)}\, M^2\, \bar I_Q^{(2)}  \Big) \,,
\label{result-tadpole-0}\\  
&& \Pi _{H\in \left[1^-\right]}^{\rm bubble }= \sum _{Q\in [8]} \sum _{R\in \left[0^-\right]} 
\Bigg(\frac{G_{QR}^{(H)}}{2\,f}\Bigg)^2 
\Bigg\{-\frac{1}{12}\,\Big(M_H^2-M_R^2+m_Q^2\Big) \,\bar I_Q  - \frac{1}{3}\,M_H^2 \,p^2_{QR}\,
\bar I_{Q R} \Bigg\}
\nonumber\\
&& \qquad \quad \;\,+\,
\sum _{Q\in [8]} \sum _{R\in \left[1^-\right]} 
\Bigg(\frac{G_{QR}^{(H)}}{2\,f}\Bigg)^2 
\Bigg\{ \frac{M_H^2+2\,M_R^2}{12\,M_R^2}\,\bar I^{(2)}_Q- \frac{(M_H^2+ M_R^2)^2}{6\,M_R^2} \,p^2_{QR}\,
\bar I_{Q R}
\nonumber\\
&& \qquad \qquad  -\,\Bigg( \frac{(M_H^2-M_R^2)\,(M_H^2+ M_R^2)^2}{24\,\,M_H^2\,M_R^2} 
+ \frac{M_R^4+ 6\,M_R^2\,M_H^2-3\,M_H^4}{24\,M_H^2\,M_R^2}\, m_Q^2  \Bigg) \,\bar I_Q   \Bigg\} \,,
\label{result-loop-1_full} \\
&& \Pi _{H\in \left[1^-\right]}^{\rm tadpole }= \frac{1}{4\,f^2}\sum _{Q\in [8]}\,\Big( G_{H Q}^{(\chi) }\,\bar I_Q-G_{H Q}^{(S)} \,m_Q^2\,\bar I_Q
- G_{H Q}^{(V)}\, (M+ \Delta)^2\, \bar I_Q^{(2)}  \Big) \,,
\label{result-tadpole-1}
\end{eqnarray}
where the loop functions are expressed in terms of physical meson masses. The sums in (\ref{result-loop-0_full}, \ref{result-loop-1_full}) extend over intermediate Goldstone bosons ($Q$) and pseudo-scalar or vector $D$ 
mesons ($R$) with either $R \in[0^-]$ or $R \in [1^-]$. The Clebsch coefficients $G_{QR}^{(H)}$ are specified in Tab. \ref{Cf_Q3}. 
In the contributions from the tadpole diagrams
the sums in (\ref{result-tadpole-0}, \ref{result-tadpole-1}) extend over the intermediate Goldstone bosons $Q$. The coefficients $G_{H Q}^{(\chi) }$, $G_{H Q}^{(S)}$, $G_{H Q}^{(V)}$ are listed in 
Tab. \ref{Cf_Q4}.

\begin{table}[t]
\begin{tabular}{cc|c |c|c}
\hline
 $H$ & $Q$ & $G_{H Q}^{(\chi) }/B_0$ & $G_{H Q}^{(S)}$ & $M^2\,G_{H Q}^{(V)}$ \\
 \hline
 $D$ & $ \pi$  & $-48\,\big(2\,c_0-c_1\big)\,m-24\,c_1\,m$ & $24\,\big(2\,c_2+c_3\big)-12\,c_3$ & $24 \,(2\,c_4+c_5)-12 \,c_5$ \\
  & $ K$ & $-32\,\big(2\,c_0-c_1\big)\,\big(m_s+m\big)-8\,c_1\,\big(m_s+m\big)$ & $32\,\big(2\,c_2+c_3\big)-8\,c_3$ & $32\, (2\,c_4+c_5)-8 \,c_5$ \\
  & $ \eta$  & $-\frac{16}{3} \,\big(2\,c_0-c_1\big)\,\big(2\, m_s+m\big)-\frac{8}{3}\,c_1\,m$ & $8\,\big(2\,c_2+c_3\big)-\frac{4}{3}\,c_3$ & $8 \,(2\,c_4+c_5)-\frac{4}{3} \,c_5$ \\
 $D_s$ & $ \pi$  & $-48\,\big(2\,c_0-c_1\big)\,m$ & $24\,\big(2\,c_2+c_3\big)$ & $24 \,(2\,c_4+c_5 )$ \\
  & $ K$ & $-32\,\big(2\,c_0-c_1\big)\,\big(m_s+m\big)-16 \,c_1\,\big(m_s+m\big)$ & $32\,\big(2\,c_2+c_3\big)-16 \,c_3$ & $32\, (2\,c_4+c_5 )-16 \, c_5$ \\
  & $ \eta$ & $-\frac{16}{3} \,\big(2\,c_0-c_1\big)\big(2\,m_s+m\big)-\frac{32 }{3}\,c_1\,m_s$ & $8\,\big(2\,c_2+c_3\big)-\frac{16}{3}\,c_3$
 & $8 (2\,c_4+c_5)-\frac{16}{3} \,c_5$\\ \hline
\end{tabular}
\caption{Coefficients $G_{H Q}^{(\chi) }$, $G_{H Q}^{(S)}$ and $G_{H Q}^{(V)}$. The corresponding results for the $D^*$ and $D^*_s$ follow
by the replacement $c_i \to \tilde c_i$ and $M \to M+\Delta$. }
\label{Cf_Q4}
\end{table}

The results (\ref{result-loop-0_full}, \ref{result-loop-1_full}) deserve a detailed discussion. First let us emphasize that a chiral expansion of the loop function as they are given confirms  
the leading chiral power as expected from dimensional counting rules. All power-counting violating contributions are subtracted owing to the $\chi$-MS approach. Here we adopted the conventional counting rules
\begin{eqnarray}
m_Q \sim Q \quad {\rm and} \qquad  M_{1^-} -M_{0^-} \sim \Delta \sim Q \,,
\end{eqnarray}
which is expected to be effective for $\Delta \sim m_Q$. Our results (\ref{result-loop-0_full}, \ref{result-loop-1_full}) are model dependent, as there are various subtraction schemes available 
to obtain loop expression that are compatible with dimensional counting rules. Most prominently there is the infrared regularization of  Becher and Leutwyler \cite{Becher:1999he} and 
the minimal subtraction scheme proposed by Gegelia and collaborators \cite{Gegelia:1999gf}. Following our previous work on the chiral extrapolation of the baryon masses we 
will attempt to extract a model independent part of such loop expressions. This goes in a few consecutive steps. The driving strategy behind this attempt is to keep the physical masses inside 
the loop function.

Consider first the terms that are proportional to the tadpole loop function $\bar I_Q$. There are two distinct classes of terms. The coefficient in front of any $\bar I_Q $ is either proportional to $m_Q^2 $ 
or to $M^2_H-M_R^2 $. The terms proportional to $m_Q^2\,\bar I_Q$ or also to $\bar I^{(2)}_Q$ in (\ref{result-loop-0_full}, \ref{result-loop-1_full}) have the same form as the corresponding structures 
in (\ref{result-tadpole-0}, \ref{result-tadpole-1}) and therefore 
renormalize the low-energy parameters $c_n$ and $\tilde c_n$ with 
\allowdisplaybreaks[1]
\begin{eqnarray}
&& c^{\,r}_2=c_2 + \frac{1}{8} \,g_P^2 \,, \qquad \qquad c^{\,r}_4 = c_4 \,,\qquad \qquad \tilde{c}^{\,r}_4 =\tilde{c}_4-\frac{1}{8}\,\tilde{g}_P^2\,,
\nonumber\\
&& c^{\,r}_3 =c_3 - \frac{1}{4} \,g_P^2\,, \qquad  \qquad c^{\,r}_5 = c_5 \,,\qquad \qquad \tilde{c}^{\,r}_5 =\tilde{c}_5+\frac{1}{4}\,\tilde{g}_P^2\,,
\nonumber\\
&& \tilde{c}^{\,r}_2 =\tilde{c}_2 + \frac{1}{12}\,\tilde{g}_P^2 + \frac{1}{24} \,g_P^2\, , 
\nonumber\\
&& \tilde{c}^{\,r}_3 = \tilde{c}_3 -\frac{1}{6}\,\tilde{g}_P^2 - \frac{1}{12} \,g_P^2\,.
\end{eqnarray}
We conclude that the terms proportional to $m_Q^2\,\bar I_Q$ or $\bar I^{(2)}_Q$ in (\ref{result-loop-0_full}, \ref{result-loop-1_full}) may be dropped if we use the renormalized low-energy parameters 
$c^{\,r}_n$ and $\tilde c^{\,r}_n$ in the tadpole contributions  (\ref{result-tadpole-0}, \ref{result-tadpole-1}) but also in (\ref{res-Gamma-dn}). Note, however, that by doing so some higher order terms proportional to 
\begin{eqnarray}
\left( 1 -\frac{M^2_R}{M_H^2}\right)^n\,m_Q^2\,\bar I_Q \;\to \; 0 \,,
\end{eqnarray}
with $n \geq 1$ are neglected in $\Pi_{H\in [1^-]} $. We argue that the latter terms would cause a renormalization scale dependence that can not be absorbed into the available counter terms at the considered accuracy level. 
In order to avoid a model dependence such terms should be dropped. 

We are left with the terms proportional to $(M_R^2-M^2_H)\,\bar I_Q$. If the charm meson masses are decomposed into their chiral moments 
the leading renormalization scale dependence of such terms can be absorbed into the $Q^2$ counter terms $c_{0,1}$ and $\tilde c_{0,1}$. Similarly the components of order $Q^4$ can 
be matched with counter terms $d_{n}$ and $\tilde d_{n}$.
Most troublesome, however, are the subleading contributions proportional to $m_{Q}^5\,\log \mu$ in such a  strict chiral expansion of the vector $D$ meson masses. There is no counter term 
available to remove such a scale dependence. In fact, only within a two-loop computation this issue is resolved in a conventional approach. 
Instead we keep the charm meson masses unexpanded in the terms $(M_R^2-M^2_H)\,\bar I_Q$ and follow the strategy proposed in \cite{Lutz:2018a}.  For those terms 
we provide the following decomposition
\begin{eqnarray}
(M_R^2 - M_H^2) \,\bar I_Q =\Big( M_R^2 - M_H^2\Big)\underbrace{ \frac{m_Q^2}{(4 \pi)^2} \, \log \frac{m_Q^2}{M_R^2}}_{= \, \bar I_Q|_{\mu = M_R}} + \frac{M_R^2 - M_H^2 }{M_R^2} \,m_Q^2 \,\underbrace{\bar I_R}_{\to \,0} \, ,
\end{eqnarray}
where the second term depending on the heavy-meson tadpole $\bar I_R$ can be systematically dropped without harming the chiral Ward identities. We end up with the following renormalized bubble-loop 
expressions 
\allowdisplaybreaks[1]
\begin{eqnarray}
&&\bar \Pi _{H\in [0^-]}^{\rm bubble}= \sum _{Q\in [8]} \sum _{R\in \left[1^-\right]} \left(\frac{G_{QR}^{(H)}}{2f}\right)^2 \Bigg\{\alpha_{QR}^H\, -M_H^2\,p_{QR}^2 \bar{I}_{Q R}  
+ \frac{1 }{4} \,\Big( M_R^2 - M_H^2\Big)\,\frac{m_Q^2}{(4 \pi)^2} \, \log \frac{m_Q^2}{M_R^2}  \Bigg\}
\nonumber\\
&& \bar \Pi _{H\in [1^-]}^{\rm bubble} = \sum _{Q\in [8]} \sum _{R\in \left[0^-\right]} \left(\frac{G_{QR}^{(H)}}{2\,f}\right)^2\Bigg\{ \frac{1}{3}\,\alpha_{QR}^H
-\frac{1}{3}\,M_H^2\,p_{QR}^2 \,\bar{I}_{Q R}
  \nonumber\\
&& \qquad  \qquad \qquad  \qquad \qquad  \qquad \qquad  + \,\frac{1 }{12 } \,\Big(M_R^2 - M_H^2\Big)\,\frac{m_Q^2}{(4 \pi)^2} \, \log \frac{m_Q^2}{M_R^2}
  \Bigg\}
 \nonumber\\
&& \qquad \quad \;\, + \sum _{Q\in [8]} \sum _{R\in \left[1^-\right]} \left(\frac{G_{QR}^{(H)}}{2\,f}\right)^2
\Bigg\{-\frac{ \left(M_H^2+M_R^2\right)^2}{6 \,M_R^2} \, p_{QR}^2\, \bar{I}_{Q R} 
  \nonumber\\
&& \qquad  \qquad \qquad  \qquad \qquad  \qquad \qquad  + \,\frac{ (M_H^2 + M_R^2)^2}{24 \,M_H^2\, M_R^2}  \,\Big( M_R^2 - M_H^2\Big)\,\frac{m_Q^2}{(4 \pi)^2} \, \log \frac{m_Q^2}{M_R^2}\, \Bigg\}\,,
\label{result-loop}
\end{eqnarray}
which will be the basis for our following studies. Note yet the additional subtraction terms $\alpha^{H}_{QR}$ in (\ref{result-loop}).  
Such terms were suggested in \cite{Lutz:2018a} for the analogous case of a baryon self-energy computation.
In order to arrive at consistent results for $m_Q \ll \Delta$ the terms $\alpha^{H}_{QR}$ are instrumental:
\begin{eqnarray}
&& \alpha^{H\in\,[0^-]}_{QR}\, = \frac{\alpha_1\,\Delta^2}{32\,\pi^2} \Bigg\{ 
 \Big( M^2_H - M^2 \Big)\, \Big( \frac{\Delta\,\partial}{\partial\,M} -\frac{\Delta\,\partial}{\partial\,\Delta} -
 \frac{M+ \Delta}{M} \Big)
\nonumber\\
&& \qquad \qquad +\, \Big( M^2_R - (M +\Delta)^2  \Big)\, \frac{M}{M+\Delta}\, \Big( \frac{\Delta\,\partial}{\partial\,\Delta} 
+ 1 \Big) \, \Bigg\}\,\gamma_1 
   +\, \frac{\Delta\, M\,m_Q^2}{16\,\pi^2}\,\alpha_1\,\gamma_2 \,, 
\nonumber\\
&& \alpha^{H\in[1^-]}_{QR} = \frac{\tilde \alpha_1\,\Delta^2}{32\,\pi^2} \Bigg\{ 
\Big( M^2_H - (M + \Delta)^2\Big)\, \frac{M}{M+\Delta} \, \Big( \frac{\Delta\,\partial}{\partial\,\Delta} + 1\Big)
\nonumber\\
&& \qquad \qquad +\,\Big( M^2_R - M^2 \Big)\, \Big( \frac{\Delta\,\partial}{\partial\,M} - \frac{\Delta\,\partial}{\partial\,\Delta} -
 \frac{M+ \Delta}{M}\Big)
\Bigg\}\,\tilde \gamma_1 
 +\, \frac{\Delta\,M\,m_Q^2}{16\,\pi^2} \,\tilde \alpha_1\,\tilde \gamma_2  \,,
\label{def-alphaBR}
\end{eqnarray}
where the functions $\alpha_i, \tilde \alpha_i$ and $\gamma_i, \tilde \gamma_i$ depend on the ratio $\Delta/M$ only. They are listed in Appendix A 
and Appendix B. While the rational functions $\alpha_i$ and $\tilde \alpha_i$ all approach the numerical value one in the limit $\Delta \to 0$, 
the functions $\gamma_i$ and $ \tilde \gamma_i$ show a logarithmic divergence  in that limit. 
We summarize the convenient implications of our subtraction scheme
\begin{itemize}
\item
the chiral limit values of the $D$ meson masses are not renormalized 
\item
the low-energy parameters $c_{0,1}$ and $\tilde c_{0,1}$ are not renormalized 
\item
the wave-function factor of the $D$ mesons are not renormalized in the chiral limit 
\end{itemize}

We close this section with a brief discussion on the role of the renormalization scale $\mu$. Given our scheme a scale dependence arises from the tadpole terms only.
Such terms need to be considered in combination with the tree-level contribution $\Pi_H^{(4-\chi)}$. This leads to the condition
\begin{eqnarray}
&& \qquad \qquad \qquad \mu^2 \,\frac{d}{d\,\mu^2}\,d_i = -\frac{1}{4}\,\frac{\Gamma_{d_i}}{(4\,\pi\,f)^2}\,,
\nonumber\\
&&  \Gamma_{d_1}= \frac{1}{6}\, \big(4\, c_1+12 \,c_3+3\, c_5\big)\,,\qquad \qquad 
\Gamma_{d_2}= \frac{1}{9}\, \big(44 \,c_1-52 \,c_3-13\, c_5\big),\nonumber\\
&& \Gamma_{d_3}= \frac{1}{18}\, \big(240\, c_0-84 \,c_1+240 \,c_2+68\, c_3+60 \,c_4+17 \,c_5\big)\,,\nonumber\\
&& \Gamma_{d_4}= \frac{1}{27} \,\big(264\, c_0-132\, c_1+264 \,c_2+140\, c_3+66 \,c_4+35\, c_5\big)\,,
\label{res-Gamma-dn}
\end{eqnarray}
where identical results hold for the $\tilde c_i$ and $\tilde d_i$ coupling constants. However, it is evident that scale invariant results follow with (\ref{res-Gamma-dn}) only if 
the meson masses in the tadpole contributions are approximated by the leading order Gell-Mann Oakes Renner relations with $m_\pi^2= 2\,B_0\,m$ and $m_K^2= B_0\,(m+ m_s)$ for instance. 
This is unfortunate since we wish to use physical masses inside all loop contributions. Recalling our previous work \cite{Lutz:2018a} there may be an efficient remedy of this issue. 
Indeed the counter term contributions can be rewritten in terms of physical masses such that scale invariance follows without insisting on the Gell-Mann Oakes Renner relations for the 
meson masses. Such a rewrite is most economically achieved in terms of suitable linear combinations of the low-energy constants
\begin{eqnarray}
&& d^{c}_1 = - \frac{1}{23} (26 \,d_1 + 9 \,d_2)\,, \qquad \qquad \qquad
d^{c}_3 = \frac{1}{345}\,\Big( 43 \,d_1 + 60 \,d_2 + 69 \,d_3\Big)\,,
\nonumber\\
&& d^{c}_2 = \frac{1}{276}\Big(-132 \,d_1 + 18 \,d_2\Big) \,,\qquad \quad  \;\;
d^{c}_4   = \frac{1}{45}\,\Big(-11 \,d_1 + 15 \,d_2 - 33 \,d_3 + 45 \,d_4\Big) \,.
\label{def-dc}
\end{eqnarray}
With Tab. \ref{tab:zz1} our rewrite is specified in detail. We assure that replacing the meson masses in the table by their leading order expressions the original 
expressions as given in (\ref{res-tree-level}) are recovered identically. 
We note a particularity: at leading order the effects of $c_0$ in $G^{(\chi )}_{HQ}$ cannot be discriminated from $c_2$ in $G^{(S)}_{HQ}$. Scale invariance requires to consider 
the particular combinations $ c_2 + c_0 $ in $G^{(S)}_{HQ}$ and in turn use $c_0 = 0 $ in $G^{(\chi)}_{HQ}$.

\begin{table}[t]
\setlength{\tabcolsep}{1.5mm}
\renewcommand{\arraystretch}{1.}
\begin{center}
\begin{tabular}{c|cc}\hline
$\Pi_H^{(4-\chi)}$           & $H=D$\,                               & $H=D_s$                                \\  \hline \hline
$m_\pi^4$                    & $-9 \,d^{c}_2 + 18 \,d^{c}_3 $        & $-18 \,d^{c}_2 + 18 \,d^{c}_3 $        \\
$m_K^4$                      & $-18 \,d^{c}_2 + 24 \,d^{c}_3 $       & $ -12 \,d^{c}_2 + 24 \,d^{c}_3$        \\
$m_\eta^4$                   & $-5 \,d^{c}_2 + 6 \,d^{c}_3 $         & $ -2 \,d^{c}_2 + 6 \,d^{c}_3$          \\ \hline
$B_0\,m\,m_\pi^2$            & $ 9 \,d^{c}_1$                        & $ 18 \,d^{c}_1$                        \\ 
$B_0\,(m+m_s)\,m_K^2$        & $9 \,d^{c}_1 $                        & $ 6 \,d^{c}_1$                         \\
$B_0\,m\,m_\eta^2$           & $ d^{c}_1$                            & $ 2 \,d^{c}_1$                         \\ 
$B_0\,m_s\,m_\eta^2$         & $ 4 \,d^{c}_1$                        & $ 0$                                   \\ \hline
$B_0^2\,(2\,m+m_s)^2$        & $ 4 \,d^{c}_4$                        & $ 4 \,d^{c}_4$                                            
\end{tabular}
\caption{A rewrite of $\Pi_H^{(4-\chi)}$ in (\ref{res-tree-level}).  }
\label{tab:zz1}
\end{center}
\end{table}

\clearpage

\section{Self consistent summation approach}

The renormalized loop functions depend  on the 
physical masses of the $D$ mesons. In a conventional chiral expansion scheme the meson masses inside the loop would be expanded to a given order so that 
a self-consistency issue does not arise. This is fine as long as the expansion is rapidly converging. For a slowly converging system 
such a summation scheme is of advantage even though this may bring in some model dependence \cite{Semke:2005sn,Semke:2011ez,Lutz:2014oxa,Lutz:2014oxa,Lutz:2018a}.

Let us be specific on how the summation scheme is set up in detail. There is a subtle point emphasized recently in \cite{Lutz:2018a} which 
needs some discussions. The coupling constant $g_P$ was determined in \cite{Lutz:2007sk} from the pion-decay width of the $D^*$ meson 
using a tree-level decay amplitude. Alternatively the decay width can be extracted from the $D^*$ meson propagator in the presence of 
the one-loop polarization $\Pi^{\rm bubble}_{D^*}$. The latter has imaginary contributions proportional to same coupling constant 
$g_P^2$ that reflect the considered decay process. In the absence of wave-function renormalization effects one would identify
a Breit-Wigner width by
\begin{eqnarray}
 M_{D^*}\,\Gamma_{D^* \to D\,\pi} = - \Im \,\Pi^{\rm bubble}_{D^*}\,,
 \label{width-argument}
\end{eqnarray}
where the loop function is evaluated at the $D^*$ meson mass $M_{D^*}$. Both determinations would provide identical results. 
However, in the presence of a wave-function renormalization effect from the loop function 
\begin{eqnarray}
Z_H - 1= \frac{\partial }{\partial M^2_H}\,\bar \Pi_H\,,
\label{def-ZH}
\end{eqnarray}
this would no longer be the case. Following \cite{Lutz:2018a} we will therefore use the following form of the Dyson equation 
\begin{eqnarray}
&& M^2_H - \bar \Pi^{(0)}_H - \bar \Pi_H^{(2)}  - \bar \Pi^{(4-\chi)}_H -  \bar \Pi^{\rm tadpole}_H - \bar \Pi^{\rm bubble}_H/ Z_H  = 0\,, \qquad  
\label{gap-equation-A}
\end{eqnarray}
where we take $\bar \Pi^{(0)}_{H} = M^2 $ and $\bar \Pi^{(0)}_{H} = (M+ \Delta)^2$ for the pseudo-scalar and vector $D$ mesons respectively.
The second order terms $ \bar \Pi_H^{(2)}$ are the tree-level contributions (\ref{res-tree-level}) proportional to the quark masses as 
written in terms of the parameters $c_0, c_1$ and $\tilde c_0, \tilde c_1$. The fourth order terms $ \bar \Pi_H^{(4-\chi)}$ are the 
tree-level contributions (\ref{res-tree-level}) proportional to the product of two quark masses. Here the parameters $d_i$ and $\tilde d_i$
are probed. We recall that the  wave-function renormalization $Z_H$ has a quark-mass dependence which 
cannot be fully moved into the counter terms of the chiral Lagrangian.

\begin{table}[t]
\setlength{\tabcolsep}{3.5mm}
\renewcommand{\arraystretch}{1.2}
\begin{center}
\begin{tabular}{c|rrl||c|rr }\hline
$H$           & $\bar \Pi^{(2) }_H/(2\,M_H) $  & $\bar \Pi^{\rm bubble }_H/(2\,M_H)$   & $Z_H$\,       &                & {\rm with\,bubble}  & {\rm tree\,level}                \\ \hline \hline
$D$           &   4.7 \, MeV                   &  -50.2  \, MeV                 & 1.108         &  $ M $         & 1907.4  \,MeV     & 1862.7  \,MeV    \\
$D_s$         &  106.2  \, MeV                 &  -65.5 \, MeV                  & 1.418         &  $\Delta$      & 191.7\, MeV       & 141.3  \,MeV  \\
$D^*$         &  5.0  \, MeV                   &  -113.4  \, MeV                & 1.163         &  $ c_1 $       & 0.440             & 0.426           \\
$D^*_s$       &  114.1  \, MeV                 &  -166.1  \, MeV                & 1.643         &  $\tilde c_1$  & 0.508             & 0.469          \\ \hline
\end{tabular}
\caption{The loop functions (\ref{result-loop}) are evaluated with the coupling constants $g_P=\tilde g_P \simeq 0.57$  and the physical isospin averaged meson masses. In addition the
large-$N_c$ relations (\ref{large-Nc-recall}) are assumed.}
\label{tab:ZH}
\end{center}
\end{table}

We provide a first numerical estimate of the importance of the various terms in (\ref{gap-equation-A}). We put $\bar \Pi^{(4-\chi)}_H =  \bar \Pi^{\rm tadpole}_H =0$ since 
the associated counter terms are not known reliably. Insisting on the large-$N_c$ relations 
\begin{eqnarray}
 2\,c_0 = c_1\,,\qquad \qquad 2\,\tilde c_0 = \tilde c_1\,,
 \label{large-Nc-recall}
\end{eqnarray}
we adjust the four parameters $c_1, \tilde c_1$ and $M, \Delta$ to the four isospin averaged pseudo-scalar and vector $D$ meson masses \cite{PDG}. The results of this procedure are collected in the second last 
column of Tab. \ref{tab:ZH}. In the third column  we show the size of the loop contribution $\bar \Pi^{\rm bubble }_H$ and the wave function renormalization factor $Z_H$. From those numbers we conclude 
that the loop terms are as important as the contributions of the $Q^2$ counter terms (shown in the second column). Note also the significant size of the wave function factor for the strange $D$ mesons. 
It is instructive to compare the values of the four parameters 
$c_1, \tilde c_1$ and $M, \Delta$ with their corresponding values that follow in a scenario where all loop effects are neglected. Such values are shown in the last column of Tab. \ref{tab:ZH}. 
A reasonable spread of the parameters as compared to the initial scenario is observed.

While with (\ref{gap-equation-A}) we arrive at a renormalization scale invariant and self consistent approach for a chiral extrapolation of the $D$ meson masses that 
considers all counter terms relevant at N$^3$LO, there is an important issue remaining. Is it possible to decompose the renormalized loop function $\bar \Pi^{\rm bubble}_H$ into 
its chiral moments and therewith shed more light on the convergence properties of such a chiral expansion.  It is known that a conventional chiral expansion 
has not too convincing convergence properties at physical values of the strange quark mass. Does a resumed scheme that is formulated in terms of physical 
meson masses show  an improved convergence pattern?

\section{Power-counting decomposition of the loop function}

\begin{figure}[t]
\center{
\includegraphics[keepaspectratio,width=0.5\textwidth]{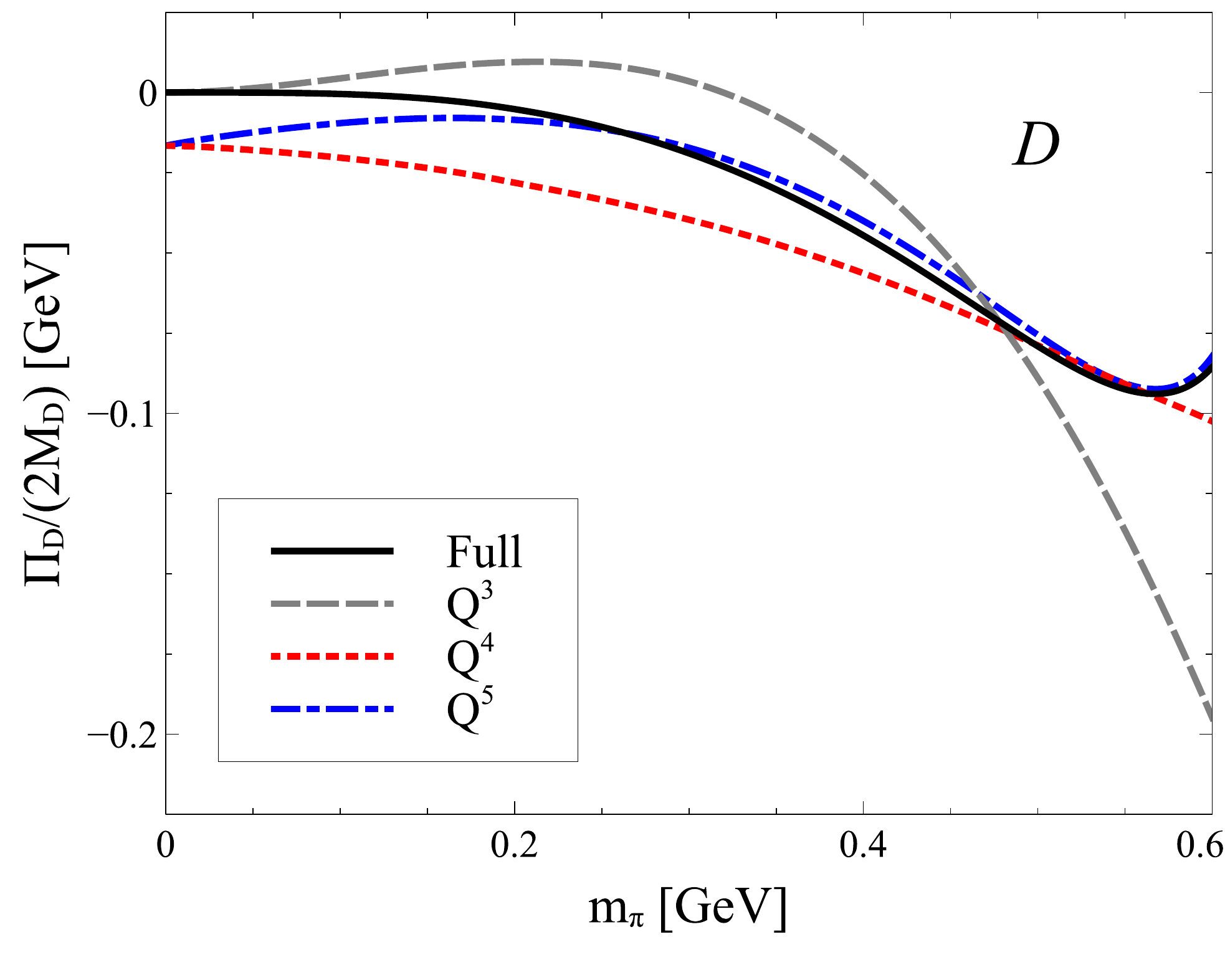}\includegraphics[keepaspectratio,width=0.5\textwidth]{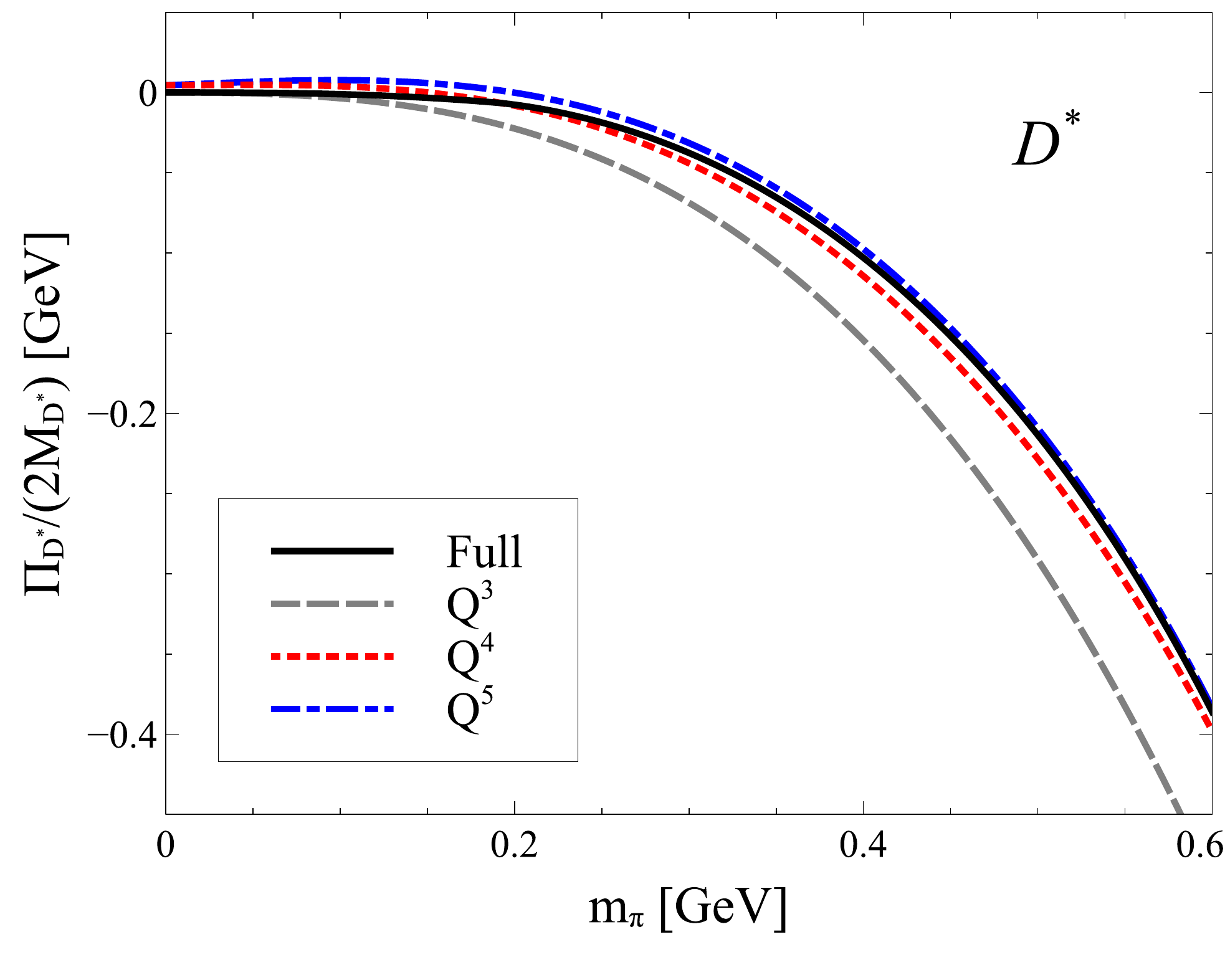} }
\vskip-1.5cm
\caption{\label{fig:1} $D$ and $D^*$ meson masses in the flavour limit as a function of the pion mass. The power-counting decomposed loop functions 
of (\ref{result-loop-B}) are used with the parameter set of Tab. \ref{tab:ZH}.  }
\end{figure}

At sufficiently small quark masses a linear dependence of the $D$ meson masses is expected as recalled in (\ref{res-tree-level}).
The associated slope parameters $c_0, c_1$  and $\tilde c_0, \tilde c_1$ are scale independent. This is an effect of chiral order $Q^2$.  
With increasing quark masses additional terms in the chiral expansion turn relevant. While there is no controversy on how to count the $Q^4$ contributions 
$\bar \Pi^{(4-\chi)}_H $ and  $\bar \Pi^{\rm tadpole}_H $, it is less obvious how to further decompose the loop contribution $\bar \Pi^{\rm bubble}_H $ into its power counting moments. 
The loop functions depend on the physical masses $m_Q$, $M_H$ and $M_R$. In 
any power counting ansatz based on chiral dynamics we would assign
\begin{eqnarray}
 \frac{m_Q}{M_R} \sim Q \sim \frac{m_Q}{M_H} \,,
 \label{def-count-A}
\end{eqnarray}
for the ratios of the Goldstone boson masses over the D meson masses. The mass differences of either pseudo-scalar or vector mesons
\begin{eqnarray}
 \frac{M_H -M_R}{m_Q} \sim Q \,,\qquad \qquad \frac{M_R - M_H }{M_H} \sim Q^2\,\qquad {\rm for}\qquad  H \parallel R \,,
 \label{def-count-B}
\end{eqnarray}
can also be counted without controversy. In (\ref{def-count-B}) we use a notation $H \parallel R$ requesting  $H, R \in[0^-]$ or $H,R \in [1^-]$.
Less obvious is how to treat the mass differences of a pseudo-scalar and a vector $D$ meson.

\begin{table}[t]
\setlength{\tabcolsep}{3.5mm}
\renewcommand{\arraystretch}{1.2}
\begin{center}
\begin{tabular}{c|rrrrr }\hline
$H$             & $\bar \Pi^{\rm bubble }_H/(2\,M_H)$  & $\bar \Pi^{\rm bubble -3}_H/(2\,M_H)$ & $\bar \Pi^{\rm bubble -4}_H/(2\,M_H)$  & $\bar \Pi^{\rm bubble-5 }_H/(2\,M_H)$     \\ \hline \hline
$D$             &  -50.2  \, MeV                     &  -38.7  \, MeV                      &  -29.4  \, MeV                       &  22.8  \, MeV                           \\
$D_s$           &  -65.5 \, MeV                      &  -93.2 \, MeV                       &  27.3 \, MeV                         &  2.4   \, MeV                           \\
$D^*$           &  -113.4  \, MeV                    &  -135.1  \, MeV                     &  19.0  \, MeV                        &  6.3   \, MeV                           \\
$D^*_s$         &  -166.1  \, MeV                    &  -308.3  \, MeV                     &  99.8  \, MeV                        &  61.8  \, MeV                           \\ \hline
\end{tabular}
\caption{The loop functions (\ref{result-loop}) are evaluated with the coupling constants $g_P=\tilde g_P \simeq 0.57$  and the physical isospin averaged meson masses. A decomposition according to
(\ref{def-count-A}, \ref{def-count-B}) and (\ref{def-counting-A}) is performed.}
\label{tab:3-4-5}
\end{center}
\end{table} 
There are different schemes possible. Technically most straight forward is the extreme assumption
\begin{eqnarray}
 \frac{M_{R}-M_{H}}{m_Q}  \sim Q \,,\qquad \qquad \frac{M_R -M_H}{M_H} \sim Q^2\,  \qquad  {\rm for } \qquad H \perp R \,,
 \label{def-counting-A}
\end{eqnarray}
which can be motivated in the limit of a large charm quark mass where $\Delta \to 0$ and therewith $\Delta \ll  m_\pi$. In (\ref{def-counting-A}) we use a notation $H\perp R$ implying that 
either $H \in[0^-]$ and $R \in [1^-]$ or $H \in[1^-]$ and $R \in [0^-]$.  
While the counting ansatz (\ref{def-counting-A}) is expected to be faithful for 
$m_Q = m_K$ it is not so useful for $m_Q = m_\pi$. However, since the loop corrections are typically dominated  by contributions involving the 
kaon and eta meson masses, such an assumption should have some qualitative merits nevertheless. The leading order terms are readily worked out with
\allowdisplaybreaks[1]
\begin{eqnarray}
&&\bar \Pi _{H\in [0^-]}^{\rm bubble}= \sum _{Q\in [8]} \sum _{R\in [1^-]} \left(\frac{G_{QR}^{(H)}}{2\,f}\right)^2 \Bigg\{ \alpha_{QR}^H + X^{(H)}_{QR} 
\nonumber\\ 
&& \qquad \qquad \qquad  +\,\frac{\gamma^H_{R}}{16\,\pi^2} \,M_H^2\, m_Q^2 \,\bigg[1 -\bigg(\frac{m_Q}{2\, M_H} - \frac{M_R - M_H}{ m_Q} \bigg)^2 \,\bigg]  \Bigg\} + {\mathcal O}\left(Q^6\right)\,, 
\nonumber\\ \nonumber\\ 
&&\bar \Pi _{H\in [1^-]}^{\rm bubble}= \frac{2}{3}\,\sum _{Q\in [8]} \sum _{R\in [1^-]} \left(\frac{G_{QR}^{(H)}}{2f}\right)^2 X^{(H)}_{QR}  
+\frac{1}{3}\, \sum _{Q\in [8]} \sum _{R\in [0^-]} \left(\frac{G_{QR}^{(H)}}{2\,f}\right)^2  \Bigg\{ \alpha_{QR}^H+  X^{(H)}_{QR} 
\nonumber\\ 
&& \qquad \qquad \qquad  +\,\frac{\gamma^H_{R}}{16\,\pi^2} \,M_H^2\, m_Q^2\,\bigg[1 -\bigg(\frac{m_Q}{2 \,M_H} - \frac{M_R - M_H}{ m_Q} \bigg)^2\, \bigg]  \Bigg\} + {\mathcal O}\left(Q^6\right) \,,
\nonumber\\ \nonumber\\
&& X^{(H)}_{QR} =M_H^2\,\frac{m_Q^2}{16\, \pi ^2}\bigg(\frac{m_Q^2}{M_H^2} +2\, \frac{M_R-M_H}{M_H} \bigg) 
- M_H^2\, \frac{m_Q^2}{32\, \pi^2} \bigg( \frac{m_Q^2}{M_H^2}-3\,\frac{ M_R - M_H }{M_H} \bigg) \log \frac{m_Q^2}{M_R^2} 
\nonumber\\
&& \qquad \; + \,M_H\,\frac{ m_Q^3}{16\,\pi^2 }  \, \bigg[ -\pi + \frac{3\, \pi}{2} \bigg(\frac{m_Q}{2\, M_H}-\frac{M_R-M_H}{m_Q}\bigg)^2\, \bigg]\,,
\label{result-loop-B}
\end{eqnarray}
accurate to order $Q^5$. The coefficients $\alpha_{QR}^H $ and $\gamma^H_{R}$ were given already in (\ref{def-alphaBR}) and (\ref{def-master-loop}).
In Tab. \ref{tab:3-4-5} we decompose the loop function into third, fourth and fifth order numerical values. The results are compared with the exact numbers already shown in Tab. \ref{tab:ZH}. 
While we observe a qualitative reproduction of the full loop function, owing to contributions form intermediate pion states, there is no convergence observed - as expected. 
By construction, the counting rule (\ref{def-counting-A}) fails in the chiral regime where all quark masses, in particular the strange quark mass approach zero. This is illustrated by Fig. \ref{fig:1} where we 
plot the loop function $\bar \Pi_H $ in the flavour limit with $m_\pi = m_K = m_\eta$.  Here the $D$ meson masses  $M_D = M_{D_s}$ and $M_{D^*} = M_{D^*_s}$  are obtained as the solution of the set of 
Dyson equation  (\ref{gap-equation-A})  where the full loop expression (\ref{result-loop}) is assumed.  The parameter set of Tab. \ref{tab:ZH} is applied which is based on the scenario 
$\bar \Pi^{(4-\chi)}_H =  \bar \Pi^{\rm tadpole}_H =0$. While for large pion  masses the hierarchy of dashed and dotted lines systematically approach the solid line, this is not the case for pion masses smaller 
than $m_\pi \leq  \Delta \sim 200$ MeV.

\begin{figure}[t]
\center{
\includegraphics[keepaspectratio,width=0.5\textwidth]{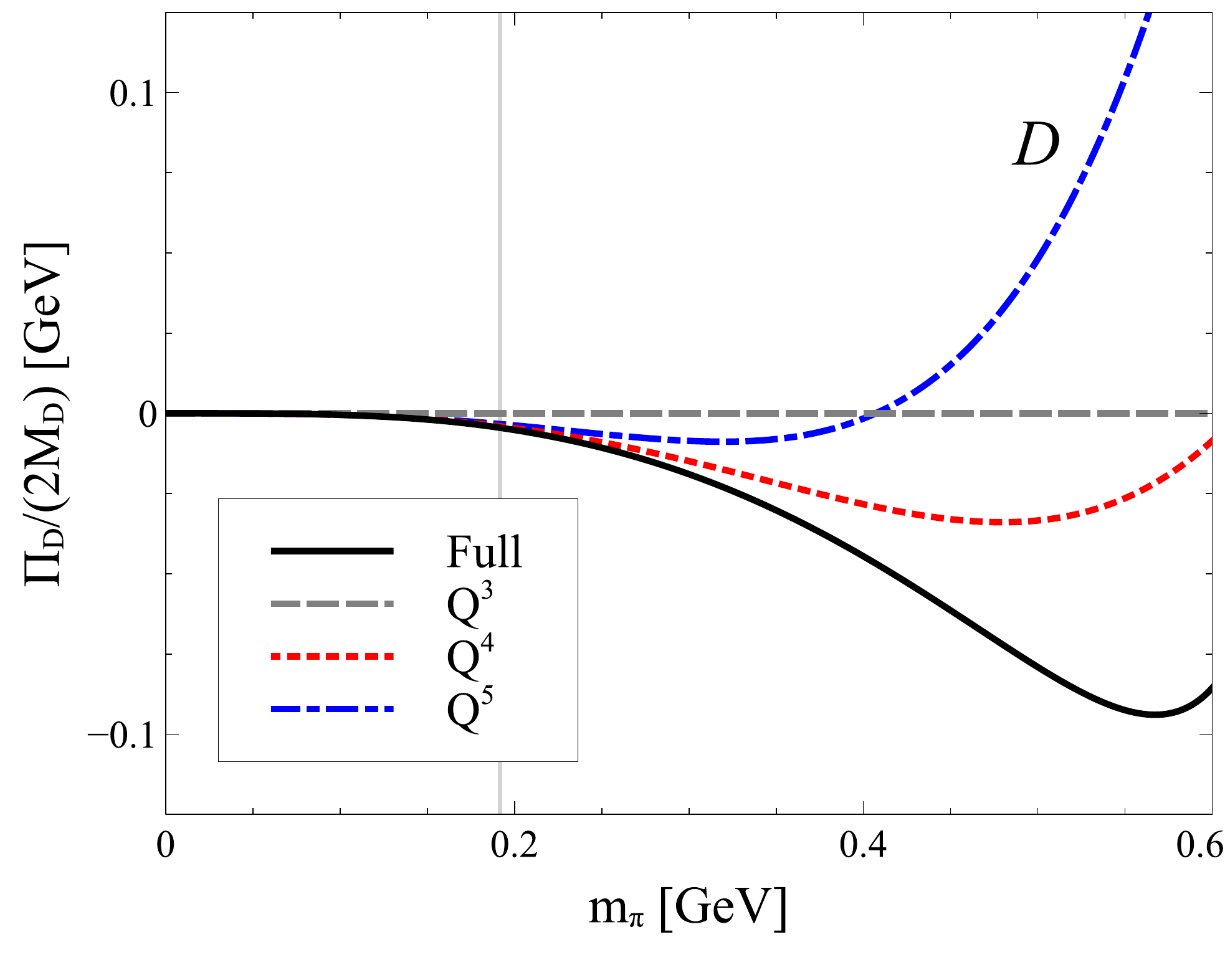}\includegraphics[keepaspectratio,width=0.5\textwidth]{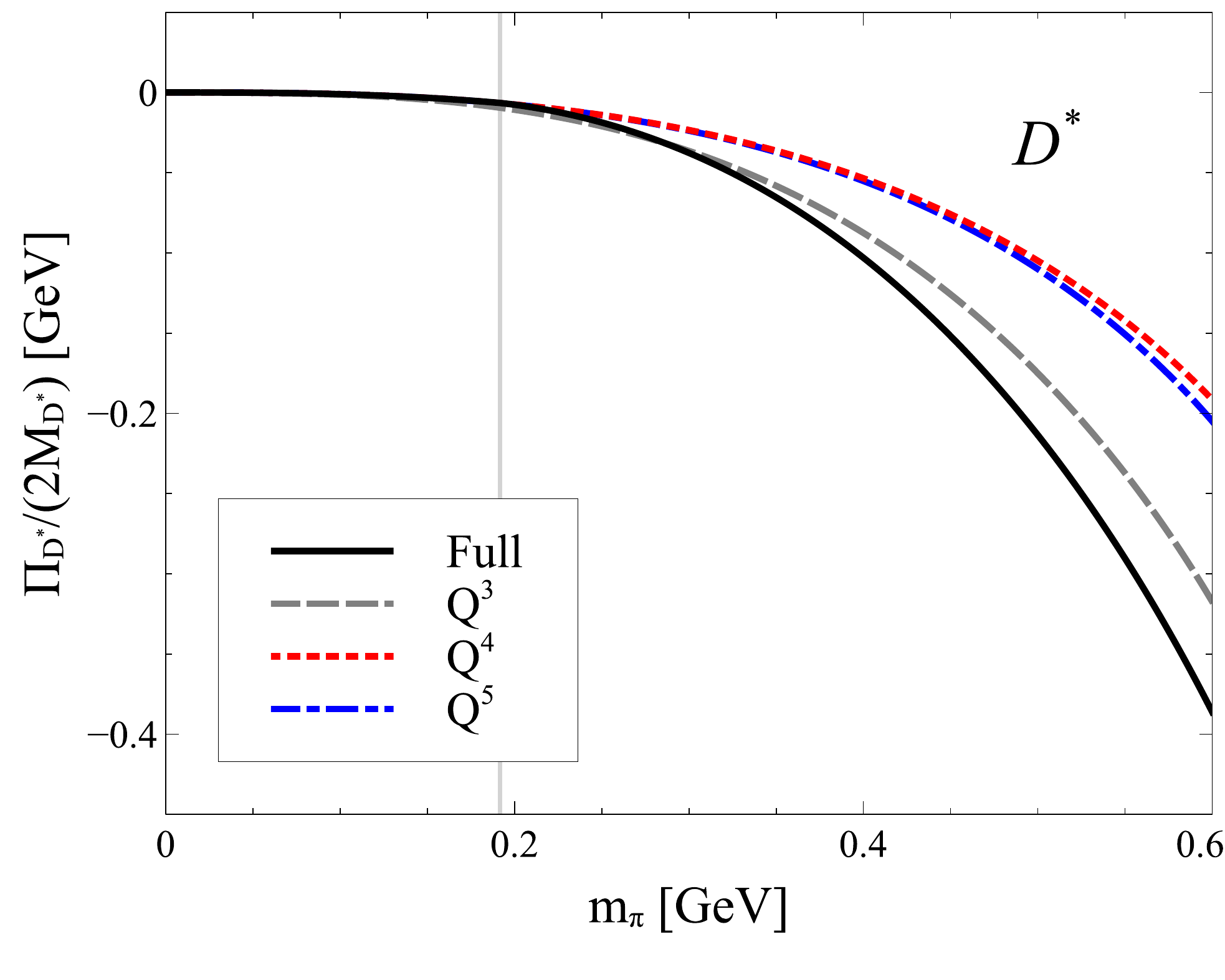} }
\vskip-1.5cm
\caption{\label{fig:2} $D$ meson masses in the flavour limit as a function of the pion mass. The power-counting decomposed loop functions 
of (\ref{def-chiral-regime-1}, \ref{def-chiral-regime}, \ref{def-chiral-regime-3}) are used with the parameter set of Tab. \ref{tab:ZH}.  }
\end{figure}

How to improve on the counting rule (\ref{def-counting-A}). Before presenting a universal approach we consider yet two further interim power counting scenarios.  First 
we work out the extreme chiral region where all Goldstone boson masses are significantly smaller than $\Delta \sim 200$ MeV. In this case the counting rules
\begin{eqnarray}
  \frac{m_Q}{\Delta}  \sim Q \,, \qquad \qquad  \frac{\Delta}{M} \sim Q^0\,,
 \label{def-counting-B}
\end{eqnarray}
are used. Since the extreme chiral region is not realized in  nature, such 
an assumption is not expected to provide any significant results for quantities measurable in experimental laboratories. 

Since at some stage lattice QCD simulations may be feasible at such low strange quark masses 
we provide the corresponding expressions for the loop function nevertheless. Here we decompose all meson masses into their chiral  moments in application of a strict chiral expansion. 
At third order 
\begin{eqnarray}
 \Pi^{\rm bubble-3}_{H \in [0^-]} = 0\,, \qquad \qquad 
 \Pi^{\rm bubble-3}_{H \in [1^-]} =- \frac{2}{3}\, \pi\, \sum_{Q\in[8]}\sum_{R\in \left[1^-\right]} \left(\frac{G_{QR}^H}{8\pi f}\right)^2 m_Q^3\, \big( M + \Delta \big) \,,
 \label{def-chiral-regime-1}
\end{eqnarray}
the vector $D$ mesons pick up a contribution only. At fourth order the expressions turn more complicated. 
We do not expand in powers of $\Delta/M$ because there are terms present proportional to $\log \Delta/M$, and also because we do not want to pollute the strict chiral expansion by a further scale assumption. 
The algebra required 
is somewhat involved and we organize it by a series of suitable dimensionless coefficients $\alpha_n, \gamma_n$ and $\tilde \alpha_n,\tilde \gamma_n$ that depend on the 
ratio $\Delta/M$ only. While the coefficients $\gamma_n, \tilde \gamma_n$ characterize the chiral expansion of the scalar bubble functions, 
the  $\alpha_n, \tilde \alpha_n $ result from a chiral expansion of the coefficients in front of the scalar loop functions. 
Altogether we derive the compact expressions
\begin{eqnarray}
&&\bar \Pi_{H\in [0^-]}^{\rm bubble-4}
    =\sum_{Q\in[8]}\sum_{R\in \left[1^-\right]} \left(\frac{G_{QR}^H}{8\,\pi f}\right)^2
    \Bigg\{ \gamma_d^{(1)} m_Q^2\, \Pi_R^{(2)} + \gamma_d^{(2)}\, m_Q^2\, \Pi_H^{(2)} + \gamma_d^{(3)}\,\Pi_R^{(2)}\,\Pi_R^{(2)} +\gamma_d^{(4)}\,\Pi_H^{(2)}\,\Pi_H^{(2)} 
\nonumber\\
    &&\qquad \qquad \quad +\,\gamma_d^{(5)}\,\Pi_R^{(2)} \,\Pi_H^{(2)}  + \frac{M }{\Delta}\,  m_Q^4\,\bigg[
 \big( \alpha_2 \,\gamma_2 -\alpha_1 \, \gamma_4 \big) 
    + \big(\alpha_2\, \gamma_3 - \alpha_1 \,\gamma_5 \big)\, \log \frac{m_Q^2}{(M+ \Delta )^2} \bigg]
     \Bigg\}\,, 
\nonumber\\
&&\bar \Pi_{H\in [1^-]}^{\rm bubble-4}
    =\sum_{Q\in[8]}\sum_{R\in \left[0^-\right]} \left(\frac{G_{QR}^H}{8\,\pi\, f}\right)^2
    \Bigg\{\tilde \gamma_d^{(1)} m_Q^2\, \Pi_R^{(2)} + \tilde \gamma_d^{(2)}\, m_Q^2\, \Pi_H^{(2)} + \tilde \gamma_d^{(3)}\,\Pi_R^{(2)}\,\Pi_R^{(2)} +\tilde \gamma_d^{(4)}\,\Pi_H^{(2)}\,\Pi_H^{(2)} 
\nonumber\\
    &&\qquad \qquad \quad +\,\tilde \gamma_d^{(5)}\,\Pi_R^{(2)}\, \Pi_H^{(2)}  + \frac{M }{3\,\Delta}\,  m_Q^4\,\bigg[
 \big( \tilde \alpha_2 \,\tilde \gamma_2 - \tilde \alpha_1 \, \tilde \gamma_4 \big) 
    + \big(\tilde \alpha_2\, \tilde \gamma_3 - \tilde \alpha_1 \,\tilde \gamma_5 \big)\, \log \frac{m_Q^2}{M^2} \bigg]
     \Bigg\}
\nonumber\\
&& \qquad \qquad \quad \;\,+\,\sum_{Q\in[8]}\sum_{R\in \left[1^-\right]} \left(\frac{G_{QR}^H}{8\pi f}\right)^2 \frac{2}{3}\,
     \Bigg\{ m_Q^2\, \Big( m_Q^2 - \Pi_H^{(2)} + \Pi_R^{(2)} \Big) 
\nonumber\\
    && \qquad \qquad \quad - \,\frac{1}{4}\,  \Big(2\,m_Q^2 +3 \,\Pi_H^{(2)} -3 \,\Pi_R^{(2)}\Big)\,m_Q^2\, \log \frac{m_Q^2}{(M+\Delta)^2} \Bigg\} \,.
\label{def-chiral-regime}
\end{eqnarray} 
The dimension less coefficients $\gamma_d^{(n)} $ and $\tilde \gamma_d^{(k)} $ are expressed in terms of the basic coefficients $\alpha_n, \gamma_n$ and $\tilde \alpha_n,\tilde \gamma_n$
in Appendix A and B. Again they depend on the ratio $\Delta/M$ only. We note that the rational functions $\alpha_n $ and $\tilde \alpha_n$ approach one in the limit $\Delta/M \to 0$.
In contrast  the $\gamma_n$ and $\tilde \gamma_n$ have contributions proportional to $\log \Delta/M$ and do not approach one in the heavy-quark mass limit. All terms in (\ref{def-chiral-regime}) 
that are proportional to $\gamma^{(n)}_d$ or $\tilde \gamma^{(n)}_d$
can be viewed as a renormalization of the low-energy parameters  $d_n$ and $\tilde d_n$. This is illustrated in Appendix A and B, where explicit expressions are provided.
We note that the fifth order terms can also be readily constructed. For the vector $D$ mesons we derive
\begin{eqnarray}
&&\bar \Pi_{H\in[1^-]}^{{\rm bubble}-5} = \sum_{Q\in [8]} \sum_{R\in [1^-]} \left(\frac{G_{QR}^{(H)}}{8\pi\,f}\right)^2 \frac{\pi\,m_Q}{12\,(M+ \Delta)} \bigg\{3  m_Q^4 
+ m^2_Q \, \big(2\, \Pi_H^{(2)} - 6\,\Pi_R^{(2)} \big) 
\nonumber\\
&& \qquad \qquad \qquad  + \,3\,\big(\Pi_H^{(2)} - \Pi_R^{(2)} \big)^2   \bigg\}  +\,\cdots \,,
\label{def-chiral-regime-3}
\end{eqnarray}
where the dots stand for additional terms extracted from (\ref{def-chiral-regime}) 
with the replacement $\Pi_{H}^{(2)} \to \Pi_{H}^{(3)}$. For the pseudo-scalar $D$ mesons the corresponding expression follow from (\ref{def-chiral-regime}) 
with the replacement $\Pi_{H}^{(2)} \to \Pi_{H}^{(3)}$ only. 

\begin{figure}[t]
\center{
\includegraphics[keepaspectratio,width=0.5\textwidth]{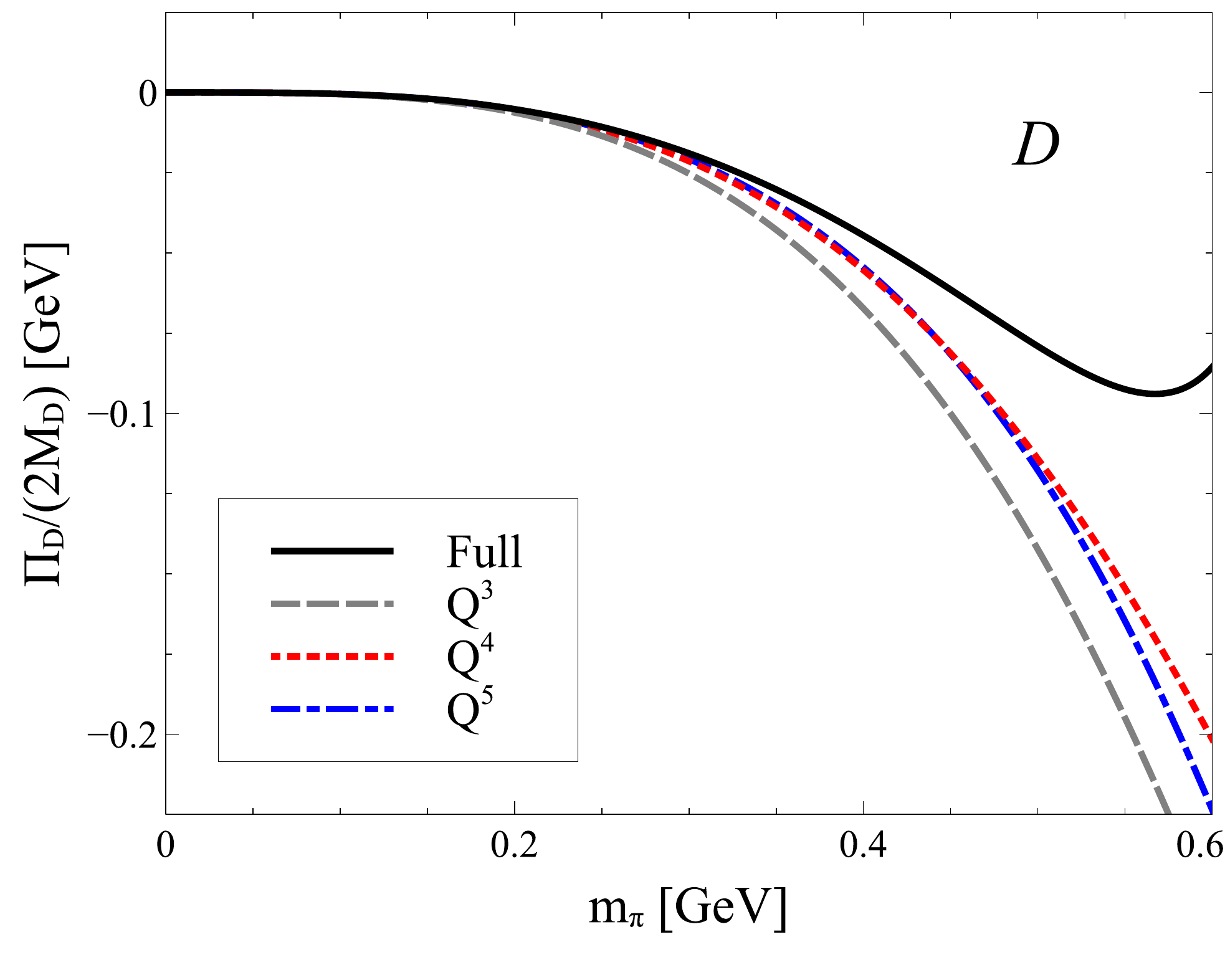}\includegraphics[keepaspectratio,width=0.5\textwidth]{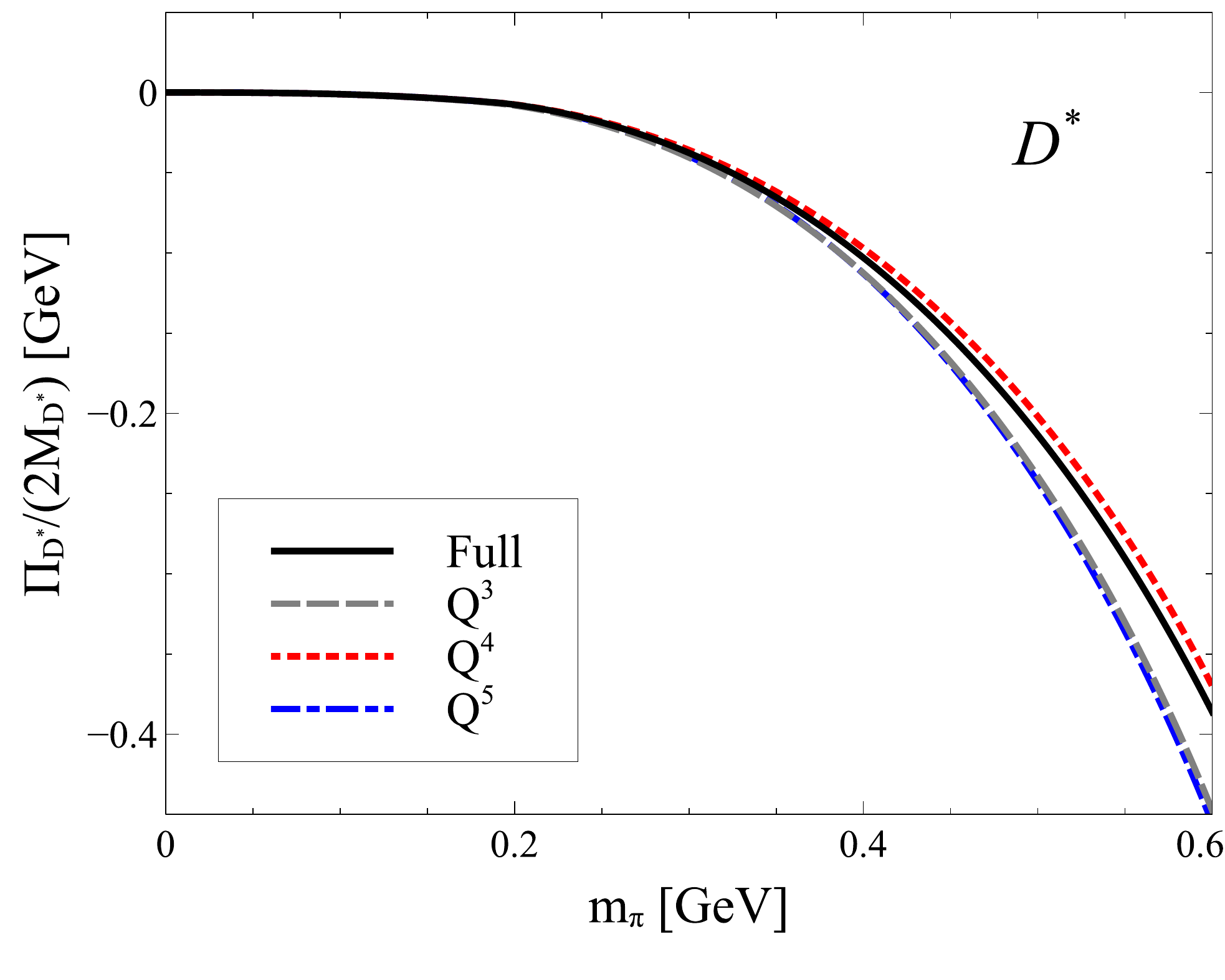} }
\vskip-1.cm
\caption{\label{fig:3} $D$ meson masses in the flavour limit as a function of the pion mass. The power-counting decomposed loop functions 
of (\ref{res-Delta-expansion-3}, \ref{res-Delta-expansion}, \ref{res-Delta-expansion-5A}, \ref{res-Delta-expansion-5B}) are used with the parameter set of Tab. \ref{tab:ZH}.  }
\end{figure}

We plot the loop function $\bar \Pi_H $ in the flavour limit with $m_\pi = m_K = m_\eta$ and $M_D = M_{D_s}= M$ and $M_{D^*} = M_{D^*_s}=M+\Delta$. Here we use our first estimate for 
the low-energy parameters $c_{0,1}$ and $\tilde c_{0,1}$  as displayed in the next to last column of  Tab. \ref{tab:ZH}. From Fig. \ref{fig:2} we conclude that for pion masses smaller than $\Delta$ the 
successive orders (dashed, dotted and dash-dotted lines) approach the exact solid line convincingly. 
Unlike the consequences of the power-counting ansatz (\ref{def-counting-A}) as illustrated in the previous Fig. \ref{fig:1} this is clearly not the case for (\ref{def-counting-B}) in the  
large pion mass domain with $m_\pi > \Delta$.

Neither of the extreme counting assumptions (\ref{def-counting-A}) nor (\ref{def-counting-B}) generates an expansion scheme that converges for physical up, down and strange quark masses. 
A step forward may be provided by the following conventional ansatz 
\begin{eqnarray}
 \Delta \sim m_Q \sim Q  \,, \qquad \qquad \Delta_Q = \sqrt{\Delta^2 - m_Q^2} \sim Q\,,\qquad \qquad \frac{\Delta}{M} \sim Q \,,
 \label{def-counting-C}
\end{eqnarray}
suggested originally by Banerjee and collaborators \cite{Banerjee:1994bk,Banerjee:1995wz} for the chiral expansion of baryon masses. Even though the authors demonstrated in a recent work \cite{Lutz:2018a} that such an expansion 
is not suitable to arrive at a meaningful expansion for the baryon octet and decuplet masses at physical values of the up, down and strange quark  masses, it deserves a closer study whether 
it may prove significant for a chiral expansion of the D meson masses. The counting rules (\ref{def-counting-C}) lead to somewhat more complicated expressions. 
Again we derive the third, fourth and fifth order terms. We find
\begin{eqnarray}
&&\bar \Pi_{H\in [0^-]}^{{\rm bubble}-3}
    =\sum _{Q\in [8]} \sum _{R\in [1^-]} \left(\frac{G_{QR}^{(H)}}{8\pi\,f}\right)^2 \Bigg\{  - \Delta_Q^3\,M\,\Big(\log \big(\Delta +\Delta_Q\big)
   -\,\log \big(\Delta -\Delta_Q\big)\Big)\nonumber\\
   &&\qquad\qquad \qquad - \,\Delta\,M \Big(\Delta_Q^2 - \frac{1}{2}\,m_Q^2\Big)\, \log\frac{ m_Q^2}{4\,\Delta^2}
     - \frac{1}{2}\,\Delta \,M\,m_Q^2 \Bigg\} \,,
\nonumber\\
&&\bar \Pi_{H\in [1^-]}^{{\rm bubble}-3}
    =\sum _{Q\in [8]} \sum _{R\in [1^-]} \left(\frac{G_{QR}^{(H)}}{8\pi\,f}\right)^2 \frac{2}{3}\,\Bigg\{   -\pi\, M \,m_Q^3  \Bigg\} 
\nonumber\\
    &&\qquad \quad \,\,+ \sum _{Q\in [8]} \sum _{R\in [0^-]} \left(\frac{G_{QR}^{(H)}}{8\pi\,f}\right)^2 \frac{1}{3}\,\Bigg\{ 
   \Delta_Q^3\,M\,\Big(\log \big(-\Delta- \Delta_Q\big)    
    -\,\log \big(-\Delta + \Delta_Q\big)\Big) \nonumber\\
   &&\qquad\qquad \qquad + \,\Delta\,M \Big(\Delta_Q^2 - \frac{1}{2}\,m_Q^2\Big)\, \log\frac{ m_Q^2}{4\,\Delta^2}
    + \frac{1}{2}\,\Delta \,M\,m_Q^2 \Bigg\}\,,
\label{res-Delta-expansion-3}
\end{eqnarray}
and
\begin{eqnarray}
&&\bar \Pi_{H\in [0^-]}^{\rm bubble-4}
    =\sum _{Q\in [8]} \sum _{R\in [1^-]} \left(\frac{G_{QR}^{(H)}}{8\pi\,f}\right)^2 \Bigg\{  \frac{1}{4}\,\Big( -3 \,\Delta^2 +4 \,m_Q^2 -4 \,\Pi^{(2)}_H +4 \,\Pi^{(2)}_R \Big)\,m_Q^2
\nonumber\\
    &&\qquad \qquad \qquad - \frac{3}{2}\,\Delta^2 \Big(\Delta_Q^2 -\Pi^{(2)}_H + \Pi^{(2)}_R -\frac{1}{2}\,m_Q^2\Big) \, \log\frac{ m_Q^2}{4\,\Delta^2} 
\nonumber\\
    &&\qquad \qquad \qquad -\,\frac{1}{4}\,\Big(2 \,m_Q^2 +3\, \Pi^{(2)}_H -3 \,\Pi^{(2)}_R \Big)\,m_Q^2\,\log \frac{  m_Q^2}{M^2}\,  -\, \Delta_Q\,M\,\Big(\log \big(\Delta +\Delta_Q\big)
\nonumber\\
    &&\qquad \qquad \qquad  \qquad    -\,\log \big(\Delta -\Delta_Q\big)\Big) \, \frac{3\,\Delta}{2\, M} \left(\Delta_Q^2- \Pi^{(2)}_H + \Pi^{(2)}_R\right) \Bigg\} \,,
\nonumber\\
&&\bar \Pi_{H\in [1^-]}^{\rm bubble-4}
    =\sum _{Q\in [8]} \sum _{R\in [1^-]} \left(\frac{G_{QR}^{(H)}}{8\pi\,f}\right)^2 \frac{2}{3}\,\Bigg\{   -\pi\, \Delta \,m_Q^3 + \Big(m_Q^2 -\Pi^{(2)}_H +\Pi^{(2)}_R\Big) \,m_Q^2
\nonumber\\
    && \qquad \qquad \qquad - \,\frac{1}{4} \,\Big(2\, m_Q^2 + 3 \Pi^{(2)}_H - 3\Pi^{(2)}_R \Big)\,m_Q^2\,\log\frac{ m_Q^2}{M^2}  \Bigg\} 
\nonumber\\
    &&\qquad \quad \,\,+ \sum _{Q\in [8]} \sum _{R\in [0^-]} \left(\frac{G_{QR}^{(H)}}{ 8\pi\,f}\right)^2 \frac{1}{3}\,\Bigg\{ 
    \frac{1}{4}\,\Big( - \Delta^2 +4\, m_Q^2 -4 \,\Pi^{(2)}_H +4 \,\Pi^{(2)}_R \Big) \,m_Q^2
\nonumber\\
    &&\qquad \qquad \qquad - \,\frac{\Delta^2}{2}  \Big(\Delta_Q^2 -3\, \Pi^{(2)}_H +3 \,\Pi^{(2)}_R -\frac{1}{2}\,m_Q^2\Big) \, \log\frac{  m_Q^2}{4\,\Delta^2} 
\nonumber\\
    &&\qquad \qquad \qquad
    -\,\frac{1}{4}\,\Big(2\, m_Q^2 +3\, \Pi^{(2)}_H -3\, \Pi^{(2)}_R \Big)\,m_Q^2\,\log \frac{  m_Q^2}{M^2} - \Delta_Q\,M\,\Big(\log \big(-\Delta- \Delta_Q\big)
\nonumber\\
    &&\qquad \qquad \qquad  \qquad       
    -\,\log \big(-\Delta + \Delta_Q\big)\Big) \, \frac{\Delta}{2\, M} \left(\Delta_Q^2 -3\, \Pi^{(2)}_H +3\, \Pi^{(2)}_R\right) \Bigg\}\,,
\label{res-Delta-expansion}
\end{eqnarray}
with $\Delta_Q$ of (\ref{def-counting-C}). Since the fifth order contributions are quite lengthy they are delegated to Appendix A and B. 
In Tab. VI we decompose the loop function into third, fourth and fifth order numerical values. The results are compared with the exact numbers already shown in Tab. \ref{tab:ZH}. 
The conclusions of that table are unambiguous: the power counting ansatz (\ref{def-counting-C}) is not  suitable for a chiral extrapolation of the D meson masses. 
We note that (\ref{def-counting-C}) neither reproduces the results of (\ref{def-counting-A}) nor those of (\ref{def-counting-B}). We further 
demonstrate our claim by a plot of the loop function $\bar \Pi_H $ in the flavour limit with 
$m_\pi = m_K = m_\eta$ as was done in the previous figures \ref{fig:1} and \ref{fig:2}. Fig. \ref{fig:3} demonstrates that for  $m_\pi > \Delta$ no quantitative reproduction 
of the solid line is obtained. 

\begin{table}[t]
\setlength{\tabcolsep}{3.5mm}
\renewcommand{\arraystretch}{1.2}
\begin{center}
\begin{tabular}{c|rrrrr }\hline
$H$             & $\bar \Pi^{\rm bubble }_H/(2\,M_H)$  & $\bar \Pi^{\rm bubble -3}_H/(2\,M_H)$ & $\bar \Pi^{\rm bubble -4}_H/(2\,M_H)$  & $\bar \Pi^{\rm bubble-5 }_H/(2\,M_H)$     \\ \hline \hline
$D$             &  -50.2  \, MeV                     &  -67.7  \, MeV                     &  15.0  \, MeV                        &  -8.9  \, MeV                           \\
$D_s$           &  -65.6 \, MeV                      &  -152.8 \, MeV                     &  27.8 \, MeV                         &  26.6   \, MeV                           \\
$D^*$           &  -113.4  \, MeV                    &  -111.7  \, MeV                    &  -57.1  \, MeV                       &  18.6   \, MeV                           \\
$D^*_s$         &  -166.1  \, MeV                    &  -252.0  \, MeV                    &  84.3  \, MeV                        &  -69.5  \, MeV                           \\ \hline
\end{tabular}
\caption{The loop functions (\ref{result-loop}, \ref{res-Delta-expansion}) are evaluated with the coupling constants $g_P=\tilde g_P \simeq 0.57$  and the physical isospin averaged meson masses. 
A decomposition according to (\ref{def-counting-C}) is performed.}
\end{center}
\label{tab:3-4-5-B}
\end{table}

We finally present our counting ansatz that is expected to be applicable from small to medium size quark masses uniformly. It is an adaptation of the 
framework developed recently for the chiral extrapolation of the baryon octet and decuplet masses \cite{Lutz:2018a} and implements the driving idea to 
formulate the expansion coefficients in terms of physical masses. It is supposed to interpolate the two extreme counting rules (\ref{def-counting-A}) and (\ref{def-counting-B}). 
The counting rules are 
\begin{eqnarray}
&&  \frac{M_R -M_H}{m_Q} \sim Q \,,\qquad \qquad \frac{M_R - M_H }{M_H} \sim Q^2\,\qquad \qquad \;\;\, \;\,{\rm for}\qquad  H \parallel R \,,
\nonumber\\
&& \frac{M_{R}-M_{H}}{m_Q}  \sim Q^0 \,,\qquad \qquad \!\! \frac{M_R - M_H \pm  \Delta_H }{M_H}  \sim Q^2\,\,\,\qquad {\rm for} \qquad H \perp R
\nonumber\\
&& \Delta_Q = \sqrt{ (M_H - M_R )^2- m_Q^2 }\,  \sim Q \qquad {\rm with}\qquad \Delta_H = \Delta\, M_H \lim_{m_{u,d,s}\to 0}\frac{1}{M_H}, \qquad
 \label{def-counting-D}
\end{eqnarray}
where the sign $\pm$ is chosen such that the last ratio in (\ref{def-counting-D}) vanishes in the chiral limit. The implications of (\ref{def-counting-D}) 
are more difficult to work out. The counting rules (\ref{def-counting-D}) as they are necessarily imply
\begin{eqnarray}
Q  \sim\frac{\Delta_H}{M_H}  = \left\{\begin{array}{cl}
\frac{\Delta}{M} \qquad &{\rm for }\qquad  H\in[0^-] \\
\frac{\Delta}{M+\Delta} \qquad &{\rm for } \qquad H\in[1^-]
                         \end{array} \right. \,,
\end{eqnarray}
which is at odds with the assumption in (\ref{def-counting-B}). Therefore we supplement (\ref{def-counting-D}) by the request that the implications 
of (\ref{def-counting-D}) are recovered in the chiral regime. This requires a particular summation of terms proportional to $(\Delta/M)^n$ with $n =1,2,3, \cdots$.

There is yet another issue pointed out in \cite{Lutz:2018a}. The chiral expansion of the scalar bubble function is characterized by an alternating feature.  
We recall from \cite{Lutz:2018a} the following approximation hierarchy
\begin{eqnarray}
&& (4\pi)^2\,\bar I_{QR} = -\Big\{ 1 - \frac{1}{8}\,x^2 - \frac{1}{128}\, x^4 - \frac{1}{1024} \,x^6
+ {\mathcal O} (x^8)\Big\} \,\pi \,\sqrt{x^2}
\nonumber\\
&& \qquad \qquad \;\;\;\;\, +\, \Big\{ 1 - \frac{1}{12}\,x^2 -\frac{1}{120}\, x^4  - \frac{1}{840}\,x^6 + {\mathcal O} (x^8) \Big\} \,x^2
\nonumber\\
&& \qquad \qquad  \;\;\;\;\,
-  \,\frac{1}{2}\,x^2\,\log x^2    \,,
\label{fn-exp}
\end{eqnarray}
where we denoted $x = m_Q/M_H$ and $M_R = M_H$.  As was discussed in \cite{Lutz:2018a} the terms with even and odd powers in $x$ have opposite signs always.  
This implies a systematic cancellation effect amongst terms proportional to $x^n$ and $x^{1+n}$, where the effect is most striking for $n= 1$.
Therefore it is useful to always group such terms together. Even though the need of such an reorganization is not very strong for the D meson systems under consideration we adapt this strategy in the following.  
Note that the convergence domain of (\ref{fn-exp}) was proven to be limited by $|x|< 2$ only, a surprisingly large convergence circle. 
Given this scheme accurate results can be obtained by a few leading order terms. We construct the third order contributions from the one-loop 
diagrams.
\allowdisplaybreaks[1]
\begin{eqnarray}
&&\bar \Pi^{{\rm bubble}-3}_{H \in [0^-]} = \sum_{Q\in [8], R\in [1^-]}
\left(\frac{G_{QR}^{(H)}}{4 \pi \,f}  \right)^2 \frac{\alpha_1}{4}\,
	\Bigg\{ \delta_7\, \Delta\, M \,m_Q^2 + \delta_6\, \Delta^2 \,M \,\big(M_R - M_H - \Delta_H\big) 
\nonumber\\
&& \qquad \qquad \quad + \,M_H \,\bigg[ \, \Delta_Q^2 \Big( \gamma_1 \,\Delta_H - \delta_1 (M_R -M_H) \Big) 
\nonumber\\
&& \qquad \qquad \qquad -\,\frac{2\,M + \Delta}{2\, M} \, \bigg( (M_R - M_H)\Big(\Delta_Q^2 - \frac{1}{2}\,m_Q^2\Big) \log \frac{m_Q^2}{M_R^2} 
\nonumber\\
&& \qquad \qquad \qquad \quad +\, \Delta_Q^3 \Big[ \log \big(M_R -M_H + \Delta_Q\big)-\log \big(M_R - M_H - \Delta_Q\big)\Big] \bigg) 
\nonumber\\
&& \qquad \qquad \qquad  +\,\frac{m_Q^2}{\Delta_H}\bigg( - \delta_2 \,\Delta_Q^2 + \delta_3 \,m_Q^2 \log \frac{m_Q^2}{M_R^2} \bigg) \Bigg] \Bigg\}\,,
\nonumber\\ \nonumber\\
&& \bar \Pi^{{\rm bubble}-3}_{H \in [1^-]} =\sum_{Q\in [8], R\in [1^-]} \left(\frac{G_{QR}^{(H)}}{4\pi\,f} \right)^2 
\frac{M_H }{6}\,  \bigg\{ \frac{m_Q^2}{M_H}\, \Big( 1 -  \log \frac{m_Q}{M_R} \Big)  -\pi\, m_Q 
  \bigg\}\,\Big( m_Q^2 - (M_H-M_R)^2 \Big) 
\nonumber\\
&& \qquad +\, \sum_{Q\in [8], R\in[0^-]} \left(\frac{G_{QR}^{(H)}}{4 \pi \,f}\right)^2 \frac{\tilde \alpha_1}{12}\,
	\Bigg\{ \tilde \delta_7 \, \Delta\, M\, m_Q^2 - \tilde \delta_6\, \Delta^2 M \big(M_R - M_H + \Delta_H \big) 
\nonumber\\
&& \qquad \qquad \quad +\, M_H\, \frac{M}{M + \Delta} \Bigg[\,\Delta_Q^2 \Big( \tilde \gamma_1 \,\Delta_H - \tilde \delta_1 \,(M_H - M_R) \Big) 
\nonumber\\
&& \qquad \qquad \qquad  + \, \frac{M\, (2\,M + \Delta)}{2\, (M + \Delta )^2}  \bigg( (M_H - M_R)\Big(\Delta_Q^2 - \frac{1}{2}\,m_Q^2\Big) \log \frac{m_Q^2}{M_R^2} 
\nonumber\\
&& \qquad \qquad \qquad \quad + \,\Delta_Q^3 \Big[ \log \big(M_R -M_H - \Delta_Q\big)-\log \big(M_R - M_H + \Delta_Q\big)\Big] \bigg) 
\nonumber\\
&& \qquad \qquad \qquad +\,\frac{m_Q^2}{\Delta_H}\bigg( - \tilde \delta_2\, \Delta_Q^2 + \tilde \delta_3\, m_Q^2 \log \frac{m_Q^2}{M_R^2} \bigg) \Bigg] \Bigg\}\,,
\label{loop-HB-3}
\end{eqnarray}
with $\Delta_Q$ and $\Delta_H$ as introduced in (\ref{def-counting-D}).  The dimension less  coefficients $\alpha_i, \gamma_i, \delta_i$ and $ \tilde \alpha_i, \tilde \gamma_i, \tilde \delta_i$ depend on 
the ratio $\Delta/M$ only. They are detailed in Appendix A and Appendix B. The contributions proportional to $\alpha_i\,\delta_j$ and 
$\tilde \alpha_i\,\tilde \delta_j$ in (\ref{loop-HB-3}) are constructed to ensure that the terms proportional to $(m_Q^4/\Delta)$ and 
$(m_Q^4/\Delta)\,\log m_Q^2$ are recovered exactly. 

\begin{table}[t]
\setlength{\tabcolsep}{3.5mm}
\renewcommand{\arraystretch}{1.2}
\begin{center}
\begin{tabular}{c|rrrrr }\hline
$H$             & $\bar \Pi^{\rm bubble }_H/(2\,M_H)$  & $\bar \Pi^{\rm bubble -3}_H/(2\,M_H)$ & $\bar \Pi^{\rm bubble -4}_H/(2\,M_H)$  & $\bar \Pi^{\rm bubble-5 }_H/(2\,M_H)$     \\ \hline \hline
$D$             &  -50.2  \, MeV                     &  -48.5  \, MeV                     &  -2.8  \, MeV                        &  1.1   \, MeV                           \\
$D_s$           &  -65.6 \, MeV                      &  -88.3 \, MeV                      &  20.1  \, MeV                        &  2.9   \, MeV                           \\
$D^*$           &  -113.4  \, MeV                    &  -99.5  \, MeV                     &  -17.1 \, MeV                        &  3.1   \, MeV                           \\
$D^*_s$         &  -166.1  \, MeV                    &  -197.5 \, MeV                     &  26.3  \, MeV                        &  6.6   \, MeV                           \\ \hline
\end{tabular}
\caption{The loop functions (\ref{result-loop}) are evaluated with the coupling constants $g_P=\tilde g_P \simeq 0.57$  and the physical isospin averaged meson masses. A decomposition according to
(\ref{def-counting-D}) is performed. This leads to (\ref{loop-HB-3}, \ref{loop-HB-4}, \ref{loop-HB-5A}, \ref{loop-HB-5B}).}
\label{tab:3-4-5-C}
\end{center}
\end{table}

We advance to the fourth order terms. The following explicit expressions are obtained
\begin{eqnarray}
&&\bar \Pi_{H\in[0^-]}^{{\rm bubble}-4} = \sum_{ Q\in [8], R\in[1^-]} \bigg( \frac{G_{QR}^{(H)}}{4 \pi\, f} \bigg)^2 
	\Bigg\{ - \frac{1}{4} \, \alpha_1\, M \,\Delta^2 \,\delta_6 + \frac{M_H}{4} \Bigg[ \alpha_1\, \Delta^2 \,\frac{\partial \,\Delta\, \gamma_1}{\partial\, \Delta}  + \Delta_Q^2\, \beta_4 
\nonumber\\
&& \qquad \qquad \qquad - \,\Delta_Q^2\, \beta_5 \,\frac{M_R - M_H}{\Delta_H} 
	  - \frac{\beta_1 }{\Delta_H} \, \bigg( (M_R - M_H)\Big(\Delta_Q^2 - \frac{m_Q^2}{2}\Big) \log \frac{m_Q^2}{M_R^2} 
\nonumber\\
&& \qquad \qquad \qquad + \,\Delta_Q^3 \Big[ \log \big(M_R -M_H + \Delta_Q\big)-\log \big(M_R - M_H - \Delta_Q\big)\Big] \bigg) 
\nonumber\\
	&& \qquad \qquad \qquad
	  +\,\frac{m_Q^2}{\Delta_H^2} \Big( - \beta_2\, \Delta_Q^2 + \beta_3 \,m_Q^2 \log \frac{m_Q^2}{M_R^2} \Big)
	\Bigg] \Bigg\}\big( M_R -M_H - \Delta_H \big)\,, 
\nonumber\\
&& \bar \Pi_{H\in[1^-]}^{{\rm bubble}-4} = \sum_{\substack{Q\in [8]\\ R\in [1^-]}} \left(\frac{G_{QR}^{(H)}}{4 \pi f}\right)^2 
	\frac{M_H}{3} \bigg\{ \bigg( -\frac{3\pi }{4} \frac{m_Q}{M_H} + 1 +\frac{1}{2}\log\frac{m_Q^2}{M_R^2}\bigg)
	\Big( m_Q^2 - (M_R - M_H )^2 \Big) 
\nonumber\\
	&& \qquad \qquad \qquad + \,\frac{m_Q^2 }{4} \log \frac{m_Q^2}{M_R^2} \bigg\}
	\big( M_R -M_H \big)\nonumber\\
	&& \qquad \qquad + \sum_{\substack{ Q\in [8] \\R\in[0^-]}} \left(\frac{G_{QR}^{(H)}}{4 \pi \,f}\right)^2 
	\Bigg\{ \frac{1}{12} \,\tilde \alpha_1 \,M \,\Delta^2 \,\tilde \delta_6 
	+ \frac{M_H}{12}\, \Bigg[ - \tilde \alpha_1\, \Delta^2 \frac{\partial \,\Delta\, \tilde \gamma_1}{\partial\, \Delta} - \tilde \beta_4 \, \Delta_Q^2
\nonumber\\
	&& \qquad \qquad \qquad + \,\tilde \beta_5 \, \Delta_Q^2\, \frac{M_H - M_R}{\Delta_H} 
	  - \frac{\tilde \beta_1 }{\Delta_H} \, \bigg( (M_H - M_R)\Big(\Delta_Q^2 - \frac{m_Q^2}{2}\Big) \log \frac{m_Q^2}{M_R^2} 
\nonumber\\
	&& \qquad \qquad \qquad + \,\Delta_Q^3 \Big[ \log \big(M_R - M_H - \Delta_Q\big)-\log \big(M_R - M_H + \Delta_Q\big)\Big] \bigg) 
\nonumber\\
	&& \qquad \qquad \qquad
	  -\,\frac{m_Q^2}{\Delta_H^2} \Big( - \tilde \beta_2\, \Delta_Q^2 + \tilde \beta_3\, m_Q^2 \log \frac{m_Q^2}{M_R^2} \Big) 
	\Bigg] \Bigg\} \big( M_R - M_H + \Delta_H \big)\,,
	\label{loop-HB-4}
\end{eqnarray}
with $\Delta_Q$ and $\Delta_H$ already introduced in (\ref{loop-HB-3}). 
\begin{figure}[t]
\center{
\includegraphics[keepaspectratio,width=0.5\textwidth]{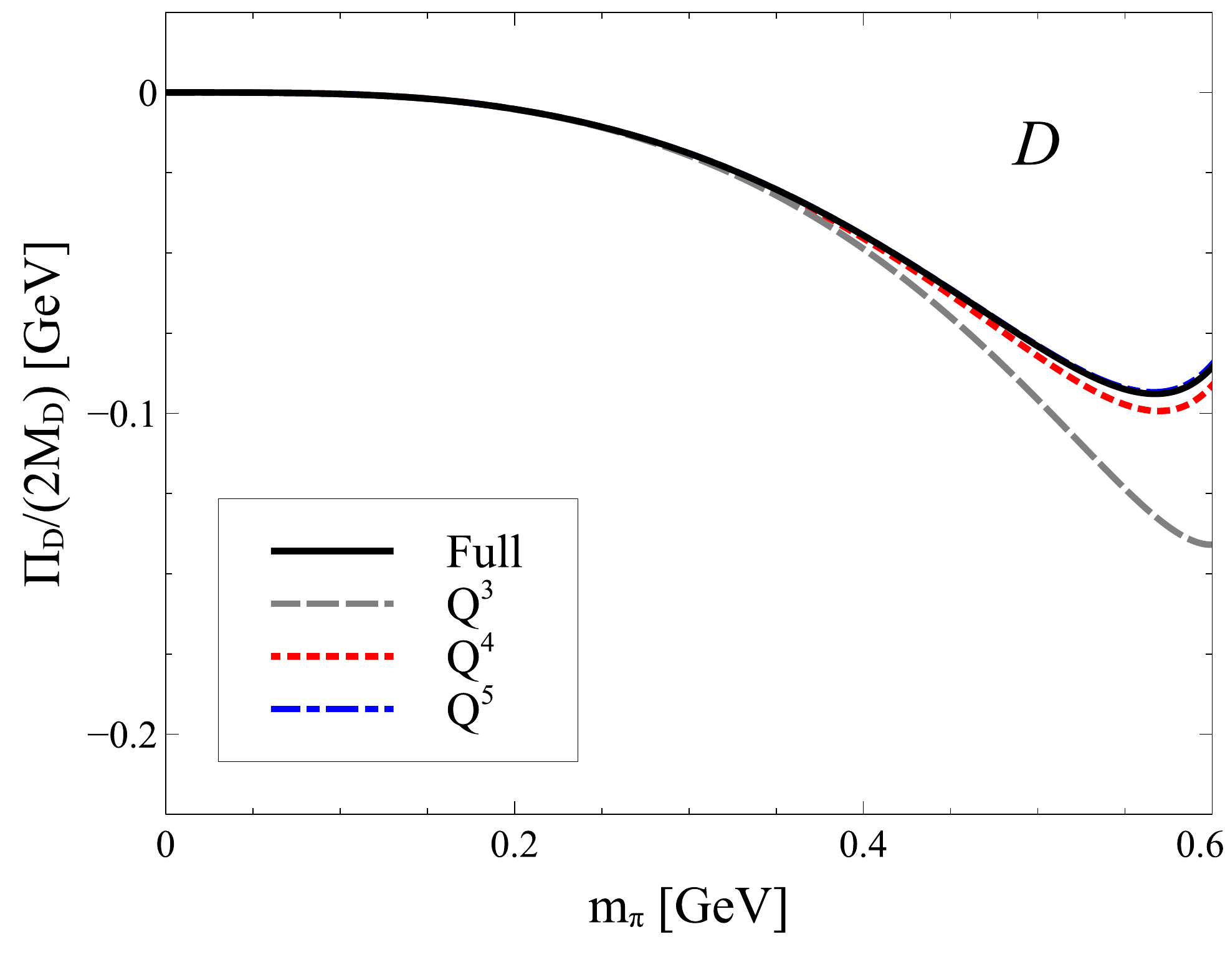}\includegraphics[keepaspectratio,width=0.5\textwidth]{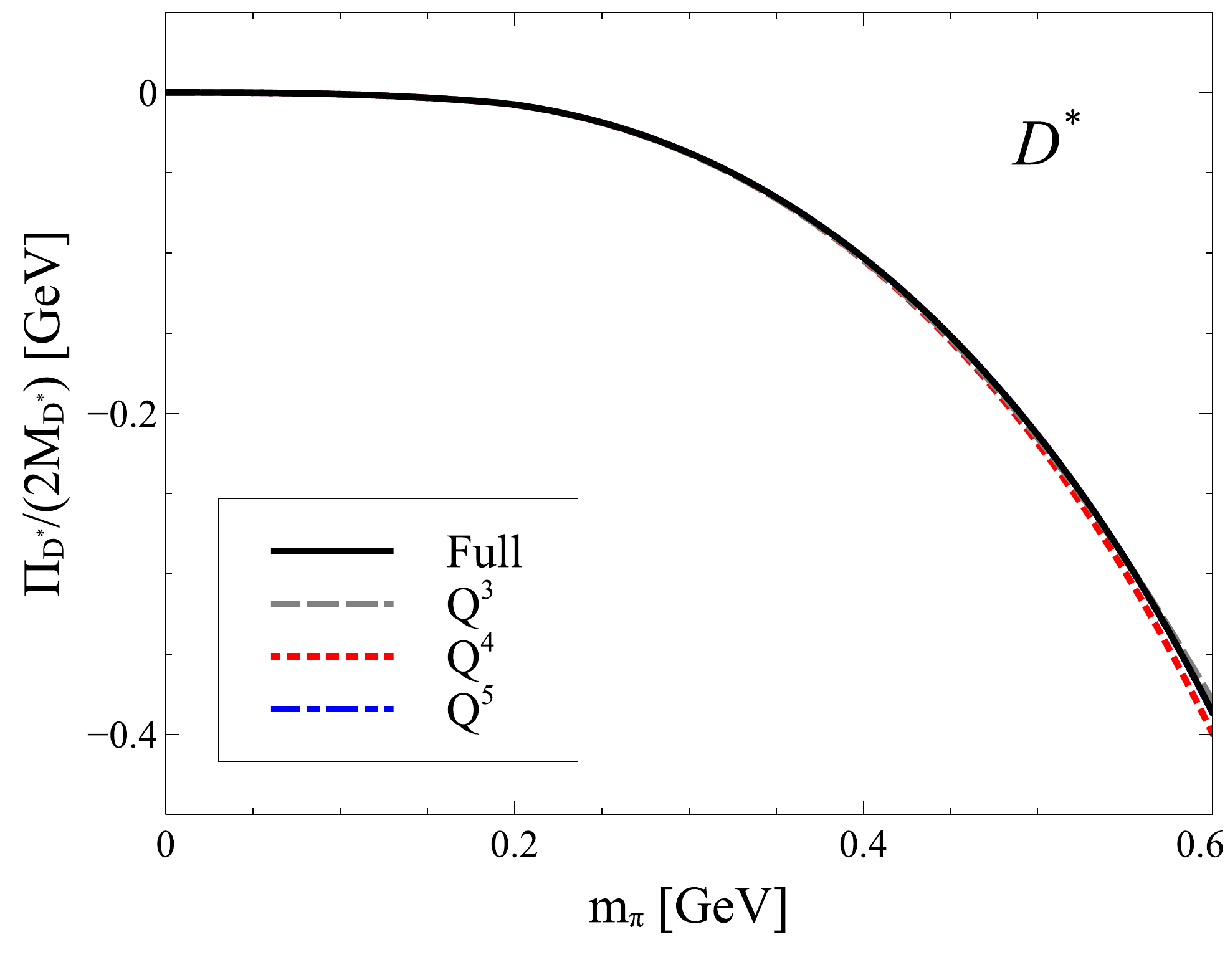} }
\vskip-1.5cm
\caption{\label{fig:4} $D$ meson masses in the flavour limit as a function of the pion mass. The power-counting decomposed loop functions of (\ref{loop-HB-3}, \ref{loop-HB-4}, \ref{loop-HB-5A}, \ref{loop-HB-5B}) are used with the parameter set of Tab. \ref{tab:ZH}.  }
\end{figure}

In Tab. \ref{tab:3-4-5-C} we decompose the loop function into third, fourth and fifth order numerical values. The results are compared with the exact numbers already shown in Tab. \ref{tab:ZH}. 
The conclusions of that table are unambiguous: the power counting ansatz (\ref{def-counting-D}) is well justified for a chiral extrapolation of the D meson masses. We note that the fifth order contributions to the 
$D$ meson masses are on average about 3 MeV only. Our novel expansion scheme is characterized by a rapid convergence property. All $D$ meson masses are reproduced at the few MeV level. We further 
substantiate our claim by Fig. \ref{fig:4}, which shows the loop function $\bar \Pi_H $ in the flavour limit with 
$m_\pi = m_K = m_\eta$. The figures are in correspondence to the previous figures \ref{fig:1},  \ref{fig:2} and \ref{fig:3} and demonstrate that for any reasonable pion mass, say $0 \leq m_\pi < 600$ MeV,
a quantitative reproduction of the solid line is obtained. 
We conclude that it is justified to identify the full loop expressions as the loop function to be used at chiral order $Q^4$ without any significant error from the incomplete 5th order terms.

\newpage

\section{Fit to QCD lattice data}

In this section we will determine the low-energy constants $c_i$ and $d_i$ of the chiral Lagrangian from lattice QCD simulations of the D meson masses.
Open-charm mesons have been extensively studied on different QCD lattices \cite{Aubin:2005ar,Na:2010uf,Mohler:2011ke,Liu:2012zya,Follana:2007uv,Bazavov:2011aa,Dowdall:2012ab,Mohler:2013rwa,Moir:2013ub,Lang:2014yfa,Bazavov:2014wgs,Kalinowski:2015bwa,Cichy:2016bci,Cheung:2016bym}. 
For a recent review we refer to \cite{Aoki:2016frl}. There exists a significant data set for $D$-meson masses at various unphysical quark masses. We consider data sets where the pion and kaon masses 
are smaller than about 600 MeV only. Once we determined the LECs in our mass formula, the $D$-meson masses can be computed at any values for the up, down and strange quark masses, 
sufficiently small as to justify the application of the chiral extrapolation. 

Though in principle such an analysis can be done at different chiral orders, we do so using the subtracted loop expressions (\ref{result-loop})  in (\ref{gap-equation-A})
with the scalar loop functions as worked out previously for the finite box case  in  \cite{Lutz:2014oxa}. It is a matter of convenience to perform  
our fits using the full one-loop functions rather than any truncated form. Therewith the finite volume corrections specific to the various chiral moments, 
whose explicit derivation would require further tedious algebra, are not required. This strategy is justified since we have demonstrated with Tab. VII that the full loop function is reproduced 
quite accurately by its N$^3$LO approximation, with a residual uncertainty for the D meson masses of about 3 MeV only. It is emphasized that such a point of view relies heavily 
on our reorganized chiral expansion approach, which is formulated in terms of physical meson masses.

While for instance in  \cite{Kalinowski:2015bwa,Mohler:2011ke}
the extrapolation towards the physical point was the focus the purpose of our study is the extraction of the low-energy constants of the chiral Lagrangian. Therefore 
a different strategy is used in our work.  We use the empirical $D$-meson masses as an additional constraint in our analysis. For a  given pion and kaon mass we infer the 
quark masses from the one-loop mass formulae for the pseudo Goldstone bosons to be used in our expressions for the $D$ meson masses. Assuming that the lattice data can be properly moved 
to the physical charm quark mass the low-energy constants are obtained by a global fit to the QCD lattice data set. Altogether there are about 80 data points considered in our analysis.

A comprehensive published data set is from Mohler and Woloshyn \cite{Mohler:2011ke,Lang:2014yfa} based on the 
PACS-CS ensembles \cite{Aoki:2008sm}. The Fermilab approach is employed in implementing the 
valence charm-quark \cite{ElKhadra:1996mp,Oktay:2008ex}. In this approach, heavy-quark mass dependent counter terms 
are added in the heavy-quark action to systematically reduce discretization effects. The valence charm-quark mass dependence is 
parameterized by  a hopping parameter $\kappa_c$, which is tuned to match the average of the physical 
kinematic $D$-meson masses. In Tab. \ref{tab:PACS} we recall the relevant results, which are the pion, kaon and the four $D$ meson masses 
in units of the lattice spacing $a$. The levels for the $D$ mesons as given in Tab. \ref{tab:PACS} are not the masses rather energies measured 
relative to some fixed reference. In turn only mass differences of $D$ mesons are constrained by that table in our studies. 

\begin{table}[t]
\setlength{\tabcolsep}{1.5mm}
\renewcommand{\arraystretch}{0.9}
\begin{center}
\begin{tabular}{l|ll|llll}
 & $a \,m_{\pi }$ & $a\, m_K$ & $a\, E_D$ & $a\, E_{D_s}$ & $a\, E_{D^*}$ & $a\, E_{D_s^*}$ \\
\hline

$32^3\times 64$ & 0.0717(32)     &  0.2317(6)     & 0.7765(12)  &  0.8197(24)   &  0.8447(27)   & 0.8850(24) \\
$32^3\times 64$ & 0.13593(140)   &  0.27282(103)  & 0.78798(82) &  0.83929(26)  &  0.85776(122) & 0.90429(43) \\
$32^3\times 64$ & 0.17671(129)   &  0.26729(110)  &   -         &  0.82848(40)  &    -          & 0.89015(69) \\
$32^3\times 64$ & 0.18903(79)    &  0.29190(67)   & 0.79580(61) &  0.84000(36)  &  0.86327(99)  & 0.90429(60) 
\end{tabular}
\vskip0.2cm
\caption{Meson masses and energy levels in units of the lattice spacing $a$ as taken from \cite{Mohler:2011ke,Lang:2014yfa} and \cite{Aoki:2008sm}. Statistical errors are given only.
The results are based on ensembles from PACS for which their estimate of the lattice spacing is $a = 0.0907\,(13)$ fm. }
\label{tab:PACS}
\end{center}
\end{table}

\begin{figure}[b]
\includegraphics[height=5cm]{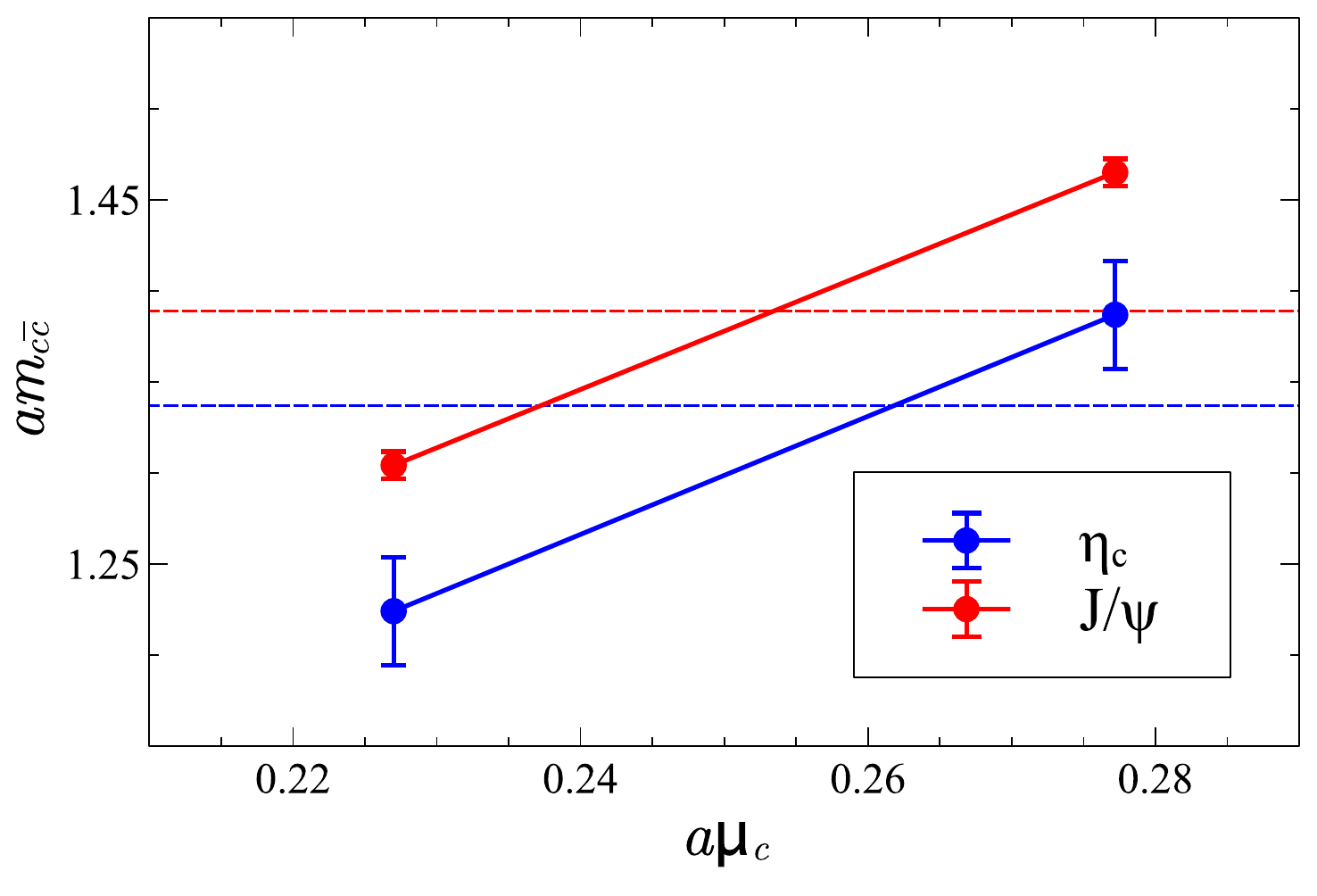}
\vskip-0.5cm
\caption{The interpolation of charmonium masses to determine $\mu_c$, at given $a=0.0885$ fm. The ensemble is chosen with $a\, m_\pi = 0.1240$. 
The physical values of $a\,M_{\eta_c}$ and $a\,M_{J/\psi}$ are indicated by the dashed lines. }
\label{fig:muc}
\end{figure}

\begin{table}[t]
\renewcommand{\arraystretch}{0.9}
\setlength{\tabcolsep}{2.5mm}
\begin{center}
 \begin{tabular}{l|l|cc|ccccccc}
                    &  $a$ [fm]   &  $a\,m_{\pi}$        & $a\,m_{K}$     & $a\,\mu_c$        & $a \,M_{\eta_c}$  & $a \,M_{J/\Psi}$  \\  \hline
 $48^3\times96$     &  0.0619     &  0.0703(4)           & 0.1697(3)      &0.2230             & 1.0595(2)         & 1.1006(3)  \\ 
                    &             &                      &		  &0.1919             & 0.9570(2)         & 1.0003(4)  \\ 
 $48^3\times96$     &  0.0619     &  0.0806(3)           & 0.1738(5)      &0.2227             & 1.0579(2)         & 1.0989(4)  \\ 
                    &             &                      & 	          &0.1727             & 0.8915(2)         & 0.9364(5)  \\ 
 $48^3\times96$     &  0.0619     &  0.0975(3)           & 0.1768(3)      &0.2230             & 1.0591(1)         & 1.1002(3)  \\ 
                    &             &                      &                &0.1727             & 0.8919(1)         & 0.9370(3)  \\  \hline 
 $32^3\times64$     &  0.0815     &  0.1074(5)           & 0.2133(4)      &0.2230             & 1.3194(2)         & 1.3835(4)  \\ 
                    &             &                      &                &0.1727             & 1.1567(2)         & 1.2233(4)  \\ 
 $32^3\times64$     &  0.0815     &  0.1549(2)           & 0.2279(2)      &0.2230             & 1.3251(1)         & 1.3903(2)  \\ 
                    &             &                      &                &0.1727             & 1.1573(1)         & 1.2253(2)  \\ 
 $24^3\times48$     &  0.0815     &  0.1935(4)           & 0.2430(4)      &0.2230             & 1.3179(3)         & 1.3837(4)  \\ 
                    &             &                      &                &0.1727             & 1.1582(3)         & 1.2273(4)  \\  \hline
 $32^3\times64$     &  0.0885     &  0.1240(4)           & 0.2512(3)      &0.2772             & 1.3869(1)         & 1.4649(3)  \\ 
                    &             &                      &                &0.2270             & 1.2241(2)         & 1.3042(4)  \\ 
 $32^3\times64$     &  0.0885     &  0.1412(3)           &0.2569(3)       &0.2768             & 1.3859(1)         & 1.4636(3)  \\ 
                    &             &                      &                &0.2389             & 1.2642(1)         & 1.3430(3)  \\ 
 $24^3\times48$     &  0.0885     &  0.1440(6)           &0.2589(4)       &0.2768             & 1.3863(2)         & 1.4645(4)  \\ 
                    &             &                      &                &0.2389             & 1.2645(2)         & 1.3442(5)  \\ 
 $24^3\times48$     &  0.0885     & 0.1988(3)            &0.2764(3)       &0.2929             & 1.4273(2)         & 1.5069(4)  \\ 
                    &             &                      &                &0.2299             & 1.2353(2)         & 1.3172(5) 
\end{tabular}
\caption{Meson masses in units of the lattice spacing $a$ based on the ensembles of the ETM collaboration. The values in the table are provided to us by the authors 
of \cite{Kalinowski:2015bwa}. Statistical errors are given only. The data correspond to three different $\beta_{QCD} =1.90,1.95,2.10 $ values for which in \cite{Alexandrou:2013joa} an estimate 
of the lattice scale is provided with $a = 0.0934(37),0.0820(37), 0.0644(26) $ fm respectively.}
\label{tab:ETMC}
\end{center}
\end{table}

\begin{table}[t]
\renewcommand{\arraystretch}{1.15}
\setlength{\tabcolsep}{2.5mm}
\begin{center}
 \begin{tabular}{ll|cccccc}
 $a\,m_{\pi}$   & $a \,m_{K}$  &  $a \,M_{D}$   &  $a\, M_{D_s}$  &$a \,M_{D^*}$     &$a\, M_{D^*_s}$  &$a M_{J/\psi}$  \\ \hline
 0.0703(4)      & 0.1697(3)    & 0.5905(52)     & 0.6236(56)      & 0.6466(86)       & 0.6770(28)      & 0.9715(20)     \\
 0.0806(3)      & 0.1738(5)    & 0.5906(64)     & 0.6234(57)      & 0.6506(26)       & 0.6763(11)      & 0.9697(21)     \\
 0.0975(3)      & 0.1768(3)    & 0.5913(50)     & 0.6229(57)      & 0.6486(28)       & 0.6764(15)      & 0.9703(21)     \\ \hline
 0.1074(5)      & 0.2133(4)    & 0.7840(122)    & 0.8159(147)     & 0.8568(44)       & 0.8905(34)      & 1.2791(55)     \\
 0.1549(2)      & 0.2279(2)    & 0.7895(128)    & 0.8183(144)     & 0.8678(47)       & 0.8950(39)      & 1.2828(55)     \\
 0.1935(4)      & 0.2430(4)    & 0.7934(148)    & 0.8175(151)     & 0.8745(38)       & 0.8965(41)      & 1.2818(58)     \\ \hline
 0.1240(4)      & 0.2512(3)    & 0.8514(181)    & 0.8953(206)     & 0.9356(28)       & 0.9806(45)      & 1.3890(75)     \\ 
 0.1412(3)      & 0.2569(3)    & 0.8544(168)    & 0.8972(208)     & 0.9363(41)       & 0.9802(45)      & 1.3895(75)     \\ 
 0.1440(6)      & 0.2589(4)    & 0.8552(159)    & 0.8978(208)     & 0.9403(23)       & 0.9844(45)      & 1.3906(77)     \\ 
 0.1988(3)      & 0.2764(3)    & 0.8599(184)    & 0.8950(219)     & 0.9487(60)       & 0.9841(66)      & 1.3882(79)    
\end{tabular}
\caption{$D$  and $J/\psi$ meson masses in units of the lattice scale $a$. 
The charm-quark mass is determined to reproduce the physical $J/\psi$ mass. This leads to 
$a\,\mu_c = 0.2535, 0.1902 $ and $0.1829$ for the three groups of ensembles. Statistical errors are given only.}
\label{tab:ETMC_JPsi}
\end{center}
\end{table}

\begin{table}[t]
\renewcommand{\arraystretch}{1.15}
\setlength{\tabcolsep}{2.5mm}
\begin{center}
\begin{tabular}{ll|cccccc}
 $a\,m_{\pi}$     & $a\,m_{K}$   &  $a\, M_{D}$     &  $a \,M_{D_s}$    &$a \,M_{D^*}$      &$a \,M_{D^*_s}$    &$a\, M_{\eta_c}$  \\ \hline
 0.0703(4)        & 0.1697(3)    & 0.5947(52)       & 0.6279(56)        & 0.6506(86)        & 0.6809(28)        & 0.9351(85)       \\ 
 0.0806(3)        & 0.1738(5)    & 0.5949(64)       & 0.6277(57)        & 0.6546(26)        & 0.6803(11)        & 0.9332(85)       \\ 
 0.0975(3)        & 0.1768(3)    & 0.5955(50)       & 0.6271(57)        & 0.6526(28)        & 0.6804(15)        & 0.9335(84)       \\ \hline
 0.1074(5)        & 0.2133(4)    & 0.7946(122)      & 0.8263(147)       & 0.8664(44)        & 0.9001(34)        & 1.2312(212)      \\ 
 0.1549(2)        & 0.2279(2)    & 0.8004(128)      & 0.8291(144)       & 0.8777(47)        & 0.9049(39)        & 1.2342(217)      \\ 
 0.1935(4)        & 0.2430(4)    & 0.8039(148)      & 0.8278(151)       & 0.8840(38)        & 0.9059(41)        & 1.2314(219)      \\ \hline
 0.1240(4)        & 0.2512(3)    & 0.8677(181)      & 0.9114(206)       & 0.9506(28)        & 0.9953(45)        & 1.3370(296)      \\ 
 0.1412(3)        & 0.2569(3)    & 0.8708(168)      & 0.9132(208)       & 0.9511(41)        & 0.9949(45)        & 1.3379(299)      \\ 
 0.1440(6)        & 0.2589(4)    & 0.8714(159)      & 0.9137(208)       & 0.9545(24)        & 0.9990(45)        & 1.3382(302)      \\ 
 0.1988(3)        & 0.2764(3)    & 0.8753(184)      & 0.9102(219)       & 0.9627(60)        & 0.9980(66)        & 1.3325(310)      
\end{tabular}
\caption{$D$ and $\eta_c$ meson masses in units of the lattice scale $a$.  The charm-quark mass was determined to reproduce the physical 
$\eta_c$ meson mass. This leads to $a\,\mu_c = 0.2618, 0.1957 $ and $0.1852$ for the three groups of ensembles. Statistical errors are given only.}
\label{tab:ETMC_EtaC}
\end{center}
\end{table}

Recently the group of Marc Wagner analyzed a large set of ensembles from the European Twisted Mass Collaboration (ETMC) \cite{Kalinowski:2015bwa,Cichy:2016bci}. 
Our analysis requires the $D$ meson masses evaluated at the physical charm quark mass. We are grateful to the authors of \cite{Kalinowski:2015bwa} for 
making available unpublished results, which allow us to independently extrapolate their lattice data to the physical charm quark mass. 
For each ensemble, the four $D$-meson masses but also the $\eta_c$ and $J/\Psi$ masses are computed at two different values of the charm valence-quark 
mass $\mu_c$. As a consequence of the discretization procedure there are corresponding pairs of meson masses that turn degenerate in the continuum limit. We use the 
notation $(\pm,\mp)$ and $(\pm,\pm)$ from \cite{Kalinowski:2015bwa,Cichy:2016bci}. In this work we focus on the $(\pm,\mp)$ states and use the masses of the partner states $(\pm,\pm)$ only as 
a rough estimate for the size of the discretization error. In the vicinity of the physical charm quark mass a linear behavior   
\begin{align}
	a\, M_H = \alpha_H + \beta_H \,a\, \mu_c\,,
	\label{ansatz-mc}
\end{align}
is expected to hold for all hadron masses. Since the chosen charm quark masses are close to the physical one the ansatz (\ref{ansatz-mc}) should be justified 
to sufficient accuracy. The parameters $\alpha_H$ and $\beta_H$ can be extracted from the data provided to us by Kalinowski and Wagner. 
In Tab. \ref{tab:ETMC} we show their results for the $\eta_c$ and $J/\Psi$ masses together with their preferred lattice spacing values $a$. 
Corresponding results for the $D$ meson masses are listed at the end of  Appendix B. It remains the 
task to determine the physical value for $\mu_c$. Since one would neither expect a significant dependence of the $\eta_c$ nor of the $J/\psi$ meson mass on the precise value of the up, 
down and strange quark masses, one may contemplate to use either of the two masses to obtain a good estimate for $\mu_c$. Both scenarios are scrutinized in the following based on the 
data of Kalinowski and Wagner. To fix the charm quark mass we always choose the ensemble with the lightest up and down quark masses. In addition 
the lattice spacing $a$ as recalled in Tab. \ref{tab:ETMC} is assumed. A typical example for this procedure is shown in Fig. \ref{fig:muc} where a sizable uncertainty for the 
extracted value of $\mu_c$ is observed.

How such an uncertainty propagates into the masses of the $D$ mesons is shown in Tab. \ref{tab:ETMC_EtaC} and Tab. \ref{tab:ETMC_JPsi} which are based on the charm quark masses from 
the $\eta_c$ and the $J/\psi$ meson respectively. As expected this uncertainty in the charm quark mass is reduced for the ensembles that correspond to even smaller lattice spacings 
with $a = 0.0815$ fm and $a = 0.0619$ fm. This can be inferred by a comparison of Tab. \ref{tab:ETMC_EtaC} and Tab. \ref{tab:ETMC_JPsi}.
While the center value of the masses in Tab. \ref{tab:ETMC_EtaC} and Tab. \ref{tab:ETMC_JPsi} are derived from the $(\pm,\mp)$ states of Appendix B, the shown error bars entail an estimate 
for the total error including the statistical error and the uncertainty from the discretization procedure. We take half of the splittings of the two modes, $(\pm,\mp)$ and $(\pm,\pm)$, for the latter. 

It is immediate from Tab. \ref{tab:ETMC_EtaC} and Tab. \ref{tab:ETMC_JPsi} that the $D$ meson masses are quite sensitive to the precise charm quark mass used but also to the 
lattice scale $a$ assumed. We note that, for instance, there exist two distinct values for the
lattice spacing for the coarsest ensembles: The value $a = 0.0885(36)$ fm obtained from the pion decay constant \cite{Carrasco:2014cwa} and $a = 0.0920(21)$ fm obtained 
from the nucleon mass \cite{Alexandrou:2013joa}. We conclude  that it may be of advantage to determine the lattice scale and the charm-quark mass from the $D$ meson masses directly. Such a procedure 
is expected to minimize the discretization errors for the $D$ meson masses. This is what we will do in the following. All information required for such a strategy is provided with 
Tab. \ref{tab:ETMC_EtaC} and Tab. \ref{tab:ETMC_JPsi}, from which the parameters $\alpha_H$ and $\beta_H$ in (\ref{ansatz-mc}) can be read off.

There are yet three further sources of QCD lattice data on the $D$ meson masses, which we will discuss briefly \cite{Liu:2012zya,Na:2012iu,Cheung:2016bym}. 
The two data sources  \cite{Liu:2012zya,Na:2012iu} are partial to the extent that not all four 
$D$ meson masses are provided. Only the pseudo-scalar masses are computed. The results of \cite{Liu:2012zya} rely on previous studies by the LHP collaboration \cite{WalkerLoud:2008bp}, who 
use a mixed action framework with domain-wall valence quarks but staggered sea-quark ensembles generated by  MILC \cite{Orginos:1999cr,Orginos:1998ue,MILC2001,MILC2004,Bazavov:2009bb}.
For the charm quark they use a relativistic heavy-quark action motivated by the Fermilab approach \cite{ElKhadra:1996mp,Oktay:2008ex}. 
In Tab. \ref{tab:LHPC} we summarize the relevant masses that are considered in our study.

The results of the HPQCD Collaboration \cite{Na:2012iu} are based on MILC ensembles together with a highly improved staggered valence quark (HISQ)
action. The HISQ action has since been used very successfully in simulations involving the charm quark such as for charmonium and for $D$ and $D_s$ meson decay 
constants. In Tab. \ref{tab:HPQCD} we collect the relevant masses in units of the lattice spacing for the configurations on three coarse and two fine lattices.

\begin{table}[t]
\setlength{\tabcolsep}{1.5mm}
\renewcommand{\arraystretch}{0.9}
\begin{center}
 \begin{tabular}{l|ll|lll}
 & ${a \,m_{\pi }}$ & ${a \,m_K}$ & ${a \,M_D}$ & ${a\, M_{D_s}}$ \\ \hline
 $20^3\times 64$ & 0.1842(7) & 0.3682(5) & 1.2081(13) & 1.2637(10) \\
 $20^3\times 64$ & 0.2238(5) & 0.3791(5) & 1.2083(11) & 1.2635(10) \\
$20^3\times 64$ & 0.3113(4) & 0.4058(4) & 1.2226(13) & 1.2614(12) \\
 $20^3\times 64$ & 0.3752(5) & 0.4311(5) & 1.2320(11) & 1.2599(12) 
\end{tabular}
\caption{$D$ meson masses based on ensembles of MILC \cite{MILC2001} as used by LHPC \cite{WalkerLoud:2008bp}. The results are recalled from \cite{Liu:2012zya}  in units of the lattice spacing $a$. 
The lattice spacing is $a \simeq 0.12$ fm.}
\label{tab:LHPC} 
\end{center}
\end{table}

Most recently the 
Hadron Spectrum Collaboration (HSC) computed the excited open-charm meson spectrum in a finite QCD box \cite{Moir:2016srx,Cheung:2016bym}. Results for the for the D mesons masses based on an ensemble with a pion 
mass of about 390 MeV are published in \cite{Moir:2016srx} and recalled in Tab. \ref{tab:HSC}. For an additional ensemble at smaller pion masses studies are on going \cite{Cheung:2016bym}.

\begin{table}[t]
\setlength{\tabcolsep}{1.5mm}
\renewcommand{\arraystretch}{0.9}
\begin{center}
 \begin{tabular}{l|ll|lll}
                    &   ${a \,m_{\pi }}$ & ${a \,m_K}$  & ${a \,M_D}$  & ${a\, M_{D_s}}$ \\ \hline
 $24^3\times 64$    & 0.1599(2)          & 0.3122(2)    & 1.1395(7)    & 1.1878(3)       \\
 $20^3\times 64$    & 0.2108(2)          & 0.3285(3)    & 1.1591(7)    & 1.2014(4)       \\
 $20^3\times 64$    & 0.2931(2)          & 0.3572(2)    & 1.1618(5)    & 1.1897(3)       \\ \hline 
 $28^3\times 96$    & 0.1344(2)          & 0.2286(2)    & 0.8130(3)    & 0.8471(2)      \\
 $28^3\times 96$    & 0.1873(1)          & 0.2458(2)    & 0.8189(3)    & 0.8434(2)       \\
\end{tabular}
\caption{$D$ meson masses from the HPQCD Collaboration in units of the lattice spacing $a$ as taken from \cite{Na:2012iu} and \cite{Na:2010uf}. The studies are based on ensembles of MILC \cite{Bazavov:2009bb}.
The lattice spacing is $a = 0.119(2)$ fm and $a = 0.0846(7)$ fm for the two sets of data respectively.  }
\label{tab:HPQCD}
\end{center}
\end{table}

We note that the charm-quark mass in \cite{Liu:2012zya}, \cite{Na:2012iu} and \cite{Moir:2016srx} was not adjusted to the $D$ meson masses. While in \cite{Liu:2012zya} 
the spin average of the physical $J/\Psi$ and $\eta_c$ meson mass was used, in \cite{Na:2012iu} the charm quark mass was tuned to the physical $\eta_c$ mass. 
In both cases we cannot exclude uncertainties significant to our analysis. In order to minimize any bias from a possibly 
imprecise charm-quark mass determination we consider only mass differences from Tab. \ref{tab:LHPC}, Tab. \ref{tab:HPQCD} and Tab. \ref{tab:HSC} in our fits. 
In addition we fine tune the lattice scales.  As we have seen in case of the ETMC results such a procedure reduces any possible bias significantly.

We introduce a  universal parameter $\Delta_c$  of the form
\begin{eqnarray}
a\,M_H \to a\, M_H + (1 + \epsilon_H )\,a\,\Delta_c\,, \qquad {\rm with }\qquad \epsilon_H \simeq 0\,,
\label{def-Deltac}
\end{eqnarray}
which is supposed to fine tune the choice of the charm quark mass. In principle the values of $\epsilon_H$ depends on the type of D meson considered but also the $\beta_{\rm QCD} $ value of the 
ensemble considered. The value $(1+ \epsilon_H)\,a\,\Delta_c$ is to be added to $a\,M_H$ as collected in Tabs. \ref{tab:ETMC_EtaC}-\ref{tab:HSC}

For the ETMC masses the magnitude of $\epsilon_H$ can be extracted from Tab.  \ref{tab:ETMC_JPsi} and Tab. \ref{tab:ETMC_EtaC}, where we insist on the normalization condition that 
$\epsilon_H = 0$ for the D meson on the ensemble with the lightest pion mass. Then values for  $|\epsilon_H |$  
of about 0.1 arise in some cases at most. Such an estimate is 
not available for the other collaborations. For these other cases we put $\epsilon_H = 0$, which would arise in the heavy-quark mass limit. We would argue that a precise determination 
of $a\,\Delta_c$ and therewith the physical charm quark mass 
for a given ensemble requires the quantitative control of the chiral extrapolation formulae for the D meson masses.

We do not implement discretization effects in our chiral extrapolation approach since this would introduce a significant number of further unknown parameters into the game.  
For each lattice group such effects have to be worked out in the context of our chiral extrapolation scheme. 
As a consequence a fully systematic error analysis is not possible yet in our present study. Here we follow the strategy suggested in \cite{Lutz:2014oxa,Lutz:2018a} where 
the statistical error given by the lattice groups is supplemented by a systematic error in mean quadrature.  
We perform fits at different ad-hoc values for the systematic error. Once this error is sufficiently large the $\chi^2$ per data point should be close to one. 
In our current studies we arrive at the estimate of 5-10 MeV. 
In anticipation of our analysis of the lattice data set we collect the result of four representative fits. Their characteristics and defining assumptions will be discussed in more detail 
in the next sections.  

For a given ensemble the statistical errors in the lattice data are correlated. However, since the statistical error for any meson mass considered here 
is typically much smaller than our estimate for the systematic error such a correlation is of no relevance in our study. In contrast, the choice of the charm quark mass 
and the lattice scale setting, both of which we treat in detail, is a significant effect.

\begin{table}[t]
\setlength{\tabcolsep}{1.5mm}
\renewcommand{\arraystretch}{0.9}
\begin{center}
 \begin{tabular}{l|ll|lllll}
 & ${a_t \,m_{\pi }}$ & ${a_t \,m_K}$ & ${a_t \,M_D}$ & ${a_t\, M_{D_s}}$ & ${a_t \,M_{D^*}}$ & ${a_t\, M_{D^*_s}}$\\ \hline
 $24^3\times 128$ & 0.06906(13) & 0.09698(9) & 0.33265(7) & 0.34426(6) & 0.35415(17) & 0.36508(88) \\
 $32^3\times 256$ & 0.03928(18) & 0.08344(7) & -  & - & - & - & - 
\end{tabular}
\caption{$D$ meson masses from HSC  in units of the temporal lattice spacing \cite{Wilson:2015dqa,Moir:2016srx}. The lattice spacing is $3.5 \,a_t =  0.123(4)$ fm. It holds 
$a = a_s\simeq 3.5 \,a_t$.}
\label{tab:HSC}
\end{center}
\end{table}

\begin{table}[t]
\setlength{\tabcolsep}{1.5mm}
\renewcommand{\arraystretch}{0.9}
\begin{center}
\begin{tabular}{l|cccc} 
                                                         &  Fit 1    &  Fit 2  &  Fit 3   & Fit 4    \\ \hline

$a_{\rm PACS-CS}\,  \hfill \mathrm{[fm]}$                &  0.0934   &  0.0940 &  0.0935  &  0.0928 \\
$a\,\Delta_{c,\rm PACS-CS}$                              &  0.1067   &  0.1110 &  0.1119  &  0.1023 \\ \hline 

$a_{\rm LHPC} \,   \hfill \mathrm{[fm]}$                 &  0.1291   &  0.1267 &  0.1291  &  0.1291\\ 
$a\,\Delta_{c,\rm LHPC}$                                 &  0.0359   &  0.0087 &  0.0443  &  0.0381\\ \hline 

$a^{\beta \simeq 6.76}_{c,\rm HPQCD}\,   \hfill \mathrm{[fm]}$   &  0.1367   &  0.1359 &  0.1336  &  0.1367\\ 
$a\,\Delta^{\beta \simeq 6.76}_{c,\rm HPQCD}$                         &  0.1500   &  0.1494 &  0.1184  &  0.1500\\

$a^{\beta \simeq 7.09}_{c,\rm HPQCD}\,   \hfill \mathrm{[fm]}$        &  0.0953   &  0.0991 &  0.0970  &  0.0992\\ 
$a\,\Delta^{\beta \simeq 7.09}_{c,\rm HPQCD}$                          &  0.0936   &  0.1336 &  0.1049  &  0.1282\\  \hline 

$a^{\beta = 1.90 }_{c,\rm ETMC}\,   \hfill \mathrm{[fm]}$        &  0.1018   &  0.0996 &  0.1025  &  0.1027\\
$a\,\Delta^{\beta = 1.90}_{c,\rm ETMC}$                          &  0.0983   &  0.0747 &  0.1041  &  0.1086\\ 

$a^{\beta = 1.95}_{\rm ETMC}\,   \hfill \mathrm{[fm]}$           &  0.0934   &  0.0925 &  0.0928  &  0.0943\\
$a\,\Delta^{\beta =1.95 }_{c,\rm ETMC}$                          &  0.0908   &  0.0817 &  0.0817  &  0.1005\\ 

$a^{\beta = 2.10}_{\rm ETMC}\,   \hfill \mathrm{[fm]}$           &  0.0695   &  0.0704 &  0.0695  &  0.0699 \\
$a\,\Delta^{\beta = 2.10}_{c,\rm ETMC}$                          &  0.0629   &  0.0728 &  0.0608  &  0.0659\\ \hline 

$a_{\rm HSC} \,   \hfill \mathrm{[fm]}$                          &  0.1211   &  0.1243 &  0.1242  &  0.1242 \\ 
$a\,\Delta_{c,\rm HSC}$                                          &  0.0050   &  0.0337 &  0.0328  &  0.0343 \\ \hline \\

$10^3\,(L_4 - 2\,L_6)\, $    & -0.1395   & -0.1112   & -0.1102  & -0.1575  \\
$10^3\,(L_5 - 2\,L_8)\, $    &  0.0406   & -0.0940   & -0.0235  & -0.0370  \\
$10^3\,(L_8 + 3\,L_7)\, $    & -0.5130   & -0.5127   & -0.4950  & -0.5207   \\
$m_s/ m $                    &  26.547   &  26.187   &  26.596  &  26.600 

\end{tabular}

\caption{Results for Fit 1 - Fit 4. The low-energy constants $L_n$ are at the renormalization scale $\mu = 0.77$ GeV.
The offset parameters $a\,\Delta_c$ is introduced in (\ref{def-Deltac}). We use $f = 92.4$ MeV throughout this work. 
A more detailed discussion of the four fit scenarios is given in Section VII and VIII.}
\label{tab:lattice-scale-Fits}
\end{center}
\end{table}

Our fit procedure goes as follows. For a given lattice ensemble we take the pion and kaon masses as given in lattice units and then determine from the one-loop expressions 
(28) in \cite{Lutz:2018a} the quark masses for that ensemble. They depend on the three particular linear combinations of the low-energy constants of Gasser and Leutwyler \cite{Gasser:1984gg}.  
One combination can be fixed by the request that the $\eta $ meson mass is reproduced at physical quark masses. The other two are determined by our fit to lattice data. With those the quark mass ratio $m_s/m$ 
is determined. This is analogous to 
\cite{Lutz:2018a} where those low-energy constants are determined from a fit to the lattice data on baryon masses. In Tab. \ref{tab:lattice-scale-Fits} we show our results for four distinct fit scenarios, which 
are reasonably close to the results of \cite{Lutz:2018a}. The quark mass ratio $m_s/m$ as given in the last row of the table is compatible with the latest result of ETMC \cite{Carrasco:2014cwa} with 
$m_s/m = 26.66(32)$. In Tab. \ref{tab:lattice-scale-Fits} also the lattice scale parameters $a$ together with the offset charm-quark mass parameters $\Delta_c$  are presented. 
All fits reproduce the D meson masses of all ensembles recalled in this work quite well. The table illustrates that the offset parameters are almost always non 
negligible. Our values for the lattice scale can be compared with the ones advocated by the various lattice groups as recalled in the tables of this section. Any deviation from 
such values may be viewed as a reflection of significant discretization effects. Those depend on the specifics of the scale setting. The aim of our work is to minimize such discretization effects 
in the open-charm  meson sector of QCD. We find interesting that in particular our values for ETMC are amazingly close to those lattice scales obtained in our previous analysis of the baryon masses  
from the identical lattice ensembles \cite{Lutz:2018a}.

\begin{figure}[t]
\center{
\includegraphics[keepaspectratio,width=0.97\textwidth]{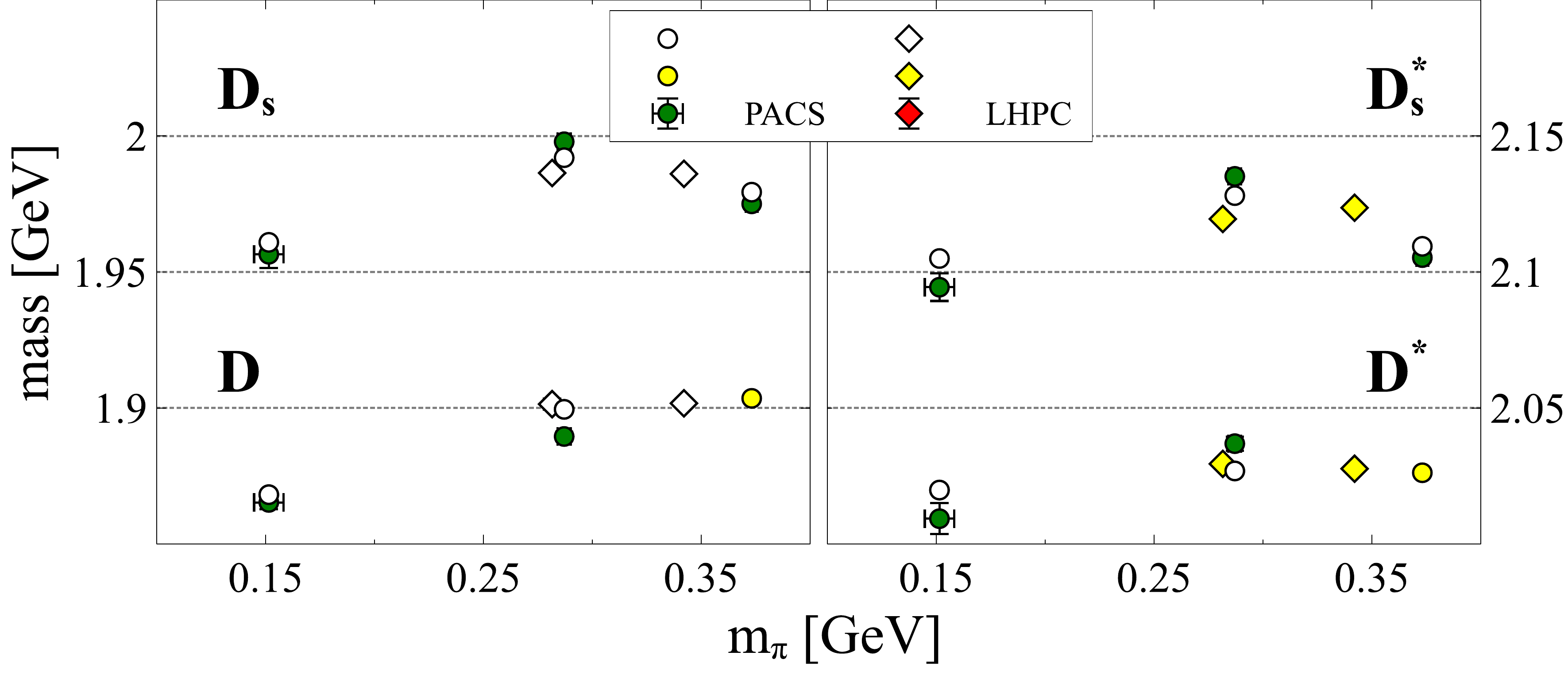} }
\vskip-0.2cm
\caption{\label{fig:PACS-LHPC} D meson masses from Fit 1  compared to results based on lattice ensembles from PACS-CS and LHPC \cite{Mohler:2011ke,Liu:2012zya,Lang:2014yfa}. The yellow symbols present 
our predictions for the case where no lattice values are available yet.  }
\end{figure}

\begin{figure}[t]
\center{
\includegraphics[keepaspectratio,width=0.97\textwidth]{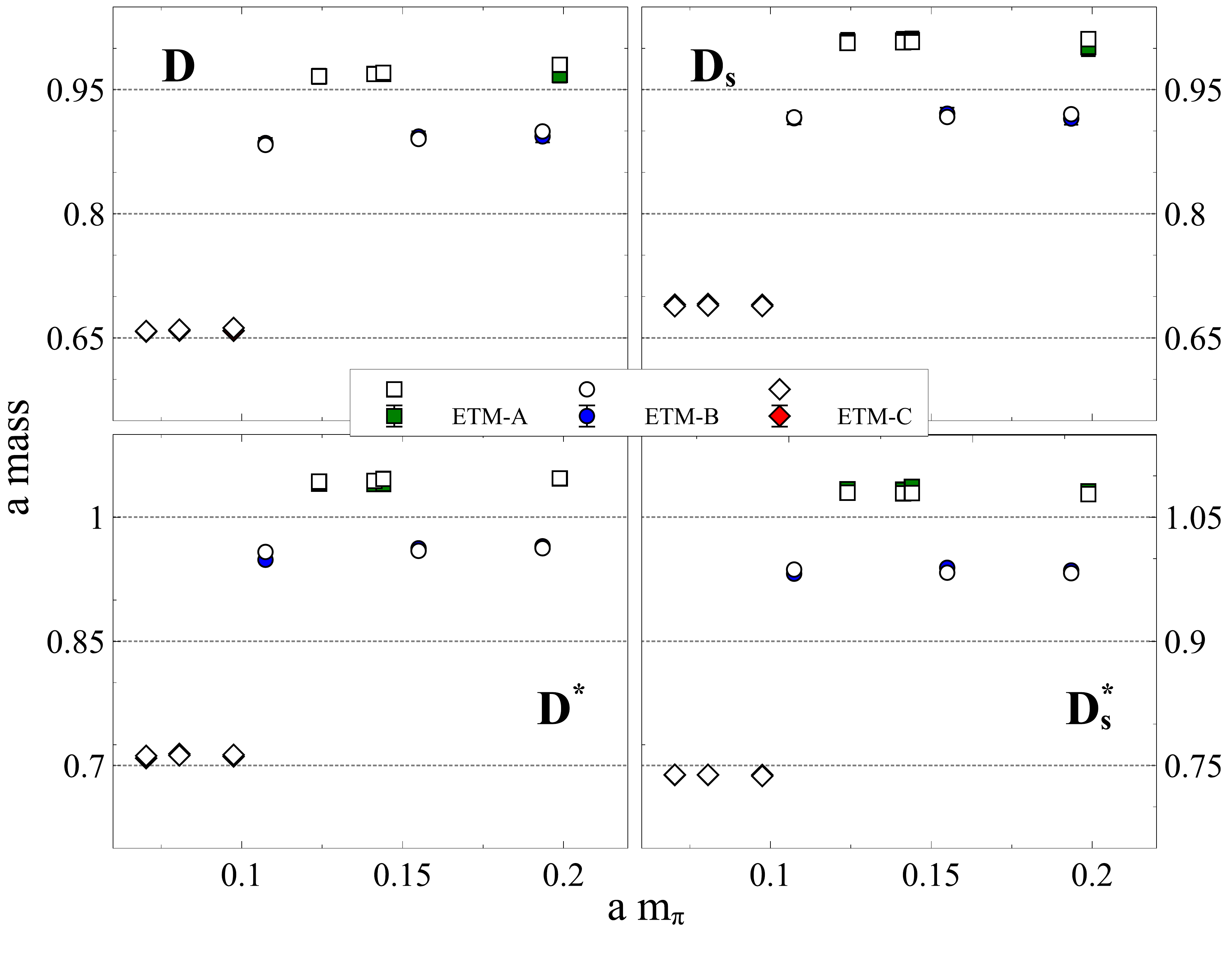} }
\vskip-0.2cm
\caption{\label{fig:ETMC} D meson masses from Fit 1 in lattice units as compared to results from  ETMC  \cite{Kalinowski:2015bwa}.  }
\end{figure}

\begin{figure}[t]
\center{
\includegraphics[keepaspectratio,width=0.97\textwidth]{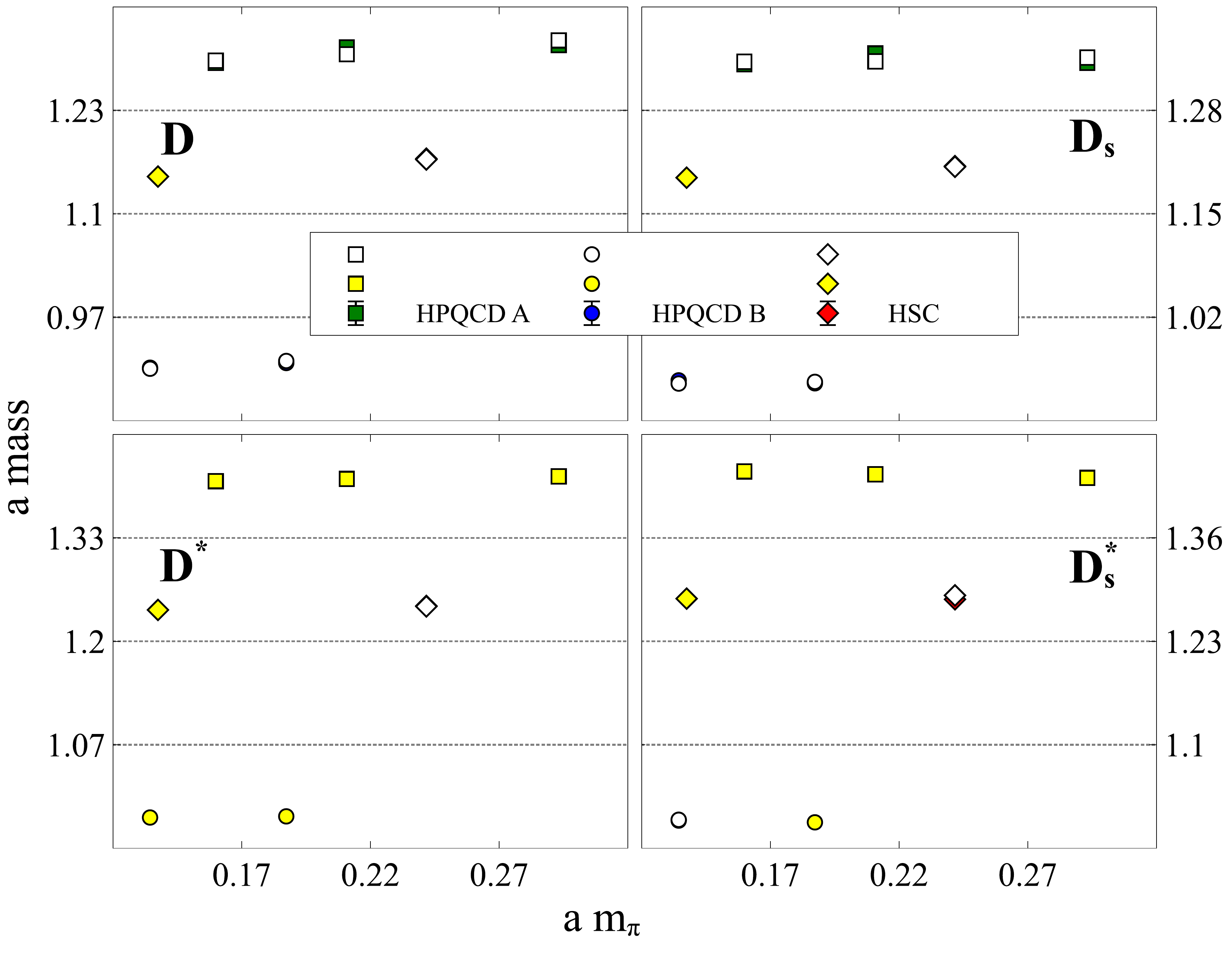} }
\vskip-0.2cm
\caption{\label{fig:HPQCD} D meson masses from Fit 1 in lattice units as compared to results from  HPQCD and HSC \cite{Na:2010uf,Na:2012iu,Moir:2016srx}. The yellow symbols present 
our predictions for the case where no lattice values are available yet. Note that we show the HPQCD data in units of their spatial lattice spacing but the HSC data in units of 3.5 times their temporal lattice spacing. }
\end{figure}

The quality of the data description is illustrated at hand of Fit 1 for which we offer a comparison with the lattice data in Fig. \ref{fig:PACS-LHPC}-\ref{fig:HPQCD}.  A more quantitative 
comparison with $\chi^2$ values  will be provided in the next section.
In all figures open symbols correspond to results from our chiral extrapolation approach. They lie always on top of the lattice points, which are shown with either green, blue or 
red filled symbols. In case that for a considered lattice ensemble there is no lattice result for the considered D meson mass available our theory prediction is presented with a yellow filled symbol. 

In Fig. \ref{fig:PACS-LHPC} we scrutinize the lattice results of \cite{Mohler:2011ke,Liu:2012zya,Lang:2014yfa} as recalled in Tab. VIII and Tab. XII. Note that the strange quark mass varies along the 
different pion masses of the figure. The D meson masses are shown in units of GeV, where the 
lattice scales for the two groups are taken from Tab. \ref{tab:lattice-scale-Fits}. In addition the effect of the fine tuned charm quark mass in terms of the appropriate $\Delta_c$ values in 
Tab. \ref{tab:lattice-scale-Fits} is considered. From Fig. \ref{fig:PACS-LHPC} we conclude that all masses from \cite{Mohler:2011ke,Liu:2012zya,Lang:2014yfa} are recovered well with an uncertainty of less
than 10 MeV. The figures include predictions of 5  meson masses shown with yellow symbols for which there do not exist so far corresponding values from the lattice  collaborations. Note that in some cases the 
lattice data point is fully covered with our chiral extrapolation symbol. This signals an almost perfect reproduction of the lattice point. 

We continue with Fig. \ref{fig:ETMC} where the predictions of ETMC are compared to our results. Here the meson masses are shown in lattice units. 
This permits an efficient presentation of the results at three distinct $\beta_{QCD}$ values. The data set of ETMC  is of particular importance for the 
chiral extrapolation since it offers masses for the $J^P = 0^-$ and $J^P = 1^-$ states consistently. The figure illustrates that such data can be reproduced accurately 
for all $\beta_{QCD}$ values. Note that the effect of a fine tuned charm quark mass is considered again in terms of the parameter $\Delta_c$ properly taken from Tab. \ref{tab:lattice-scale-Fits}

It remains a discussion of Fig. \ref{fig:HPQCD}, which combines results from HPQCD and HSC \cite{Na:2010uf,Na:2012iu,Moir:2016srx}. Again the meson masses are shown in lattice units with 
$\Delta_c$ from Tab. \ref{tab:lattice-scale-Fits}. The reproduction of the lattice data is again impressive. The reader is pointed to the fact that we predict 13 masses with yellow symbols for which 
there are not yet values available from the lattice groups. Of particular interest are the mass predictions for the second ensemble of HSC as recalled in Tab. \ref{tab:HSC}. For this ensemble the 
authors are informed that the HSC is currently computing various scattering observables. We will return to this issue below.

\begin{figure}[t]
\center{
\includegraphics[keepaspectratio,width=0.97\textwidth]{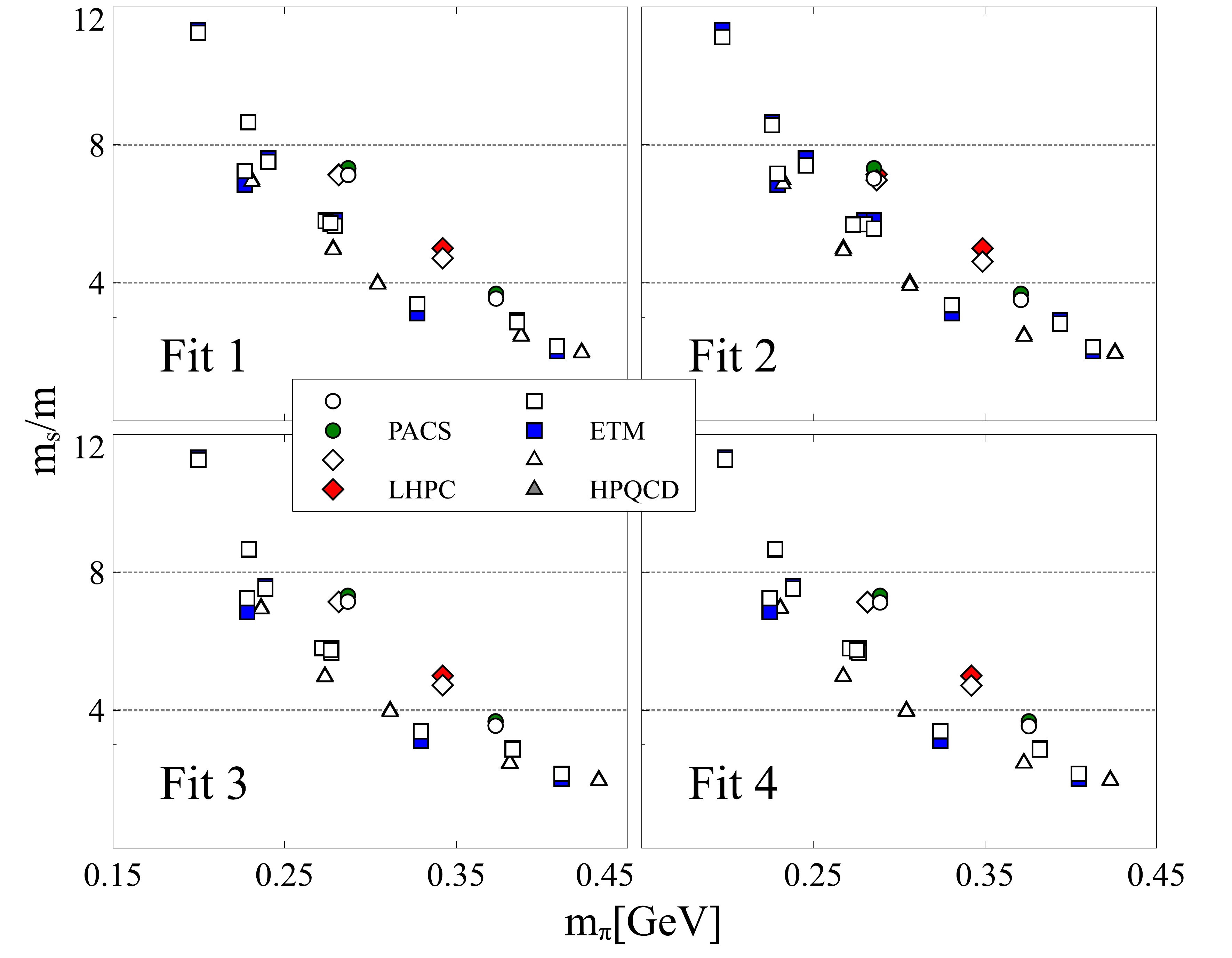} }
\vskip-0.2cm
\caption{\label{fig:ratio} 
The quark mass ratios $m_s/m$ are shown for the various lattice ensembles considered. Closed symbols show the values from the lattice collaborations, open symbols our results. }
\end{figure}

The section is closed with a brief discussion of the quark masses. Given the different fit scenarios of Tab. \ref{tab:lattice-scale-Fits} their values can be computed for any lattice ensemble 
for which the pion and kaon mass are measured on a specified lattice volume, where again here we ignore discretization effects. Within a chiral Lagrangian approach only ratios of the quark masses 
can be determined. This is so since only products of $B_0\,m$ or $B_0 \,m_s$ occur. In Fig. \ref{fig:ratio} such ratios are confronted with corresponding ratios from the various lattice groups. While 
our values are given by open symbols the lattice results by closed symbols. We follow here our convention that the open symbols are always on top of the closed symbols. An amazingly consistent 
pattern occurs. We note that the determination of the quark mass ratios depends on the action used, and may be quite involved due to non-trivial renormalization effects. 
Most straight forward are the results from HPQCD and ETMC \cite{Na:2010uf,Herdoiza:2013sla,Carrasco:2014cwa} where it is stated that the quark-mass ratio remains unrenormalized. 
The PACS and  LHPC  collaborations made significant efforts to  control their non-trivial renormalization effects in the quark masses \cite{Aoki:2008sm,Liu:2012zya}. As shown in 
our figure all quark-mass ratios appear consistent with a universal set of chiral low-energy parameters as given in Tab. \ref{tab:lattice-scale-Fits}. 
All four fit scenarios lead to almost indistinguishable results for the quark masses. The small spread in the low-energy constants is not significant.

\clearpage

\section{Low-energy constants from QCD lattice data}

We report on our efforts to adjust the low-energy parameters to the D meson masses as evaluated by the various lattice groups. 
Our first observation is that the available data set is not able to determine a unique parameter set without additional 
constraints. Therefore it would be highly desirable to evaluate the D meson masses with $J^P =0^-$ and $J^P = 1^-$ quantum numbers 
on further QCD lattice ensembles with unphysical pion and kaon masses.  

Typically solutions can be found with similar quality in the lattice data reproduction but quite different values for the low-energy parameters. This problem 
is amplified by the unknown size of the underlying systematic error from discretization effects. Almost always the size of the statistical errors 
given by the lattice groups is negligible, and it is expected that the systematic error is dominating the total error budget. In turn it is unclear whether 
a parameter set with a better $\chi^2$ value is more realistic than a solution with a worse $\chi^2$.  The D meson masses may be over fitted. 

To actually perform the fits is a computational challenge. For any set of the low-energy parameters four coupled non-linear equations are to be solved on each lattice ensemble  
considered. We apply the evolutionary algorithm of GENEVA 1.9.0-GSI \cite{Geneva} with runs of a population size 4000 on 100 parallel CPU cores.

\begin{table}[t]
\setlength{\tabcolsep}{2.5mm}
\renewcommand{\arraystretch}{1.18}
\begin{center}
\begin{tabular}{l|rrrr}
                                            &  Fit 1     &  Fit 2    & Fit 3    & Fit 4      \\ \hline
                                            
$ M\;\;$ \hfill [GeV]                       &  1.8762    &  1.9382   &  1.9089  & 1.8846  \\
$\Delta $\hfill [GeV]                       &  0.1873    &  0.1876   &  0.1834  & 0.1882   \\ \hline

$ c_0$                                      &  0.2270    &  0.3457   &  0.2957  &  0.3002  \\
$ \tilde c_0$                               &  0.2089    &  0.3080   &  0.2737  &  0.2790  \\
$ c_1$                                      &  0.6703    &  0.9076   &  0.8765  &  0.8880  \\
$ \tilde c_1$                               &  0.6406    &  0.9473   &  0.8420  &  0.8583  \\
$ c^r_2 = \tilde c^r_2$                         & -0.5625    & -2.1893   & -1.6224  & -1.3046  \\
$ c^r_3 = \tilde c^r_3$                         &  1.1250    &  4.4956   &  3.2448  &  2.9394  \\
$ c^r_4 = \tilde c^r_4$                         &  0.3644    &  2.0012   &  1.2436  &  0.9122  \\    
$ c^r_5 = \tilde c^r_5$                         & -0.7287    & -4.1445   & -2.4873  & -2.1393  \\    \hline

$d^c_1\,\hfill\mathrm{[GeV^{-2}]}$                   & 1.8331    &  1.6937   &  1.6700   & 1.9425   \\
$\tilde d^c_1\,\hfill\mathrm{[GeV^{-2}]}$            & 1.6356    &  1.6586   &  1.4701   & 1.7426   \\
$d^c_2= \tilde d^c_2\,\hfill\mathrm{[GeV^{-2}]}$     & 1.0111    &  0.9954   &  0.8684   & 1.0032   \\
$d^c_3\,\hfill\mathrm{[GeV^{-2}]}$                   & 0.1556    &  0.0679   &  0.1531   & 0.1109   \\
$\tilde d^c_3\,\hfill\mathrm{[GeV^{-2}]}$            & 0.2571    &  0.1640   &  0.2597   & 0.2143   \\
$d^c_4= \tilde d^c_4\,\hfill\mathrm{[GeV^{-2}]}$     & 0.8072    &  1.6392   &  0.8607   & 1.1255   

\end{tabular}
\caption{The low-energy constants from a fit to the pseudo-scalar and vector D meson masses based on QCD lattice ensembles of 
the PACS-CS, MILC, ETMC and HSC as described in the text. Each parameter set 
reproduces the isospin average of the empirical D meson masses from the PDG.
}
\label{tab:FitParametersA}
\end{center}
\end{table}

\begin{table}[t]
\setlength{\tabcolsep}{2.5mm}
\renewcommand{\arraystretch}{1.18}
\begin{center}
\begin{tabular}{l|cccc|c} 

                                                      &  Fit 1  &  Fit 2    &  Fit 3  &  Fit 4 &  systematic error  \\ \hline
                                                      
$\chi^2_{\rm PACS-CS}/N$                              & 0.5054  &  0.8721   & 0.5329  & 0.4824 &   10 {\rm MeV}  \\
                                                      & 1.6153  &  2.6456   & 1.9222  & 1.6726 &    5 {\rm MeV}  \\ \hline

$\chi^2_{\rm LHPC}/N$                                 & 0.0999  &  1.6006   & 0.3911  & 0.1574 &   10 {\rm MeV}\\
                                                      & 0.3659  &  5.9049   & 1.4524  & 0.5851 &    5 {\rm MeV}  \\ \hline

$\chi^2_{\rm HPQCD}/N$                                & 0.9430  &  0.9131   & 1.2962  & 1.0606 &   10 {\rm MeV}   \\
$\beta \simeq 6.76 $                                  & 3.7132  &  3.5877   & 5.1052  & 4.1814 &    5 {\rm MeV}   \\ 
                                                      
$\chi^2_{\rm HPQCD}/N$                                & 0.2468  &  0.2688   & 0.3393  & 0.4172 &   10 {\rm MeV}   \\
$\beta \simeq 7.09 $                                  & 0.9798  &  1.0662   & 1.3459  & 1.6495 &    5 {\rm MeV}   \\ \hline
                                                      
$\chi^2_{\rm ETMC}/N$                                 & 0.4584  &  1.2096   & 0.9919  & 0.8367 &   10 {\rm MeV}   \\ 
$ \beta = 1.90$                                       & 1.1053  &  2.8710   & 2.5727  & 2.1517 &    5 {\rm MeV}  \\
                                                      
$\chi^2_{\rm ETMC}/N$                                 & 0.6546  &  1.5087   & 1.0253  & 0.8279 &   10 {\rm MeV}   \\ 
$ \beta = 1.95$                                       & 1.6217  &  3.6038   & 2.5556  & 2.0590 &    5 {\rm MeV}  \\
                                                    
$\chi^2_{\rm ETMC}/N$                                 & 0.1860  &  0.4915   & 0.4431  & 0.3572 &   10 {\rm MeV}   \\ 
$ \beta = 2.10$                                       & 0.4061  &  1.1424   & 0.9964  & 0.7943 &    5 {\rm MeV}  \\ \hline 
                                                      
$\chi^2_{\rm HSC }/N$                                 & 0.1425  &  0.1710   & 0.4735  & 0.2622 &   10 {\rm MeV}   \\ 
                                                      & 0.3757  &  0.5893   & 1.8550  & 0.9965 &    5 {\rm MeV}  \\

\end{tabular}
\caption{ The table shows the impact of an ad-hoc systematic error (that is added to the statistical error in mean quadrature) on the chisquare values of the various lattice data sets. 
The set of lattice data fitted is described in the text. The corresponding low-energy parameters of Fit 1-4 are given in Tab. \ref{tab:FitParametersA}. }
\label{tab:lattice-scale}
\end{center}
\end{table}

In Tab. \ref{tab:FitParametersA} we collect four distinct fit scenarios which are constrained by additional input from first lattice results on some scattering observable. 
All four fit scenarios incorporate the s-wave scattering lengths of \cite{Liu:2012zya} into their $\chi^2$ functions. In addition Fit 2-4 are adjusted to the scattering phases shifts 
of \cite{Moir:2016srx}. In Fit 3 and Fit 4 the subleading counter terms (\ref{def- gn}) are activated.  
All parameter sets 
reproduce the D meson masses with a $\chi^2/N$ close to one given an estimate for the systematic error in the range 5-10 MeV. In all fit scenarios the four low-energy constants $c_{0,1}$ and 
$\tilde c_{0,1}$ are adjusted to recover the isospin averaged physical D meson masses with $J^P =0^-$ and $J^P = 1^-$ quantum numbers from the PDG \cite{PDG}. This implies that deviations from leading order 
large-$N_c$  or heavy-quark symmetry sum rules are considered for $c_{0,1}$ and $\tilde c_{0,1}$. In turn we must not impose the heavy quark-symmetry relations $d_n =\tilde d_n $ for all $n= 1,..., 4$. 
Scale invariant expressions request $d^c_1 \neq \tilde d^c_1$ and $d^c_3 \neq \tilde d^c_3$ but permit the assumptions $d^c_2 =\tilde d^c_2$ and $d^c_4 = \tilde d^c_4$ (see (\ref{def-dc})). 
All four fit scenarios are based on the latter. In addition we note that while Fit 1 and Fit 3 impose the leading order large-$N_c$ relations 
 \begin{eqnarray}
&&  c^r_2 = -\frac{c^r_3}{2}\,, \qquad \qquad c^r_4 = -\frac{c^r_5}{2}\,,\qquad \qquad 
 \tilde c^r_2 = -\frac{\tilde c^r_3}{2} \,,\qquad \qquad 
\tilde c^r_4 = -\frac{\tilde c^r_5}{2}\,,
\label{large-Nc-recall-r}
\end{eqnarray}
the remaining scenarios Fit 2 and Fit 4 keep those parameters unrelated.

The quality with which the four scenarios reproduce the D meson masses from the lattice ensembles is summarized in Tab. \ref{tab:lattice-scale}. 
From the fact that all chisquare values are close to one for an ad-hoc systematic error in between 5 and 10 MeV we arrive at our estimate of an intrinsic systematic error of 5-10 MeV for the D meson masses. 
All low-energy parameters are in qualitative agreement 
with the first rough estimates in (\ref{estimate-cn}). On the other hand we find significant tension with the low-energy parameters as obtained in \cite{Guo:2008gp,Liu:2009uz,Guo:2011dd,Altenbuchinger:2013vwa}. 
The parameters of Fit 2 are reasonably close to the two sets claimed in \cite{Liu:2012zya} with the notable exception of $c_1$ which differs by about a factor 2. 
Despite the considerable variations in the low-energy constants we deem all four parameter sets acceptable from the perspective of describing the D meson masses. We repeat that it is unclear whether Fit 1 
should be trusted more, only because it would be compatible with a discretization error slightly smaller than the one for Fit 4. After all a 5 MeV systematic error would be an astonishingly small value.

\begin{table}[t]
\setlength{\tabcolsep}{2.5mm}
\renewcommand{\arraystretch}{1.2}
\begin{center}
\begin{tabular}{l|cccc} 
                                                       &  Fit 1    &  Fit 2    &  Fit 3   &  Fit 4   \\ \hline
                                                      
$\chi^2_{\rm s-wave \,scattering \,lengths}/N$         &   0.9184  & 1.3849    & 2.2596   & 2.0597

\end{tabular}
\caption{Chisquare values from Fit 1-4 for the s-wave scattering length of \cite{Liu:2012zya}. The first two ensembles of Tab. \ref{tab:LHPC} with a kaon mass smaller than 600 MeV 
are considered in the chisquare function. The corresponding low-energy parameters of Fit 1-4 are given in Tab. \ref{tab:FitParametersA}. }
\label{tab:scattering-length}
\end{center}
\end{table}

We take up the additional constraints considered.  In \cite{Liu:2012zya} a set of s-wave pion and kaon scattering lengths was computed on 4 different lattice ensembles as recalled in Tab. \ref{tab:LHPC}. 
Since only for the first two ensembles the kaon mass is smaller than our cutoff choice of 600 MeV, we include into our $\chi^2$ function only the scattering lengths from the first two ensembles of that table. 
The scattering lengths are computed in the infinite volume limit based on the parameter sets collected in Tab. \ref{tab:FitParametersA}. 

We apply the coupled-channel framework 
pioneered in \cite{Kolomeitsev:2003ac,Hofmann:2003je,Lutz:2007sk} which is based on the flavour SU(3) chiral Lagrangian. It relies on the on-shell reduction scheme developed in \cite{Lutz:2001yb,Lutz:2003fm} 
which can be justified if the interaction is of short range nature or the long-range part is negligible \cite{Lutz:2011xc,Lutz:2015lca}. 
Fortunately this appears to be the case for the s-wave interactions of the Goldstone bosons off any of the D mesons. In these and the current works the coupled-channel interaction is approximated 
by tree-level expressions. Coupled-channel unitarity is implied by a particular summation scheme formulated in terms of scalar loop functions evaluated with physical meson masses and relativistic kinematics.  

An alternative chain of works  based on a somewhat different 
treatment of the coupled-channel effects are \cite{Guo:2011dd,Guo:2008gp,Liu:2012zya,Altenbuchinger:2013vwa,Cleven:2014oka,Du:2016tgp}. We did a careful comparison of the three available sources for the flavour 
structure of the coupled-channel interaction \cite{Hofmann:2003je,Liu:2012zya,Altenbuchinger:2013vwa}. We find two discrepancies amongst the original work  \cite{Hofmann:2003je} and \cite{Liu:2012zya} where we 
do take into account the different phase conventions  used in the two works for the isospin states. The two discrepancies are in the $(I,S) =(1/2,0)$ sector. One is traced as a misprint, in $C_{WT}$ of 
Tab. 2 of \cite{Hofmann:2003je}, in which the two entries $13$ and $22$ need to be interchanged (see \cite{Kolomeitsev:2003ac}). The second one we attribute to a misprint in \cite{Liu:2012zya}. Unfortunately,  we 
were not able to relate to the flavour coefficients shown in \cite{Altenbuchinger:2013vwa}. As compared to \cite{Hofmann:2003je} and \cite{Liu:2012zya} there are more than 10 unresolved contradictions. 

\begin{figure}[t]
\center{
\includegraphics[keepaspectratio,width=0.97\textwidth]{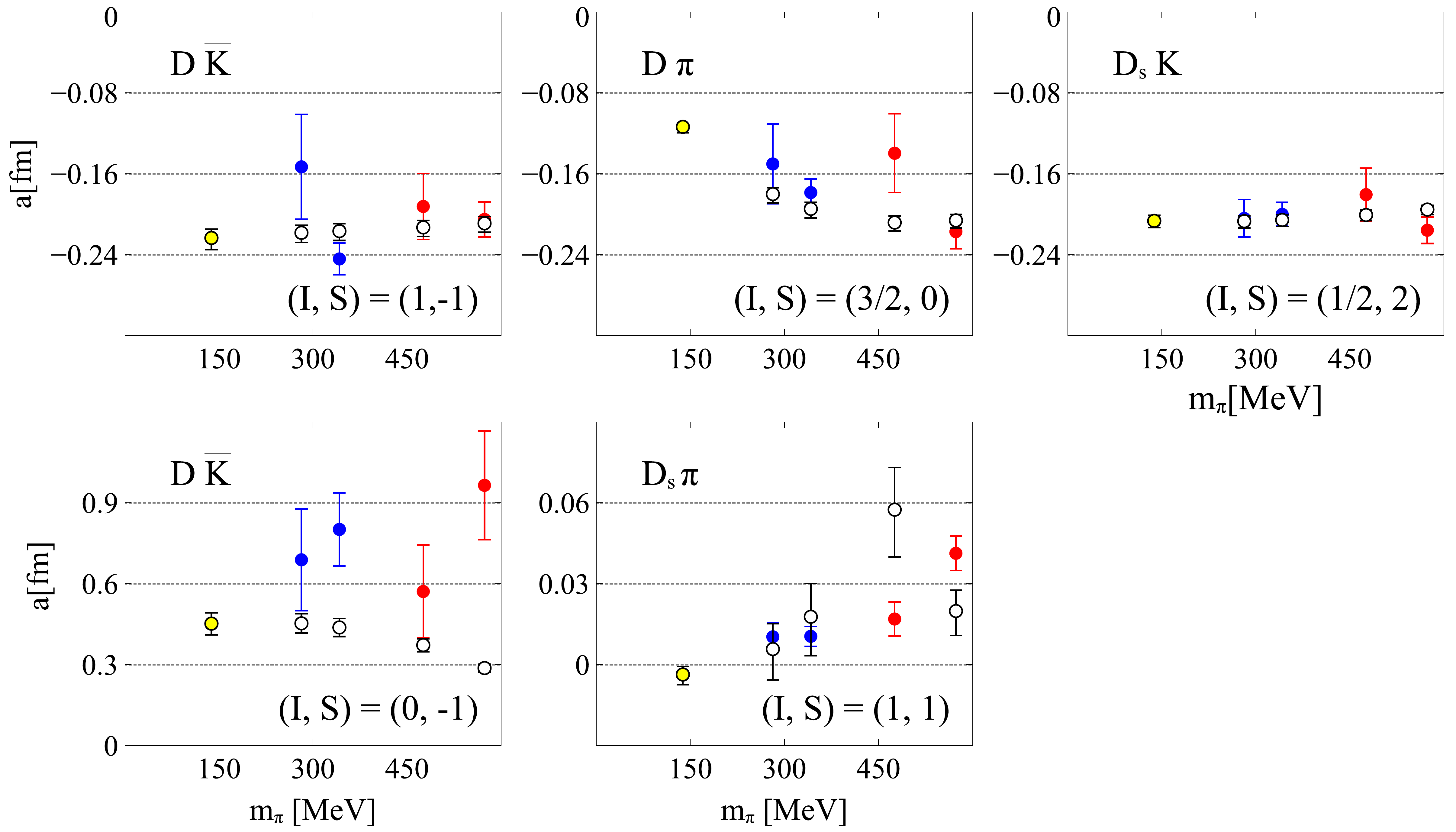} }
\vskip-0.2cm
\caption{\label{fig:length} S-wave scattering length from Fit 4 as compared to predictions from \cite{Liu:2012zya}. The blue (red) data points show the scattering lengths for the ensembles where the kaon 
mass is smaller (larger) 
than 600 MeV. The yellow points provide the physical value for the  scattering lengths.}
\end{figure}

In Tab. \ref{tab:scattering-length} we collect the $\chi^2/N$ values that characterize how well we reproduce the s-wave scattering length of \cite{Liu:2012zya} in our four fit scenarios. 
Note that we use here our estimates for the lattice scales $a_{\rm LHPC}$ as shown in Tab. \ref{tab:lattice-scale-Fits}. 
The table is complemented by Fig. \ref{fig:length} where a direct comparison of our results with the lattice data is provided for Fit 4. 
In the figure the lattice data points, shown by filled symbols, are confronted with open symbols that represent our results. 
The error bars in the latter points reflect an estimate of the systematic uncertainty in our computation of the scattering lengths, where we should state that 
the $\chi^2$ values in Tab. \ref{tab:scattering-length} are computed always in terms of the center value of our prediction. Our systematic error estimate is implied by a variation 
of the matching scale $\mu_M$ around its natural value \cite{Kolomeitsev:2003ac,Hofmann:2003je,Lutz:2007sk}. The error bars are implied by $\Delta \mu_M = \pm 100$ MeV 
with $\mu_M \to \mu_M + \Delta\mu_M$. For a detailed discussion why $\Delta \mu_M$  cannot be chosen much larger without jeopardizing the approximate implementation of crossing symmetry 
we refer to the original works \cite{Lutz:2001yb,Lutz:2003fm}. It is important to recall that dialing the matching scale slightly off its natural value does not affect our self consistent determination 
of the D meson masses. The latter is a convenient tool to estimate the uncertainties of the unitarization process.

In the upper panels of Fig. \ref{fig:length} we show the channels that are dominated by a repulsive Tomozawa-Weinberg interaction term \cite{Kolomeitsev:2003ac}. In terms of a flavour SU(3) multiplet classification  
they belong to a flavour 15plet, that can not be reached within the traditional quark-model picture. A minimal four quark state configuration is required. 
In contrast in the lower panels, channels are presented that belong to the exotic flavour sextet sector in which the leading Tomozawa-Weinberg interaction shows a weak attraction  \cite{Kolomeitsev:2003ac}. 
As pointed out in  \cite{Kolomeitsev:2003ac,Hofmann:2003je,Lutz:2007sk,Lutz:2015ejy} depending on the size of chiral correction terms exotic resonance states may be formed by the chiral dynamics. Final state interactions  
distort the driving leading order term and ultimately generate the more complicated quark mass dependence as seen in the figure.  
We discriminate results  based on ensembles with a kaon mass larger or smaller than 600 MeV by distinct colored symbols.  With red symbols we indicate that the kaon mass is larger than our cutoff value, and therefore
chiral dynamics is not expected to be reliable. A fair reproduction of all relevant scattering lengths is seen in Fig. \ref{fig:length}. Our predictions for the scattering lengths at the physical point are also 
included by the additional yellow filled points farthest to the left. 

We would conclude that with the constraints set by scattering lengths of \cite{Liu:2012zya} we cannot rule out any of our four fit scenarios in Tab. \ref{tab:FitParametersA}.

\clearpage

\section{Scattering phase shifts from QCD lattice data}

In this section we finally present an additional constraint on the low-energy parameters that provide a clear criterion which of the four fit scenarios is most reliable and should 
be used in applications. Recently HSC computed $\pi D$ phase shifts in both isospin channels. The results are based on the ensemble recalled in Tab. \ref{tab:HSC}. Given our four parameter sets 
we can compute those observable at the given unphysical pion and kaon masses. We do this for all four parameter sets. 

It is necessary to explain how we compare with those lattice results. Ultimately one should compute the various discrete levels the collaboration computed and then apply the L\"uscher method \cite{Luscher:1990ux,Luscher:1991cf} 
to extract the coupled-channel scattering amplitudes. This requires an ansatz for the form of the reaction amplitudes. In the case of a single channel problem this can be 
analyzed in a model independent manner. In turn for $\pi D$ scattering in the $I = 3/2$ channel we can compare our results with the single energy phase shifts as taken from Fig. 20 of \cite{Moir:2016srx} at different 
center-of-momentum energies $E= \sqrt{s}-m_\pi - M_D$. They are to be confronted with the four lines from our four fit scenarios. 
In the figure of Tab. \ref{tab:gn} we see that the two red lines are significantly off the lattice data points, where  
with those Fit 1 and 2 are presented. This is the case even though in Fit 2 an attempt was made to reproduce the $\pi D$ phase shifts from \cite{Moir:2016srx}. Note that in Fit 1 we ignored any of the latter. 
We assure that our conclusions are stable against a reasonable variation of the matching scale in this sector.

Based on this observation we made our ansatz for the scattering amplitudes more quantitative by the consideration of an additional set of low-energy constants relevant at chiral order three. Such terms were 
constructed in \cite{Yao:2015qia,Du:2017ttu} to take the form 
\begin{eqnarray}
&& {\mathcal L}_3 =
    4\,g_1\,{D}\,[\chi_-,\,{U}_\nu]_- \CD^\nu \,\bar{D}
  - 4\,g_2\,{D}\,\big([{U}_\mu,\,[\CD_\nu,\,{U}^\mu]_-]_- + [{U}_\mu,\,[\CD^\mu,\,{U}_\nu]_-]_-\big)\,\CD^\nu\bar{D}
\nonumber\\ 
&& \quad \;\,  - \,4\,g_3\,{D}\,[{U}_\mu,\,[\CD_\nu,\,{U}_\rho]_-]_-\,[\CD^\mu,\,[\CD^\nu,\,\CD^\rho]_+]_+\bar{D}
  + {\rm h.c.}\,.
 \label{def- gn}
\end{eqnarray}
Our motivation to consider such terms is slightly distinct to the one followed in \cite{Yao:2015qia,Du:2017ttu}. From the previous work \cite{Lutz:2007sk} we expect the light vector meson degrees of freedom
to play a crucial role for the considered physics. Ultimately we would like to consider them as active degrees of freedom. This is beyond the scope of the current work. Here we consider the low-energy 
constants as a phenomenological tool to more accurately integrate out the light vector meson degrees of freedom. 
In scenario Fit 3 and Fit 4 the contributions of the $g_n$ are worked into the coupled-channel interaction. Their values are displayed in Tab. \ref{tab:gn}, which consecutively
lead to a significantly improved reproduction of the scattering phase shift. 

\begin{table}[t]
\centering
\parbox{0.47\textwidth}{
\begin{tabular}{l|ccll} 

              &  Fit 1 \phantom{xx}   &  Fit 2  \phantom{xx}  &  Fit 3 \phantom{xx}  &  Fit 4 \phantom{xx}  \\ \hline
                                                      
$g_1$         &  0                    &  0                    &  0.2240              &  0.2338  \\
$g_2$         &  0                    &  0                    &  0.5405              &  0.4663   \\
$g_3$         &  0                    &  0                    &  0.0399              &  0.0299                                                       

\end{tabular}
\caption{While the solid lines  are from Fit 2 and 4, the dashed 
lines with respect to Fit 1 and 3. The lattice data are from \cite{Moir:2016srx}. }
\label{tab:gn}
}
\qquad
\begin{minipage}[c]{0.47\textwidth}%
\centering
    \includegraphics[width=1\textwidth]{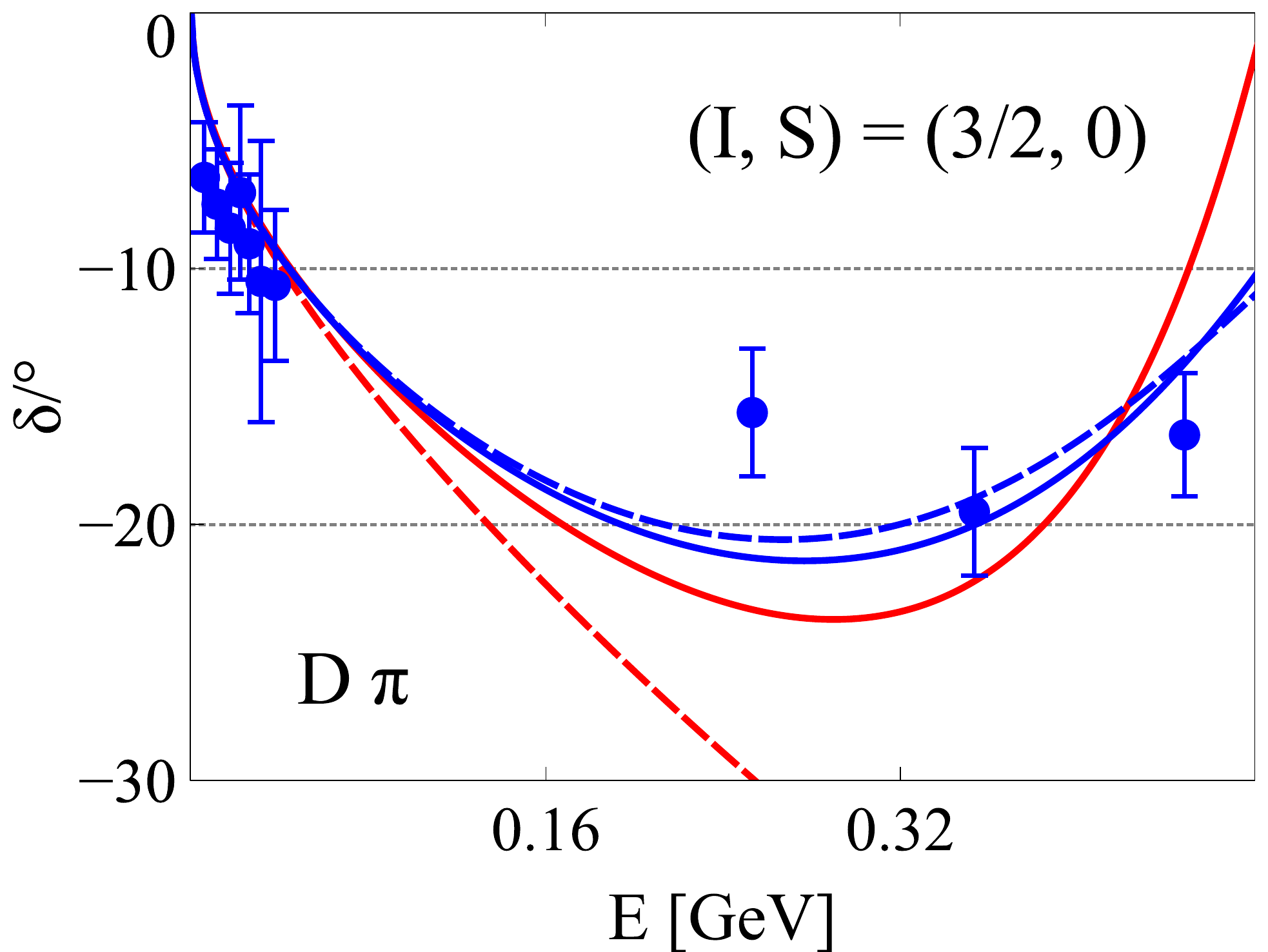}
\label{fig:figure}
\end{minipage}
\end{table}

\begin{figure}[t]
\center{
\includegraphics[keepaspectratio,width=0.92\textwidth]{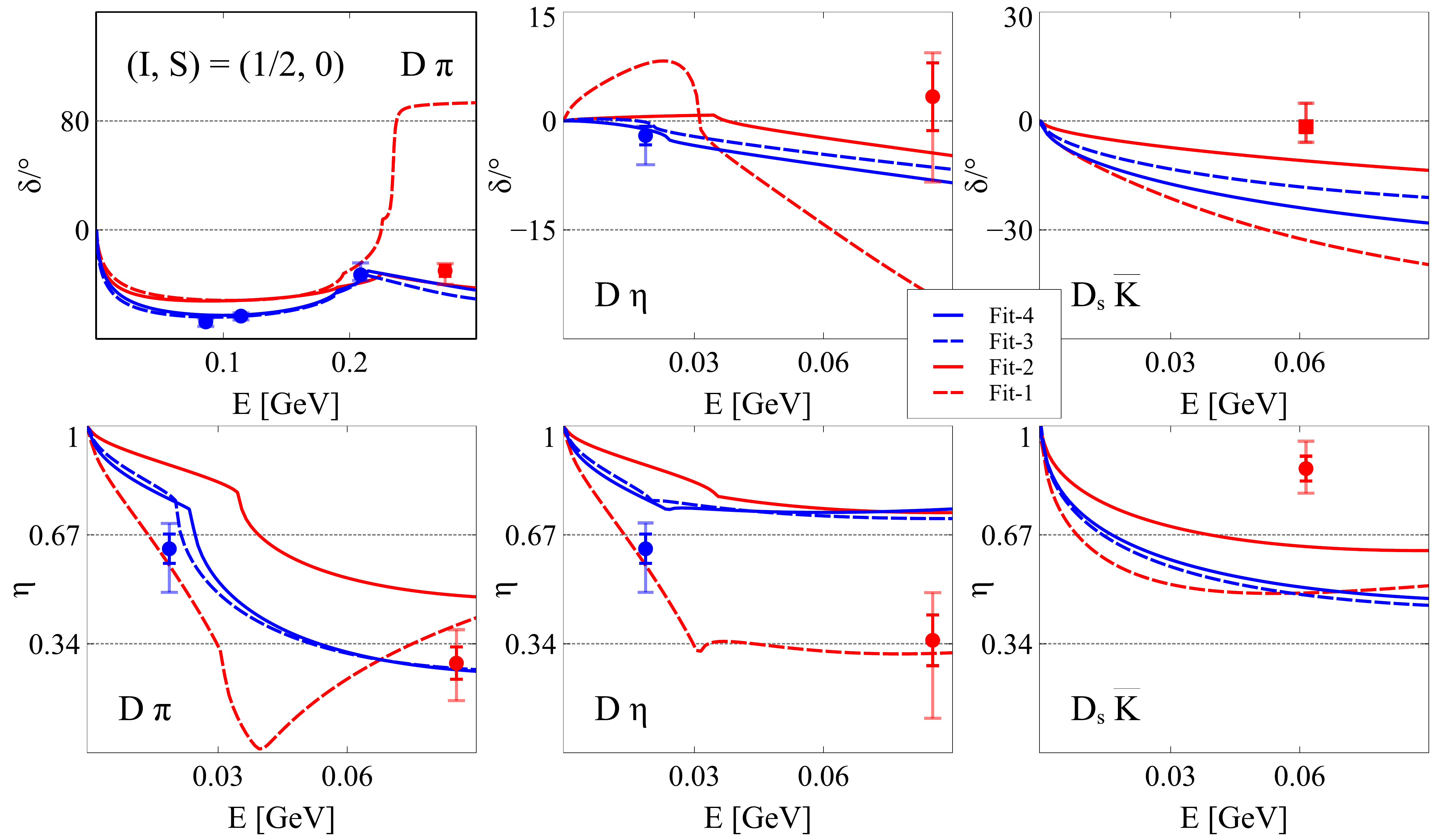} }
\vskip-0.2cm
\caption{\label{fig:phases-12} Phase shifts with $(I,S)=(1/2,0)$ from Fit 1-4 as compared to lattice data from \cite{Moir:2016srx}. 
While the solid lines  are from Fit 2 and 4, the dashed 
lines with respect to Fit 1 and 3. The two red lines present the disfavoured scenarios from  Fit 1 and 2. We apply the somewhat unusal convention of the lattice group 
where the phase shift at threshold is normalized to zero even in the presence of a bound state. }
\end{figure}

We proceed by the coupled-channel  $\pi D$ system with $I=1/2$ for which its determination of the three phase shifts and in-elasticities is more involved. Some model dependence may enter the analysis. 
In \cite{Moir:2016srx} an estimate of the latter was accessed by allowing a quite large set of different forms of the ansatz for the coupled-channel amplitudes. That then leads to two  error 
bands in their plotted phase shifts and in-elasticity parameters. The smaller one shows the statistical uncertainty, the larger one includes also the systematic error. In Fig. 9 and Fig. 10 of \cite{Moir:2016srx} it is shown in addition, on how many levels  
their results are based on in a given energy bin. Above the $\pi D$ and below the $K D_s$ threshold there are three clusters of levels. We take their center and translate those into single energy 
phase shifts and in-elasticities with error bars taken from the estimated uncertainties. In Fig. \ref{fig:phases-12} those 'lattice data' points are shown and confronted with  our results from the 
four fit scenarios. In addition a fourth lattice data point at energies above the $K D_s$ threshold is also included in the figure, but shown in red symbols. We do have some reservation 
towards those points, since the number of close-by energy levels is quite scarce. This is particularly troublesome since here it is a true three channel system that would need more rather than a fewer 
number of levels to unambiguously determine the scattering amplitude. In turn, the particular choice of ansatz is expected to play a much more significant role in the determination of the red 
lattice data points. We conclude that the error bars must be significantly underestimated for those points.

Fig. \ref{fig:phases-12} confirms our conclusions from the previous Tab. \ref{tab:gn} that only Fit 3 and Fit 4 may be expected to be faithful. The $\pi D$ and  $\eta D$ phase shift points are 
highly discriminative amongst the 4 fit scenarios. Fit 3 and Fit 4 describe the lattice data in Fig. \ref{fig:phases-12} significantly better than Fit 1 and Fit 2.  
Since Fit 4 is doing better in the D meson masses, but also in the s-wave scattering lengths one may identify Fit 4 to be the most promising candidate 
for making reliable predictions.

\begin{figure}[t]
\center{
\includegraphics[keepaspectratio,width=0.92\textwidth]{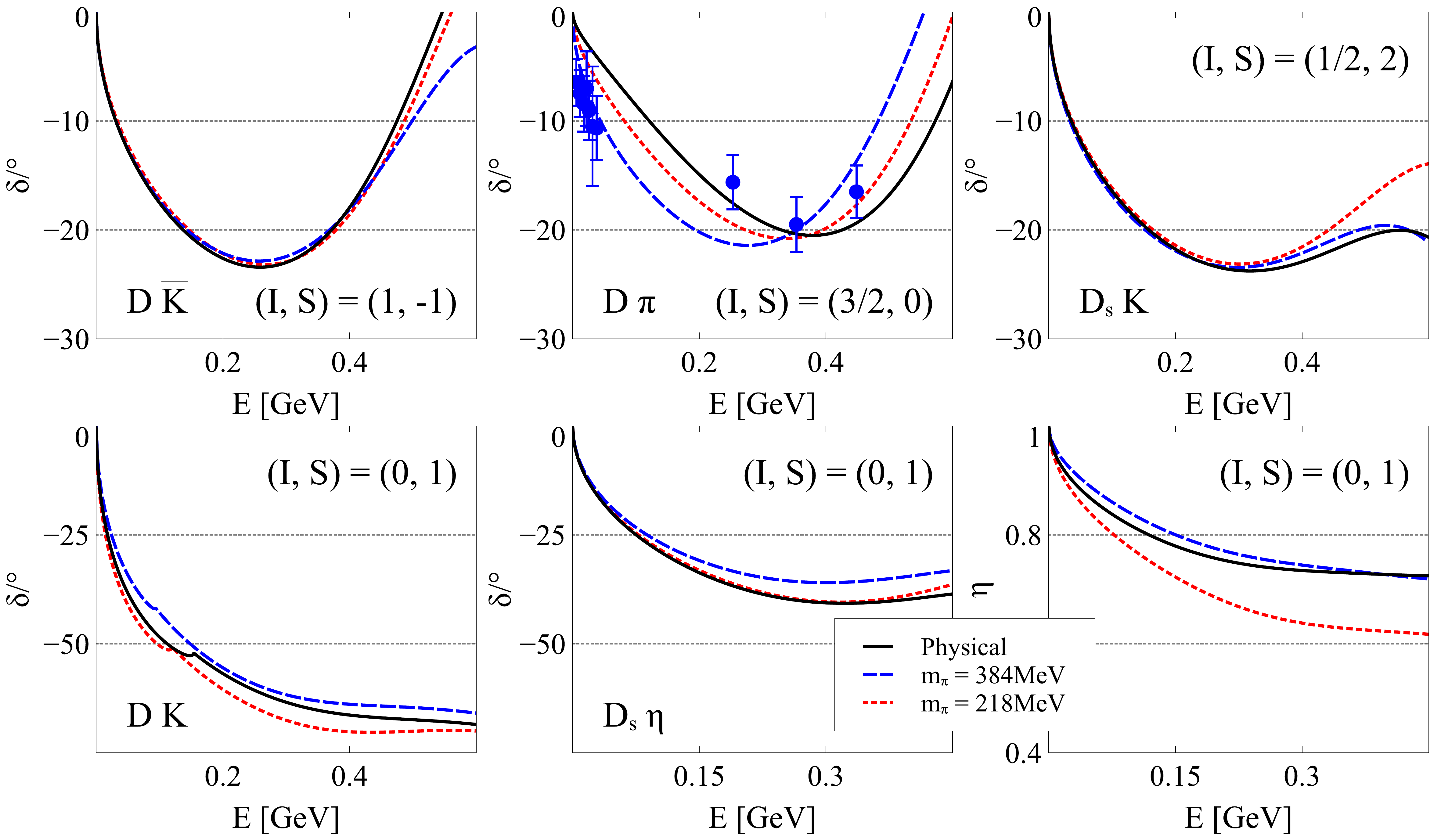} }
\vskip-0.2cm
\caption{\label{fig:phases_a} Predictions for phase shifts from Fit 4 for the physical point but also for pion and kaon masses as shown in Tab. \ref{tab:HSC}.}
\end{figure}

There is a further piece of information provided by HSC in the given ensemble. The mass $M_B$ of a bound state just below the $\pi D$ threshold is predicted. It is a member of the conventional flavour anti-triplet, which 
formation was predicted by chiral dynamics unambiguously \cite{Kolomeitsev:2003ac,Hofmann:2003je}. Within the given error it is not distinguishable from the 
$\pi D$ threshold value. The following bound is derived from data published by HSC
\begin{eqnarray}
 \epsilon_B = \frac{m_\pi + M_D}{ M_B} - 1  < 0.001 \,,
 \label{def-epsilon_B}
\end{eqnarray}
at the one sigma level. We compute this value in the four fit scenarios with 
\begin{eqnarray}
10^3\,\epsilon_B = \Bigg\{ \begin{array}{cccc}
{\rm Fit}\; 1 & {\rm Fit} \;2 & {\rm Fit} \;3 & {\rm Fit} \;4 \\
   8.0 & 5.4 & 4.3 & 5.7  
                  \end{array} \,
\end{eqnarray}
where we find discrepancies for the bound state mass of the order of our resolution of 5-10 MeV. As a consistency check we exploit the uncertainties in the unitarization process, by 
tuning the matching scale to meet the condition (\ref{def-epsilon_B}) for Fit 1 through Fit 4.  This is achieved for instance with $\Delta \mu_M \simeq 69 $ MeV and $\Delta \mu_M \simeq 86$ MeV in  
Fit 3 and Fit 4 respectively, where we emphasize that with $\Delta \mu_M$  the determination of the D meson masses is not affected. Then we reconsider the phase shifts and in-elasticities and find 
that all together the impact of such a change of the matching scale is quite moderate. While now Fit 1 goes almost perfectly through the three blue lattice data points for the $\pi D$ phase shift, 
the lines of Fit 3 and Fit 4 are slightly below those points. The crucial observation is that the significant disagreement with the single blue $\eta D$ phase shift is persistent in the Fit 1 scenario 
and therefore Fit 4 must remain our favorite choice.

\begin{figure}[t]
\center{
\includegraphics[keepaspectratio,width=0.95\textwidth]{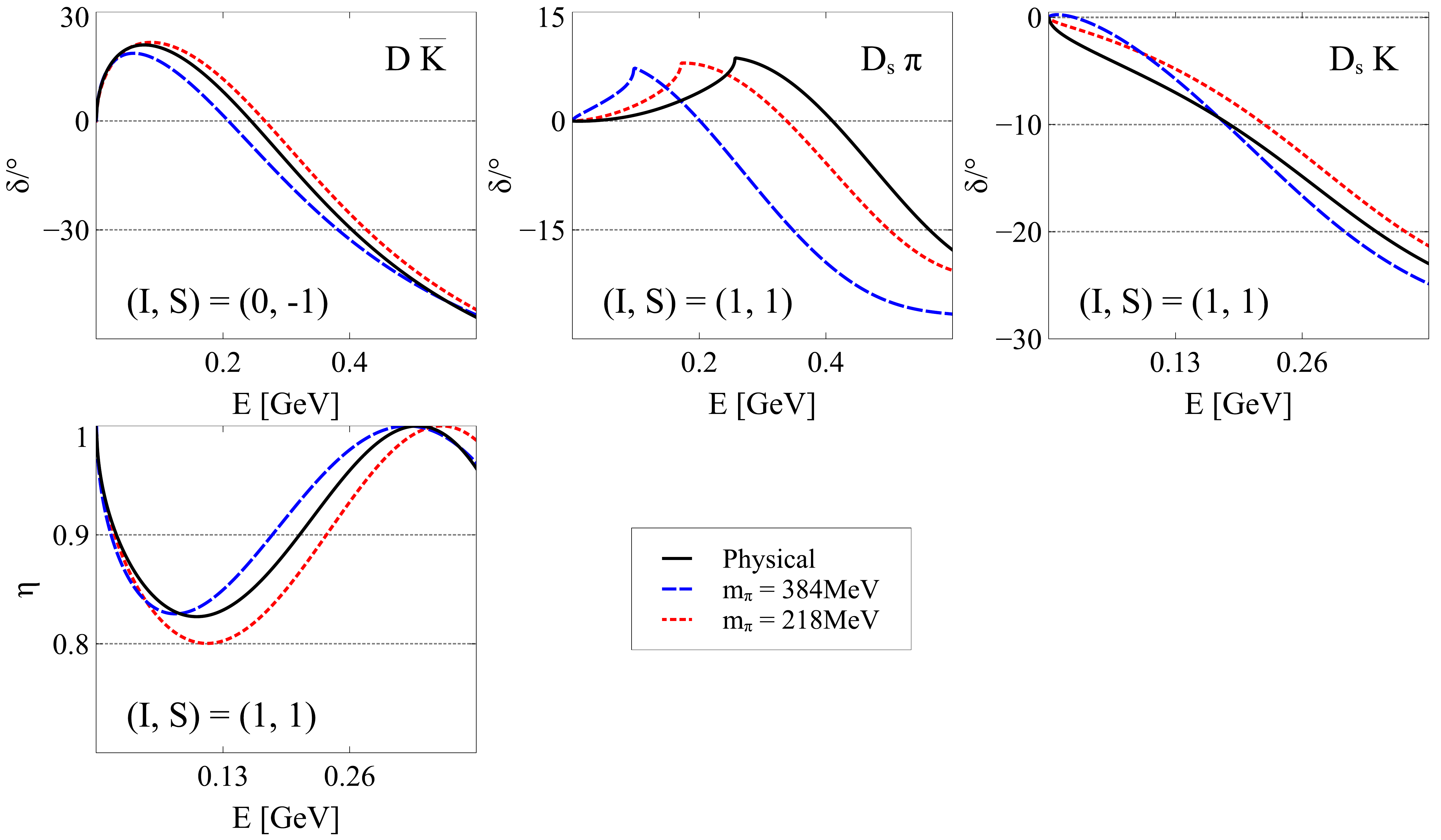} 
\includegraphics[keepaspectratio,width=0.95\textwidth]{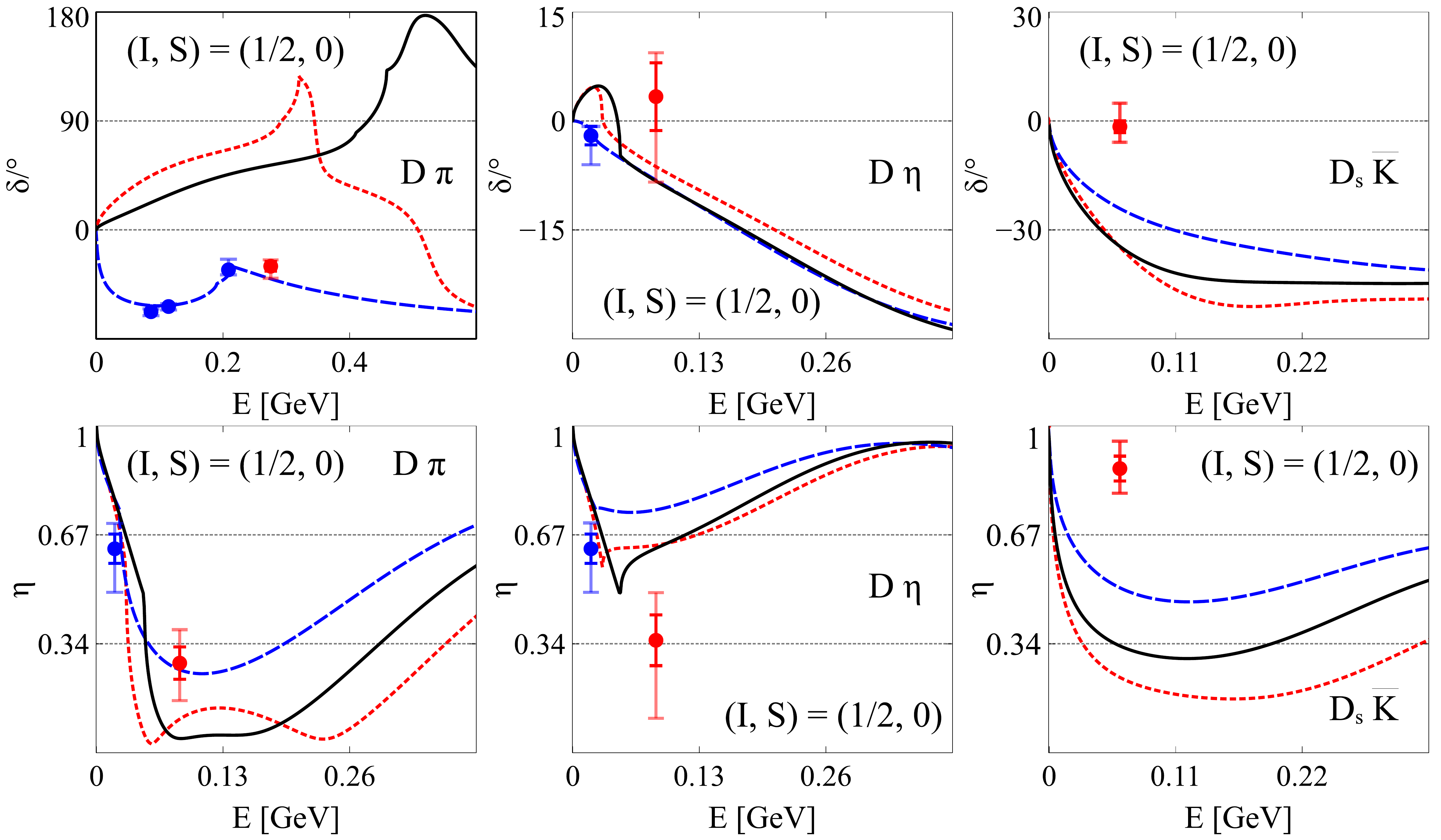}}
\vskip-0.2cm
\caption{\label{fig:phases_b} Predictions for phase shifts from Fit 4 for the physical point but also for pion and kaon masses as shown in Tab. \ref{tab:HSC}.}
\end{figure}

We wish to make one comment on Fit 1  since it is particularly interesting despite its deficiencies: a clear signal of a member of the exotic sextet state is visible in the $\pi D$ phase shift. It  
shows a significant variation a little right from the last blue lattice point. We deem it unfortunate that exactly in this region there is not yet sufficient consolidated lattice points available which may rule out 
our first fit scenario unambiguously. Note furthermore that our Fit 1 scenario, which did not take any of the scattering observables from HSC into account, is disfavoured mainly by one feature of the HSC results in 
the $(I,S) = (1/2,0)$ sector. The single blue value for the $\eta D$ phase shift. It would be interesting to make the ansatz used by HSC for the coupled-channel amplitude more flexible and allow for an exotic state 
coupling dominantly to the $\eta D$ channel. One may speculate that this exercise could show that the claimed uncertainty for this lattice  point is underestimated significantly. If this happens  
our Fit 1 scenario may come into the game again. This may be so even though HSC appears to reject our Fit 1 scenario based on their results in the $(I,S) = (3/2,0)$ sector. Here the reader should be 
cautioned that we cannot fully rule out that the phenomenological treatment of the third order effects is fooling us. More detailed studies are required to substantiate our conclusions.

In the following we take our best fit scenario Fit 4 and provide a thorough documentation of its consequences. In Fig. \ref{fig:phases_a} and Fig. \ref{fig:phases_b} all 
phase shift and in-elasticity parameters are shown for all possible combinations of $(I,S)$. In  Fig. \ref{fig:phases_a} we present the channels in which no exotic signals are expected. 
Indeed, the evolution from the two HSC ensembles of Tab. \ref{tab:HSC} with unphysical quark masses to the physical point is smooth and unspectacular. While the solid black lines correspond to the physical point, the dashed 
and dotted lines to the two HSC cases, where the dashed lines are with respect to the upper ensemble of Tab. \ref{tab:HSC}. We refrain from including our estimate of the systematic uncertainty from a variation of 
the matching scale, because, first of all it is a small effect and second it obscures the clarity of the figures.

We advance to the exotic sectors with $(I,S) = (0,-1)$ and $(1,1)$. With the upper two panels of Fig. \ref{fig:phases_b} we demonstrate that here 
the evolution from the two HSC ensembles to the physical point is still smooth but quantitatively more significant, particularly in the two-channel system with $(I,S) =(1,1)$. The 
corresponding amplitudes are characterized by strong cusp effects at threshold. The latter reflect some weak attraction present in those channels being members of the flavour sextet.

Most striking are our predictions for the quark-mass dependence of the $(I,S)=(1/2,0)$ sector, which we present with the lower two panels of Fig. \ref{fig:phases_b}. The line 
conventions are identical to the ones used in the previous figures. The largest effect is seen in the $\pi D$ phase shift. Going from the HSC ensembles to the physical point it even changes sign. 
Here we see a clear signal for a member of the exotic flavour sextet state. The $\pi D$ phase shift passes through 90$^{\rm o}$ in between the $\eta D$ and $\bar K D_s$ thresholds. We checked that 
the amplitudes $\eta D \to \eta D$ but also $\bar K D_s \to \bar K D_s $ show a well defined resonance structure, with a width significantly smaller than the 
300$\,-\,$400 MeV of the flavour anti-triplet partner at lower masses. We find this to be a spectacular confirmation of the leading order prediction of this state advocated since 15 years ago by one of the 
authors (see \cite{Lutz:2015ejy}). It is amusing to see that the clear signature of this state at the physical point may not be seen at the studied HSC ensemble with unphysically large pion masses.  
Most exciting is the most recent claim in \cite{Guo:2017jvc} that this state can be seen in data from LHCb \cite{Aaij:2014baa,Aaij:2016fma}.

\clearpage

\section{Isospin violating decay of $D^*_{s0}(2317)$ from QCD lattice data}

A most striking prediction of chiral dynamics is the formation of the $D^*_{s0}(2317)$ as a coupled-channel hadronic molecule with significant components in the $\bar K D$ and $\eta D_s$ 
two-body states \cite{Kolomeitsev:2003ac}. At leading order in a chiral expansion the coupled-channel interaction is predicted by the Tomozwa-Weinberg term that is parameterized only by the 
pion-decay or kaon-decay constants, $f_\pi$ or $f_K$, driven into their chiral flavour $SU(3)$ limit with $f_{\pi, K} \to f$. 

This term dominates the s-wave coupled-channel force of the Goldstone bosons with the pseudo-scalar and vector D mesons. The force is short ranged: it may be visualized in terms of a vector meson t-channel exchange 
process with properly adjusted coupling constants. In contrast to a widespread confusion in the field there are hadronic molecular states that are not driven by a long-range force as provided by an exchange process 
involving the pion. The challenge is to control and predict such short range forces. 

The original work \cite{Kolomeitsev:2003ac} was taken up by many authors \cite{Hofmann:2003je,Guo:2006fu,Faessler:2007gv,Lutz:2007sk,Guo:2011dd,Altenbuchinger:2013vwa,Cleven:2014oka,Du:2016tgp,Du:2017zvv,Guo:2017jvc} 
who confirm this universal picture. The challenge is to make this approach more quantitative by controlling chiral correction terms. A first attempt was made in \cite{Hofmann:2003je,Lutz:2007sk} based on 
rough assumptions on the $\pi D$ invariant mass distributions. A more sophisticated approach was pursued in \cite{Liu:2012zya,Cleven:2014oka} where first QCD lattice data on some s-wave scattering lengths were used. 
With the significantly improved and extended lattice data set the determination of the low-energy constants, as achieved in our work, is expected to be more controlled and reliable.

In this section we focus on a particular property of the $D^*_{s0}(2317)$, its isospin violating hadronic decay width. Since its mass is below the $K D$ threshold and it carries isospin zero it can decay into the 
$\pi D$ channel only via isospin violating processes. Estimates of that width within typical quark-model approaches predict such a width of less than 10 keV \cite{Colangelo:2003vg}. 
This is contrasted by estimates from chiral-coupled channel approaches. Here, already the leading order Tomozawa-Weinberg predicts a width of about 75 keV as demonstrated first in \cite{Lutz:2007sk}. 
A corresponding computation with similar physics input but less stringent framework arrived at a similar value \cite{Faessler:2007gv}.  This is to be compared to the significantly larger values of  about 
140 keV in \cite{Lutz:2007sk} and later with even an error estimate of (133$\,\pm\,$22 ) keV \cite{Liu:2012zya}. The latter two works implemented chiral correction terms, where the more sophisticated 
approach \cite{Liu:2012zya} was based on additional constraints from some early lattice data.

\begin{table}[t]
\setlength{\tabcolsep}{2.5mm}
\renewcommand{\arraystretch}{1.2}
\begin{center}
\begin{tabular}{l|cccc| c} 
                                                   &  Fit 1    &  Fit 2    &  Fit 3   &  Fit 4  & $\epsilon $ \\ \hline
                                                      
$\Gamma_{D^*_{s0}(2317)\to \pi_0 D_s }$    [keV]   &   61.1    & 54.1      & 88.6     & 80.1    & 0.0100  \\
                                                   &  74.6     &  68.4     &  115.8   & 104.4   & 0.0122

\end{tabular}
\caption{Prediction for the isospin violating decay width of the $D^*_{s0}(2317)$ in the four fit scenarios of Tab. \ref{tab:FitParametersA}. }
\label{tab:decay-width}
\end{center}
\end{table}

The results of our study for the decay width is collected in Tab. \ref{tab:decay-width} for all four fit scenarios. They are based on the framework as detailed in \cite{Lutz:2007sk}. 
Since the mass of the $D^*_{s0}(2317)$ was not  tuned in any of our fits we again use the 
uncertainty in the unitarization and adjust the matching scale as to recover the precise mass of the $D^*_{s0}(2317)$. This is achieved with 50 MeV $<\Delta \mu_M <$ 100 MeV in the four scenarios. 
Beside the low-energy constants determined in our work  the computation of the width parameter depends crucially on the mixing 
angle $\epsilon$ of the $\pi_0-\eta$ system. According to \cite{Gasser:1984gg} it is determined 
by the quark masses as follows
\begin{eqnarray}
\frac{\sin (2\,\epsilon)}{\cos (2 \,\epsilon)} =
\sqrt{3}\,\frac{m_d-m_u}{2\,m_s-m_u-m_d} \,.
\label{eps-def}
\end{eqnarray}
While in \cite{Lutz:2007sk} the value $\epsilon = 0.010(1)$ was taken from \cite{Gasser:1984gg}  an updated estimate $\epsilon = 0.0129(7)$ was used in \cite{Liu:2012zya}. Here we consider the impact 
of a recent and more precise lattice determination of the quark masses by ETMC \cite{Carrasco:2014cwa}. This leads to a significantly lower estimate $\epsilon = 0.0122(18)$ which our faithful 
results in Tab. \ref{tab:decay-width} are based on. 

Since we argued that the lattice data of HSC rule out Fit 1 and Fit 2, we estimate  
the isospin violating hadronic  width of the $D^*_{s0}(2317)$  with ($104\,-\,116$) keV, somewhat lower than the previous claimed value of (133$\,\pm\,$22) keV \cite{Liu:2012zya}. 

\clearpage

\section{Summary and conclusions}

We studied the chiral extrapolation of charmed meson masses based on the three-flavour chiral Lagrangian 
formulated with pseudo-scalar and vector charmed fields. Here the recent approach by the authors constructed for the chiral extrapolation of the 
baryon ground state masses was adapted to the charm sector successfully, where good convergence properties for the chiral extrapolation are observed. 
Within the framework the chiral expansion is formulated in terms of physical masses. While an attempt was made to remove all model dependence a residual scheme dependence 
cannot be ruled out at this stage. All D meson masses arise in a manifest scale invariant manner. The framework was applied to lattice data  such that 
an almost unique set of low-energy constants was established. While we considered finite volume effects systematically, we did not implement discretization
effects. In turn, a fully systematic error analysis was outside the realm of our present study.

The low-energy parameters were adjusted to QCD lattice data at N$^3$LO, where  large-$N_c$ sum rules or relations that follow in the heavy charm-quark mass limit 
were used systematically. We considered lattice data based on ensembles of 
PACS-CS, MILC, ETMC and  HSC with pion and kaon masses smaller than 600 MeV. Besides taking into account constraints from the D meson masses 
from the various lattice groups, we also considered first results on scattering observables in particular from HSC. Only with the latter, in particular their estimate of the 
$\eta D$ phase shift, we arrive at a rather well defined parameter set, in terms of which we make predictions. The data set on the D meson masses together with constraints from 
s-wave scattering lengths is not sufficient to nail down the set of low-energy constants. 

We computed 15 phase shifts and in-elasticities at physical quark masses 
but also for an additional HSC ensemble. Such results can be scrutinized by lattice QCD with available computing resources and technology. In addition we 
predict the isospin violating strong decay width of the $D^*_{s0}(2317)$ to be ($104\,-\,116$) keV. Given our favorite set of low-energy parameters 
we find a clear signal for a member of the exotic flavour sextet states in the $\eta D$ channel, below the 
$\bar K D_s$ threshold.

To further substantiate the claimed chiral low-energy parameters it is necessary to take additional data on QCD lattices in particular 
at unphysical quark masses. Our predictions are relevant for the PANDA experiment at FAIR, where the width of the $D^*_{s0}(2317)$ 
may be accessible by a scan experiment \cite{Lutz:2009ff}. Also the invariant $\eta D$ mass distribution, in which we expect a signal from an exotic flavour sextet state, 
may be accessed by the efficient detection of neutral particles 
with the available calorimeter.

\section*{Acknowledgments}

M.F.M. Lutz thanks Kilian Schwarz and Jan Knedlik for significant support on distributed computing issues. 
Particular thanks go to Walter Sch\"on who is operating the HPC cluster at GSI with his department in an outstanding manner. 
We are grateful to R\"udiger Berlich of Gemfony scientific UG for help with their optimization library Geneva. 
Marc Wagner and Sinead Ryan are acknowledged for stimulating discussions. We thank Feng-Kun Guo for help in correctly using Ref. [59].

\clearpage

\section*{Appendix A}

In this Appendix we collect all dimension less coefficients that are needed in the various power counting decompositions of the renormalized loop function 
(\ref{result-loop-0_full}). Here we focus on the pseudo-scalar $D$ mesons for which we find
\begin{eqnarray}
&&\alpha_1 = \frac{(2\,M+\Delta)^2}{4\,M^2}\,, \qquad  \qquad \;\;
 \alpha_2  =\frac{2\,M^2+ 2\,\Delta\,M+\Delta^2}{2\,M^2}\,,\qquad \qquad \alpha_3 =1\,, 
\nonumber\\ \nonumber\\
&& \gamma_1 = \frac{2\,M+ \Delta}{M}\,\log \frac{\Delta\,(2\,M + \Delta)}{(M+ \Delta)^2} \,,
\nonumber\\
&&  \gamma_2 = - \frac{2\,M^2+ 2\,\Delta\,M+ \Delta^2}{M\,(2\,M+ \Delta)}\,\log \frac{\Delta\,(2\,M + \Delta)}{(M+\Delta)^2}
-\frac{M}{2\,M+ \Delta}\,, \qquad 
  \gamma_3 = \frac{M}{2\,M+ \Delta}\,,
\nonumber\\
&&  \gamma_4 = -2\,\frac{M\,(M+ \Delta)^2}{(2\,M+ \Delta)^3}\,\log \frac{\Delta\,(2\,M + \Delta)}{(M+\Delta)^2}
+\frac{M^3}{2\,(2\,M+ \Delta)^3} \,, \qquad 
  \gamma_5 = \frac{M\, (M + \Delta)^2}{(2\,M + \Delta)^3} \,, 
\nonumber\\
\nonumber\\
&& \delta_1 = \gamma_1 - \frac{2\,M+ \Delta}{M}\,\log \frac{2\,\Delta }{(M+ \Delta)} \,,
\nonumber\\
&& \delta_2 = \gamma_2 + \frac{2\,M^2+ 2\,\Delta\,M+ \Delta^2}{M\,(2\,M+ \Delta)}\,\log \frac{2\,\Delta}{M} + \frac{2\,M + \Delta}{4\, M} 
\nonumber\\
&& \qquad \qquad + 2\, (\delta_3 - \gamma_3 ) \log \frac{M + \Delta}{M} \,,
\nonumber\\
&& \delta_3 = \gamma_3 - \frac{2\,M^2 + 2\, \Delta\, M + \Delta^2}{2\,M\,(2\,M+ \Delta)}\,,
\qquad \qquad \delta_5 = 0 \, ,
\nonumber\\
&& \delta_4 = \gamma_4 + \frac{2\,M\,(M+ \Delta)^2}{(2\,M+ \Delta)^3}\,\log \frac{2\,\Delta }{M} 
\nonumber\\[3pt]
&& \qquad - \frac{4\,M^2 + \Delta ( 4\,M + 5\,\Delta)}{32\, M\,(2\,M+ \Delta)} +  2\, (\delta_5 -\gamma_5 )\log \frac{M + \Delta}{M} \,,
\nonumber\\ \nonumber\\
&& \delta_6 = \frac{2\,M + \Delta}{2\,M} \frac{\partial}{\partial\, \Delta} \frac{2\, M\, \Delta}{2\,M + \Delta}\,\big( \gamma_1 - \delta_1 \big) + \delta_1 \,, 
\qquad    \delta_7 = \, \gamma_2 + \frac{1}{2} \big( \gamma_1 - \delta_1 \big) \,\frac{\Delta^2}{(2M + \Delta)^2} \,,
\nonumber\\ \nonumber\\
&&\beta_1 = \Delta \,\frac{\partial}{\partial \Delta} \,\alpha_1 \frac{2\,M + \Delta}{2\,M}\,, 
\nonumber\\
&& \beta_2 = \Delta^2 \,\frac{\partial}{\partial \Delta} \,\frac{\alpha_1\, \delta_2}{\Delta}\,, \qquad \qquad 
	 \beta_3 = \Delta^2 \frac{\partial}{\partial \Delta} \,\frac{\alpha_1 \,\delta_3}{\Delta}\,, 
\nonumber\\
&& \beta_4 = \Delta \,\gamma_1 \,\frac{\partial}{\partial \Delta} \,\alpha_1\,, \qquad  \qquad 
           \beta_5 = \Delta\,\frac{\partial}{\partial \Delta}\,\alpha_1 \,\delta_1\,,
\nonumber\\
&& \beta_6 = \frac{\Delta^2 \partial^2}{\partial \Delta\, \partial \Delta}\Bigg( \alpha_1 \frac{2\, M + \Delta}{2\,M} \Bigg),\qquad \qquad  
   	 \beta_7 = \Delta \,\frac{\Delta^2 \partial^2}{\partial \Delta\, \partial \Delta}\frac{\alpha_1\, \delta_2}{\Delta}\,
\nonumber\\
&& \beta_8 = \Delta \,\frac{\Delta^2 \partial^2}{\partial \Delta\, \partial \Delta}\frac{\alpha_1\, \delta_3}{\Delta} \,, \qquad \qquad 
 \beta_9 = \gamma_1\,\frac{\Delta^2 \partial^2}{\partial \Delta\, \partial \Delta}\,\alpha_1\,, \qquad 
\nonumber\\
&& \beta_{10} = \frac{\Delta^2 \partial^2}{\partial \Delta\, \partial \Delta}\,\alpha_1\, \delta_1\,, \qquad 
\beta_{11} = -\frac{1}{4}\, \alpha_1 \,\frac{M}{2\, M + \Delta} + \Big( \alpha_1 - \alpha_2 \Big) \,\frac{(2\, M + \Delta) \,M}{2 \,\Delta^2}\,.
\end{eqnarray}
While the $\alpha_i$ characterize the chiral expansion of the coefficients in front of $\bar I_{QR}$ and $\bar I_Q$ in  (\ref{result-loop-1_full}), the 
$\gamma_i$ and $\delta_i$ follow from a chiral expansion of $\bar I_{QR}$ with $M_H = M$ and $M_R = M+ \Delta$ and $m_Q < \Delta$. The coefficients $\beta_i$ are 
required in (\ref{loop-HB-3}, \ref{loop-HB-4}, \ref{loop-HB-5A}).

We turn to the chiral domain (\ref{def-counting-B}), in which the bubble-loop contributions to the D meson masses generate a renormalization of the low-energy parameters $d_i$. 
Such terms are proportional to the product of two quark masses (\ref{res-tree-level}). We provide detailed results with 
\begin{eqnarray}
&&\Pi^{(4-\chi) }_H \to  \sum_{Q\in[8]}\sum_{R\in \left[1^-\right]} \left(\frac{G_{QR}^H}{8\,\pi f}\right)^2
    \Bigg\{ \gamma_d^{(1)} m_Q^2\, \Pi_R^{(2)} + \gamma_d^{(2)}\, m_Q^2\, \Pi_H^{(2)} + \gamma_d^{(3)}\,\Pi_R^{(2)}\,\Pi_R^{(2)} 
\nonumber\\
    &&\qquad \qquad \quad +\,\gamma_d^{(4)}\,\Pi_H^{(2)}\,\Pi_H^{(2)}  +\gamma_d^{(5)}\,\Pi_R^{(2)} \,\Pi_H^{(2)}  
     \Bigg\}\,,
\nonumber\\
&&  d_i \to   \frac{1}{4} \,g_P^2\,\sum_{k=1}^5 \frac{\Gamma_{d_i}^{(k)}}{(4\pi f)^2} \,\gamma_d^{(k)}\,,
\label{Finite_Renorm_d}
\end{eqnarray}
and   
\begin{eqnarray}
&& \gamma_d^{(1)} = \frac{M}{2\,(M+\Delta)}\,\Bigg[
\frac{\partial }{\partial \Delta } \Big( \alpha_2\,\Delta\,\gamma_1 -\alpha_1\,\Delta\, \gamma_2 \Big) - \Delta\,\gamma_1\,
\frac{\partial\, \alpha_2 }{\partial \Delta } \Bigg]\,,
\nonumber\\
&& \gamma_d^{(2)} =
\frac{\Delta}{2\,M} \,\Bigg[ \frac{\partial }{\partial M } \Big( \alpha_2\,M\,\gamma_1 -\alpha_1\,M\, \gamma_2 \Big) - \frac{1}{M}\,\gamma_1\,
\frac{\partial }{\partial M } \,\big(\alpha_2\,M^2 \big) \Bigg] - \frac{ M+ \Delta}{M} \,\gamma_d^{(1)},
\nonumber\\
&& \gamma_d^{(3)} = -\frac{M}{4\, (M + \Delta )^2}\,\bigg(\frac{\partial\,\alpha_1\, \Delta^2}{\partial \Delta } \bigg)
\left(\frac{\partial  \,\gamma _1\,\Delta}{\partial \Delta }\right)
 - \frac{\alpha_1 \,\Delta^2\, M}{4\, (M + \Delta)}\,
\frac{\partial }{\partial \Delta }\,\Bigg[\frac{1}{2\, (M +\Delta )}\,
\bigg(\frac{\partial\,\gamma_1\, \Delta }{\partial \Delta }\bigg)\Bigg]\,,
\nonumber\\
&& \gamma_d^{(4)} = -\frac{1}{8\,M^2}\, \bigg(\frac{\partial }{\partial M}- \frac{\partial }{\partial \Delta} \bigg)^2\,
\Big(\alpha_1\,M\,\Delta^3 \,\gamma_1 \Big) +
\frac{\gamma_1\,\Delta}{8\,M^3}\, \bigg(\frac{\partial }{\partial M}- \frac{\partial }{\partial \Delta} \bigg)^2\,
\Big(\alpha_1\,M^2\,\Delta^2  \Big)
\nonumber\\
&& \qquad  +\,
\frac{1}{8\,M^3}\, \bigg(\frac{\partial }{\partial M}- \frac{\partial }{\partial \Delta} \bigg)\,
\Big(\alpha_1\,M\,\Delta^3 \,\gamma_1 \Big) -
\frac{\gamma_1\,\Delta}{8\,M^4}\, \bigg(\frac{\partial }{\partial M}- \frac{\partial }{\partial \Delta} \bigg)\,
\Big(\alpha_1\,M^2\,\Delta^2  \Big) \,,
\nonumber\\
&& \gamma_d^{(5)} = -\frac{1}{2\,M}\,\frac{\partial}{\partial M}\, \frac{1}{2\,(M + \Delta)}\,
\frac{\partial }{\partial \Delta}\, \Big( \alpha_1\,M\,\Delta^3\, \gamma_1 \Big)
\nonumber\\
&& \qquad  +\, \frac{\gamma_1 \,\Delta}{2\,M^2}\,\frac{\partial}{\partial M}\, \frac{1}{2\,(M + \Delta)}\,
\frac{\partial }{\partial \Delta}\, \Big( \alpha_1\,M^2\,\Delta^2 \Big) - 2\,\frac{M+\Delta}{M}\, \gamma_d^{(3)}\,,
\end{eqnarray}
and

\begin{eqnarray}
&& \Gamma_{d_1}^{(1)} = -\frac{32}{3}\, \tilde{c}_1\, ,\qquad
\Gamma_{d_2}^{(1)} = \frac{16}{9}\, \big(15\, \tilde{c}_0 -2\, \tilde{c}_1 \big)\,  ,\qquad
\Gamma_{d_3}^{(1)} = 8\, \tilde{c}_1\,  ,\qquad
\Gamma_{d_4}^{(1)} =\frac{88}{9}\, \big(2\, \tilde{c}_0-\tilde{c}_1\big)\,  ,
\nonumber\\
&& \Gamma_{d_1}^{(2)} = \frac{40}{3}\, c_1 \, ,\qquad \qquad\,
\Gamma_{d_2}^{(2)} = \frac{16}{9}\, \big(15\, c_0-2 \, c_1\big)\,  ,\qquad \qquad
\Gamma_{d_3}^{(2)} = 0, \qquad \qquad\;
\nonumber\\
&& \Gamma_{d_4}^{(2)} =\frac{88}{9}\, \big(2\, c_0-c_1\big)\,  , \qquad \qquad
 \Gamma_{d_1}^{(3)} = -\frac{32}{3}\, \tilde{c}_1^2  \,,\qquad \qquad
\Gamma_{d_2}^{(3)} = -\frac{64}{3} \,\big(2 \,\tilde{c}_0-\tilde{c}_1 \big) \,\tilde{c}_1\,  ,\qquad
\nonumber\\
&& \Gamma_{d_3}^{(3)} = 32 \,\tilde{c}_1^2 \, , \qquad \qquad
\Gamma_{d_4}^{(3)} = \frac{64}{3}\, \big(16\, \tilde{c}_0^2 -10\, \tilde{c}_1\, \tilde{c}_0 +  \tilde{c}_1^2\big)\,  ,\qquad \qquad
\Gamma_{d_1}^{(4)} = \frac{256}{3}\, c_1^2\,  ,\qquad \qquad
\nonumber\\
&& \Gamma_{d_2}^{(4)} = \frac{512}{3}\, \big(2\, c_0 -c_1 \big)\,c_1\,  ,\qquad \qquad
\Gamma_{d_3}^{(4)} = 0\,, \qquad \quad
\Gamma_{d_4}^{(4)} = \frac{256}{3}\, \big(2\, c_0-c_1 )^2 \,  , \qquad 
\nonumber\\
&&  \Gamma_{d_1}^{(5)} = -\frac{32}{3}\, c_1\,\tilde{c}_1\,  ,\qquad \qquad 
\Gamma_{d_2}^{(5)} = -\frac{64}{3}\, \big(-8\, c_1\,\tilde{c}_0 +  (c_0 +2\,c_1) \,\tilde{c}_1\big)\,  ,
\nonumber\\
&&\Gamma_{d_3}^{(5)} = 0 \,,\qquad \qquad \Gamma_{d_4}^{(5)} = \frac{32}{3}\, \big(2\,c_0- c_1\big)\,\big( 16\,\tilde{c}_0-5\,\tilde{c}_1\big)   \,.
\label{res-Gamma-d}
\end{eqnarray}

We turn to the conventional counting ansatz (\ref{def-counting-C}), for which the third and fourth order contributions to the D meson polarization tensor are already 
given with (\ref{res-Delta-expansion-3}, \ref{res-Delta-expansion}). The fifth-order term is
\begin{eqnarray}
&&\bar \Pi_{H\in [0^-]}^{{\rm bubble}-5}
    =\sum_{Q\in [8]} \sum_{R\in [1^-]} \left(\frac{G_{QR}^{(H)}}{8\,\pi\,f}\right)^2 \Bigg\{ \bigg[-\,\frac{3}{4}\,\Big( \Pi^{(3)}_H - \Pi^{(3)}_R \Big)\,m_Q^2 \nonumber\\
    &&\qquad \qquad \qquad -\frac{\Delta }{2 M}\,\Big( \,m_Q^4  +2 m_Q^2 \big(2 \Pi_H^{(2)} - \Pi_R^{(2)} \big) +\big(\Pi_H^{(2)} - \Pi_R^{(2)}\big)^2 \Big) \bigg]\,\log \frac{  m_Q^2}{M^2}\nonumber\\
    && \qquad \qquad \qquad  -\Big( \Pi^{(3)}_H - \Pi^{(3)}_R \Big)\,m_Q^2 +\frac{\Delta}{8M}\bigg( 14 m_Q^4 -m_Q^2 \big(3 \Delta^2 -16 \Pi_H^{(2)} +36 \Pi_R^{(2)}\big)\nonumber\\
    && \qquad \qquad \qquad \quad -10 \big( \Pi_H^{(2)} - \Pi_R^{(2)}\big)^2 \bigg) 
     -\frac{\Delta_Q}{M}\,\Big(\log \big(\Delta +\Delta_Q\big) \nonumber\\
    && \qquad \qquad \qquad\qquad -\,\log \big(\Delta -\Delta_Q\big)\Big)\,\bigg[ - \frac{3}{2}\Delta\,M \left(\Pi^{(3)}_H - \Pi^{(3)}_R\right) + \frac{3}{4}\, \Delta_Q^4 \nonumber\\
     && \qquad \qquad \qquad \qquad + \frac{1}{2} \Delta_Q^2 \big(3 \Pi_R^{(2)} -5 \Pi_H^{(2)} \big)  +\frac{3}{4}\, \big(\Pi_H^{(2)}-\Pi_R^{(2)}\big)^2 \nonumber\\
     && \qquad \qquad \qquad \qquad \quad +\frac{3}{8} m_Q^2 \bigg( \Delta_Q^2 + 2\,\big(\Pi_R^{(2)} -3 \Pi_H^{(2)} \big) +\frac{(\Pi_H^{(2)} - \Pi_R^{(2)})^2}{\Delta_Q^2} \bigg) \bigg]\nonumber\\
     && \qquad \qquad \qquad + \bigg[ \frac{3\,\Delta^2}{2} \, \Big( \Pi^{(3)}_H - \Pi^{(3)}_R \Big) - \frac{\Delta }{4 M} \Big( 3\Delta_Q^4 -2 m_Q^4 -2 \big(\Delta^2 + m_Q^2\big) \,( \Pi_H^{(2)} + \Pi_R^{(2)} ) \nonumber\\ 
     && \qquad \qquad \qquad \qquad \qquad +\big( 8 \Delta^2  + \Pi_R^{(2)} - \Pi_H^{(2)} \big)(\Pi_R^{(2)} - \Pi_H^{(2)}) \Big) \bigg] \log \frac{m_Q^2}{4\Delta^2}\, \Bigg\} \,.
 \label{res-Delta-expansion-5A}    
\end{eqnarray}
It remains to specify the fifth order term with respect to the novel counting ansatz (\ref{def-counting-D}). We find
\begin{eqnarray}
&&\bar \Pi_{H\in[0^-]}^{{\rm bubble}-5} = \sum_{\substack{ Q\in [8] \\R\in[1^-]}} \bigg(\frac{m_Q}{4 \pi f}G_{QR}^{(H)}\bigg)^2 
	\Bigg\{ - \frac{\alpha_1}{4}\, \frac{M^2}{2M + \Delta } \frac{ \Delta ^3  }{4\, M^2} \log \frac{4 \Delta^2}{(M + \Delta)^2} 
\nonumber\\
&& \qquad \qquad \quad +\, \frac{\alpha_1}{8} \frac{\Delta^2}{m_Q^2} \big(M_R - M_H - \Delta_H \big)^2 \frac{M}{M+\Delta} \frac{\partial}{\partial \Delta} \big( \gamma_1\, \Delta \big) 
\nonumber\\
&& \qquad \qquad + \,\frac{M_H}{4} \Bigg[ (\alpha_1 -\alpha_2 )\bigg( \frac{2M + \Delta}{2M} \frac{m_Q^2}{\Delta_H^2} \big(M_R - M_H\big) \log \frac{m_Q^2}{M_R^2} - (\delta_1 - \gamma_1) \frac{\Delta_Q^2}{\Delta_H} 
\nonumber\\
&& \qquad \qquad \qquad - \,\delta_1 \frac{\Delta_Q^2}{\Delta_H^2} (M_R - M_H - \Delta_H ) \bigg) - \frac{\beta_{11}}{M_H^2} \bigg( (M_R - M_H)^3 \log \frac{m_Q^2}{M_R^2} 
\nonumber\\
&& \qquad \qquad \qquad + \Delta_Q^3 \Big[ \log \big(M_R - M_H + \Delta_Q\big)-\log \big(M_R - M_H - \Delta_Q\big)\Big] \bigg) 
	 \nonumber\\
&& \qquad \qquad \quad + \frac{m_Q^2\, \Delta_Q^2}{\Delta_H^3} \Bigg( (\alpha_2 - \alpha_1 ) \Big(\delta_2 + \delta_3 \log \frac{m_Q^2}{M_R^2} \Big) - \alpha_1 \Big( \delta_4 + \delta_5 \log \frac{m_Q^2}{M_R^2} \Big) \Bigg)\,\Bigg] 
	\nonumber\\
&& \qquad \qquad + \frac{M_H}{8} (M_R - M_H - \Delta_H )^2 \Bigg[ \beta_9\, \frac{\Delta_Q^2}{m_Q^2 \Delta_H} - \beta_{10}\, \frac{\Delta_Q^2}{m_Q^2 \Delta_H^2} (M_R - M_B) 
\nonumber\\
&& \qquad \qquad \quad -\,\frac{\beta_6}{m_Q^2\,\Delta_H^2} \bigg( (M_R - M_H)\Big(\Delta_Q^2 - \frac{m_Q^2}{2}\Big) \log \frac{m_Q^2}{M_R^2} 
\nonumber\\
&& \qquad \qquad \qquad + \,\Delta_Q^3 \Big[ \log \big(M_R - M_H + \Delta_Q\big)-\log \big(M_R - M_H - \Delta_Q\big)\Big] \bigg)
\nonumber\\
&& \qquad \qquad \quad + \,\bigg( -\beta_7 \,\frac{\Delta_Q^2}{\Delta_H^3} + \beta_8 \,\frac{m_Q^2}{\Delta_H^3} \log \frac{m_Q^2}{M_R^2} \bigg)\,
	\Bigg] \Bigg\}\,.
\label{loop-HB-5A}
\end{eqnarray}

\newpage

\section*{Appendix B}
In this Appendix we collect all dimension less coefficients that are needed in the various power counting decompositions of the renormalized loop function 
(\ref{result-loop-1_full}). Here we focus on the vector $D$ mesons for which we find
\begin{eqnarray}
&&\tilde \alpha_1 =  \frac{(2\,M+\Delta)^2}{4\,M^2}\,, \qquad \qquad  \quad
 \tilde \alpha_2  =\frac{2\,M^2+ 2\,\Delta\,M+\Delta^2}{2\,M^2}\,,\qquad \qquad 
\tilde \alpha_3 = 1\,,
\nonumber\\ \nonumber\\
&&\tilde  \gamma_1 = -\frac{M\,(2\,M+ \Delta)}{(M+\Delta)^2}\,\log \frac{\Delta\,(2\,M+\Delta)}{M^2}\,,
\nonumber\\
&&\tilde  \gamma_2 = \frac{M}{2\,M+ \Delta } +M\,\frac{2\,M^2 + 2\,\Delta\,M+\Delta^2}{(2\,M + \Delta )\,(M+ \Delta )^2}\,\log \frac{\Delta\,(2\,M+ \Delta)}{M^2}\,,\qquad 
\tilde  \gamma_3 = - \frac{M}{2\,M+\Delta }\,,
\nonumber\\
&&\tilde  \gamma_4 =- \frac{M\,(M+\Delta)^2}{2\,(2\,M+ \Delta)^3 }
+ \frac{2\,M^3 }{(2\,M + \Delta )^3}\,\log \frac{\Delta\,(2\,M+ \Delta)}{M^2}\,, \qquad 
\tilde  \gamma_5 = - \frac{M^3}{(2\, M +\Delta)^3} \,,
\nonumber\\
\nonumber\\
&& \tilde \delta_1 = \tilde  \gamma_1 + \frac{M (2\,M + \Delta)}{(M+ \Delta)^2}\log \frac{2\Delta}{M}\, 
\nonumber\\
&& \tilde \delta_2 = \tilde  \gamma_2 - \frac{M\, (2\, M^2 + 2\, \Delta\, M + \Delta^2 )}{(2\, M + \Delta )(M + \Delta )^2}\log \frac{2\, \Delta}{M + \Delta} 
\nonumber\\
&& \qquad - \frac{M\, (2\, M + \Delta )}{4(M + \Delta )^2} -2 (\tilde \delta_3 - \tilde  \gamma_3 ) \log \frac{M + \Delta}{M}\,,
\nonumber\\
&& \tilde \delta_3 = \tilde  \gamma_3 + \frac{M\, (2\, M^2 + 2\, \Delta\, M + \Delta^2 )}{ 2(M + \Delta )^2 (2\, M + \Delta )} \,,
\qquad \qquad \tilde \delta_5 = 0\, , \nonumber\\
&& \tilde \delta_4 = \tilde  \gamma_4 - \frac{2\, M^3}{(2\, M + \Delta )^3} \log \frac{2\, \Delta}{M + \Delta}
\nonumber\\
&& \qquad + \frac{M\, (4\, M^2 + 4\, \Delta\,M + 5\, \Delta^2 )}{32\, (M + \Delta )^2 (2\, M + \Delta )} -2\, (\tilde\delta_5 - \tilde  \gamma_5 )\log \frac{M + \Delta}{M}\,,
\nonumber\\
&& \tilde \delta_6 = \frac{2\,M + \Delta}{2\,M }\,\frac{\partial}{\partial \Delta} \frac{2\,(M + \Delta )}{2\,M + \Delta} \,\Delta\, \big(\tilde \gamma_1 - \tilde \delta_1 \big) + \tilde \delta_1 \,, 
\qquad  \tilde \delta_7 = \tilde \gamma_2 + \frac{1}{2} \,\big(\tilde \gamma_1 - \tilde \delta_1 \big)\, \frac{\Delta^2}{(2\,M + \Delta )^2} \, ,
\nonumber\\ \nonumber\\
&& \tilde \beta_1 = \frac{M + \Delta}{M}\, \frac{\Delta\,\partial}{\partial \Delta} \,\tilde \alpha_1 \frac{(2\,M + \Delta)\,M^2}{2\, (M + \Delta )^3}\,, 
\nonumber\\
	&& \tilde \beta_2 = \Delta^2\, \frac{\partial}{\partial \Delta}\, \frac{\tilde \alpha_1 \,\tilde \delta_2}{\Delta}\,, \qquad 
	 \tilde \beta_3 = \Delta^2\, \frac{\partial}{\partial \Delta} \,\frac{\tilde \alpha_1\, \tilde \delta_3}{\Delta}\,, 
\nonumber\\[3pt]
	&& \tilde \beta_4 = \frac{\Delta}{M}\, \frac{(M + \Delta)^2}{M}\, \tilde \gamma_1\, \frac{\partial}{\partial \Delta}\, \frac{M^2 \, \tilde \alpha_1}{(M+ \Delta)^2} \,,  \qquad 
	\tilde \beta_5 = \frac{M + \Delta}{M}\, \frac{\Delta \,\partial}{\partial \Delta}\, \frac{\tilde \alpha_1\, \tilde \delta_1\, M}{M+ \Delta}\,,
\nonumber\\
&&  \tilde \beta_6 = D_{\Delta \Delta} \frac{(2M + \Delta) M^2}{2(M + \Delta)^3} \tilde \alpha_1 \,,\qquad \qquad
	 \tilde \beta_7 = \frac{\Delta}{M + \Delta} D_{\Delta \Delta}\frac{M}{\Delta} \tilde \alpha_1 \,\tilde\delta_2\,, 
\nonumber\\
&&  \tilde \beta_8 = \frac{\Delta}{M + \Delta} D_{\Delta \Delta} \frac{M}{\Delta} \tilde \alpha_1 \,\tilde\delta_3\,, \qquad 
 \tilde \beta_9 = \tilde \gamma_1 \,\frac{M + \Delta}{M} D_{\Delta \Delta} \frac{M^2}{(M + \Delta)^2} \,\tilde \alpha_1\,,\qquad 
\nonumber\\
&& \tilde \beta_{10} = D_{\Delta \Delta} \frac{M}{M + \Delta} \tilde \alpha_1 \tilde\delta_1\,, \qquad 
\tilde \beta_{11} = - \frac{1}{4}\,\tilde \alpha_1 \,\frac{M}{2\,M + \Delta} + (\tilde \alpha_1 - \tilde \alpha_2)\frac{(2\,M + \Delta)M}{2\, \Delta^2}\,,
\nonumber\\ \nonumber\\
&& \qquad {\rm with }\qquad D_{\Delta \Delta} = \frac{(M+\Delta)^2}{M^2}\,\Big( \frac{\Delta^2\, \partial^2 }{\partial \Delta \,\partial \Delta} 
+\frac{2\,\Delta}{M+\Delta}\,\frac{\Delta \,\partial }{\partial \Delta} \Big) \,.
\end{eqnarray}
While the $\tilde \alpha_i$ characterize the chiral expansion of the coefficients in front of $\bar I_{QR}$ and $\bar I_Q$ in  (\ref{result-loop-1_full}), the 
$\tilde \gamma_i$ and $\tilde \delta_i$ follow from a chiral expansion of $\bar I_{QR}$ with $M_H= M+ \Delta$ and $M_R = M$ and  $m_Q < \Delta$. 
The coefficients $\tilde \beta_i$ are required in (\ref{loop-HB-3}, \ref{loop-HB-4}, \ref{loop-HB-5B}).

We turn to the chiral domain (\ref{def-counting-B}), in which the bubble-loop contributions to the D$^*$ meson masses generate a renormalization of the low-energy parameters $\tilde d_i$. 
Such terms are proportional to the product of two quark masses (\ref{res-tree-level}). We provide detailed results with 
\begin{eqnarray}
&&\Pi^{(4-\chi) }_H \to  \sum_{Q\in[8]}\sum_{R\in \left[0^-\right]} \left(\frac{G_{QR}^H}{8\,\pi f}\right)^2
    \Bigg\{ \tilde \gamma_d^{(1)} m_Q^2\, \Pi_R^{(2)} + \tilde\gamma_d^{(2)}\, m_Q^2\, \Pi_H^{(2)} + \tilde\gamma_d^{(3)}\,\Pi_R^{(2)}\,\Pi_R^{(2)} 
\nonumber\\
    &&\qquad \qquad \quad +\,\tilde\gamma_d^{(4)}\,\Pi_H^{(2)}\,\Pi_H^{(2)}  +\tilde\gamma_d^{(5)}\,\Pi_R^{(2)} \,\Pi_H^{(2)}  
     \Bigg\}\,,
\nonumber\\
&&  \tilde d_i \to  \frac{1}{4} \, g_P^2\,\sum_{k=1}^5 \frac{\tilde \Gamma_{d_i}^{(k)}}{(4\pi f)^2} \,\tilde \gamma_d^{(k)}\,,
\label{Finite_Renorm_d}
\end{eqnarray}
with    
\begin{eqnarray}
&& \tilde \gamma_d^{(1)} =
\frac{\Delta}{6\,M} \,\Bigg[ \frac{\partial }{\partial M } \Big( \tilde \alpha_2\,M\,\tilde \gamma_1 -\tilde \alpha_1\,M\, \tilde \gamma_2 \Big) - \frac{1}{M}\,\tilde \gamma_1\,
\frac{\partial }{\partial M } \,\big(\tilde \alpha_2\,M^2 \big) \Bigg] - \frac{ M+ \Delta}{M} \,\tilde \gamma_d^{(2)},
\nonumber\\
&& \tilde \gamma_d^{(2)} = \frac{M}{6\,(M+\Delta)}\,\Bigg[
\frac{\partial }{\partial \Delta } \Big( \tilde \alpha_2\,\Delta\,\tilde \gamma_1 -\tilde \alpha_1\,\Delta\, \tilde \gamma_2 \Big) - \Delta\,\tilde \gamma_1\,
\frac{\partial }{\partial \Delta }\, \tilde \alpha_2 \Bigg]\,,
\nonumber\\
&& \tilde \gamma_d^{(3)} = -\frac{1}{24\,M^2}\, \bigg(\frac{\partial }{\partial M}- \frac{\partial }{\partial \Delta} \bigg)^2\,
\Big(\tilde \alpha_1\,M\,\Delta^3 \,\tilde \gamma_1 \Big) +
\frac{\tilde \gamma_1\,\Delta}{24\,M^3}\, \bigg(\frac{\partial }{\partial M}- \frac{\partial }{\partial \Delta} \bigg)^2\,
\Big(\tilde \alpha_1\,M^2\,\Delta^2  \Big)
\nonumber\\
&& \qquad  +\,
\frac{1}{24\,M^3}\, \bigg(\frac{\partial }{\partial M}- \frac{\partial }{\partial \Delta} \bigg)\,
\Big(\tilde \alpha_1\,M\,\Delta^3 \,\tilde \gamma_1 \Big) -
\frac{\tilde \gamma_1\,\Delta}{24\,M^4}\, \bigg(\frac{\partial }{\partial M}- \frac{\partial }{\partial \Delta} \bigg)\,
\Big(\tilde \alpha_1\,M^2\,\Delta^2  \Big) \,,
\nonumber\\
&& \tilde \gamma_d^{(4)} = -\frac{M}{12\, (M + \Delta )^2}\,\bigg(\frac{\partial\,\tilde \alpha_1\, \Delta^2}{\partial \Delta } \bigg)
\left(\frac{\partial  \,\tilde \gamma _1\,\Delta}{\partial \Delta }\right)
 - \frac{\tilde \alpha_1 \,\Delta^2\, M}{12\, (M + \Delta)}\,
\frac{\partial }{\partial \Delta }\,\Bigg[\frac{1}{2\, (M +\Delta )}\,
\bigg(\frac{\partial\,\tilde \gamma_1\, \Delta }{\partial \Delta }\bigg)\Bigg]\,,
\nonumber\\
&& \tilde \gamma_d^{(5)} = -\frac{1}{6\,M}\,\frac{\partial}{\partial M}\, \frac{1}{2\,(M + \Delta)}\,\frac{\partial }{\partial \Delta}\,
\Big( \tilde \alpha_1\,M\,\Delta^3\,\tilde \gamma_1 \Big)
\nonumber\\
&& \qquad  +\,
\frac{\tilde \gamma_1\,\Delta}{6\,M^2}\,\frac{\partial}{\partial M}\, \frac{1}{2\,(M + \Delta)}\,\frac{\partial }{\partial \Delta}\,
\Big( \tilde \alpha_1\,M^2\,\Delta^2 \Big)- 2\,\frac{M+\Delta}{M}\, \tilde \gamma_d^{(4)}\,.
\end{eqnarray} 
 where the other $\Gamma_{\tilde{d}_i}^{(k)}$ with $k =1,2, ...,5$ follow from the corresponding $\Gamma_{d_i}^{(k)}$ in (\ref{res-Gamma-d}) upon 
the interchange $c_i \leftrightarrow \tilde{c}_i$.

We turn to the conventional counting ansatz (\ref{def-counting-C}), for which the third and fourth order contributions to the D$^*$ meson polarization tensor are already 
given with (\ref{res-Delta-expansion-3}, \ref{res-Delta-expansion}). The fifth-order term is
\begin{eqnarray}
&&\bar \Pi_{H\in [1^-]}^{{\rm bubble}-5}
    =\sum_{Q\in [8]} \sum_{R\in [1^-]} \left(\frac{G_{QR}^{(H)}}{8\pi\,f}\right)^2 \frac{1}{12} \bigg\{ - 8 \Big( \Pi^{(3)}_H - \Pi^{(3)}_R\Big) \,m_Q^2 \nonumber\\
    && \qquad \qquad \qquad \quad +\, \frac{4\,\Delta}{M}\, m_Q^2 \big( 2 m_Q^2 -3 \Pi_R^{(2)} + 3 \Pi_H^{(2)} \big) - 6 \,\Big( \Pi^{(3)}_H - \Pi^{(3)}_R \Big)\,m_Q^2\,\log\frac{ m_Q^2}{M^2}   \nonumber\\
    && \qquad \qquad \qquad \qquad +\, \pi\,\frac{m_Q}{M} \Big[ 3\,m_Q^4  +2\, m_Q^2 \big( \Pi_H^{(2)} -3\, \Pi_R^{(2)} \big) +3\, \big(\Pi_R^{(2)} - \Pi_H^{(2)}\big)^2 \Big]
    \bigg\}
    \nonumber\\
    &&\qquad \qquad + \sum_{Q\in [8]} \sum_{R\in [0^-]} \left(\frac{G_{QR}^{(H)}}{8\pi\,f}\right)^2 \frac{1}{3} \Bigg\{ \bigg[ -\,\frac{3}{4}\,\Big(\Pi^{(3)}_H - \Pi^{(3)}_R \Big)\,m_Q^2 \nonumber\\
    && \qquad \qquad \qquad + \frac{\Delta }{2 M}\,\Big( \,m_Q^4  +2 m_Q^2 \big(2 \Pi_H^{(2)} - \Pi_R^{(2)} \big) +\big(\Pi _H-\Pi _R\big)^2 \Big)\bigg]\, \log \frac{  m_Q^2}{M^2}\nonumber\\
    && \qquad \qquad \qquad  -\Big(  \Pi^{(3)}_H - \Pi^{(3)}_R \Big) \,m_Q^2 +\frac{\Delta}{8M}\Big[ -6\, m_Q^4 +m_Q^2 \big(3 \Delta^2 -4 \Pi_H^{(2)} +24 \Pi_R^{(2)}\big) \nonumber\\
    && \qquad \qquad \qquad \qquad +10 \, \big(\Pi_H^{(2)} - \Pi_R^{(2)}\big)^2 \Big]   + \frac{\Delta_Q}{M}\,\Big(\log \big(- \Delta -\Delta_Q\big)\nonumber\\
    && \qquad \qquad \qquad \qquad \quad
     -\,\log \big( -\Delta +\Delta_Q\big)\Big)\,\bigg[  \frac{3}{2}\Delta M \left( \Pi^{(3)}_H - \Pi^{(3)}_R\right) + \frac{3}{4} \, \Delta_Q^4   \nonumber\\
     && \qquad \qquad \qquad \qquad \quad +\frac{1}{2} \, \Delta_Q^2\, \big(3 \Pi_R^{(2)} -5 \Pi_H^{(2)} \big) +\frac{3}{4} \, \big(\Pi_H^{(2)}-\Pi_R^{(2)}\big)^2 \nonumber\\
     && \qquad \qquad \qquad \qquad + \frac{3}{8} m_Q^2 \bigg( \Delta_Q^2 + 2\,\big(\Pi_R^{(2)} -3 \Pi_H^{(2)} \big) +\frac{(\Pi_H^{(2)} - \Pi_R^{(2)})^2}{\Delta_Q^2} \bigg) \bigg]  \nonumber\\
     &&  \qquad \qquad \qquad + \bigg[ \frac{3}{2} \Delta^2 \Big( \Pi^{(3)}_H - \,\Pi^{(3)}_R \Big) + \frac{\Delta }{4 M} \Big( 3\Delta_Q^4 -2 m_Q^4 -2 \big(\Delta^2 + m_Q^2\big) \,\big( \Pi_H^{(2)} + \Pi_R^{(2)} \big) \nonumber\\
     && \qquad \qquad \qquad \qquad \qquad +\big( 8 \Delta^2  + \Pi_R^{(2)} - \Pi_H^{(2)} \big)(\Pi_R^{(2)} - \Pi_H^{(2)}) \Big) \bigg]\log \frac{m_Q^2}{4\Delta^2}\Bigg\} \,.
\nonumber\\
\label{res-Delta-expansion-5B}
\end{eqnarray}
It remains to specify the fifth order term with respect to the novel counting ansatz (\ref{def-counting-D}). We find
\begin{eqnarray}
&& \bar \Pi_{H\in[1^-]}^{{\rm bubble}-5}= \sum_{\substack{Q\in [8]\\ R\in [1^-]}} \Bigg(\frac{m_{QR}}{4 \pi f}G_{QR}^{(H)}\Bigg)^2
	\frac{1}{3} \bigg\{ \frac{3 \pi }{16}\, \frac{m_Q^3}{M_H}-  \frac{m_Q^4}{M_H^2} \bigg(\frac{1}{6} - \frac{1}{8}  \log \frac{m_Q }{M_R }\bigg) 
\nonumber\\
&& \qquad \qquad \qquad +\, (M_R-M_H)^2 \bigg( \frac{\pi }{4 }\, \frac{M_H}{m_Q} + 1 + \frac{3}{2}\, \log \frac{m_Q }{M_R } \bigg) \bigg\}
	\nonumber\\
&& \qquad \qquad + \sum_{\substack{ Q\in [8] \\R\in[0^-]}} \Bigg(\frac{m_Q}{4 \pi f}G_{QR}^{(H)}\Bigg)^2 
	\Bigg\{ \, \frac{\tilde \alpha_1}{12}\, \frac{M^2}{ (2 M + \Delta )} \frac{ \Delta ^3  }{4 (M + \Delta )^2} \log \frac{4 \Delta^2}{ M^2} 
\nonumber\\
&& \qquad \qquad \qquad -\, \frac{\tilde \alpha_1}{24}\, \frac{\Delta^2}{m_Q^2} \big(M_R - M_H + \Delta_H \big)^2 \frac{M+\Delta}{M} \frac{\partial}{\partial \Delta} \big( \tilde \gamma_1\, \Delta \big) 
\nonumber\\
&& \qquad \qquad \quad +\, \frac{M_H}{12} \frac{M}{M + \Delta} \Bigg[ - \big(\tilde \alpha_1  - \tilde \alpha_2 \big) \Bigg( \frac{M\,(2M + \Delta)}{2 (M + \Delta )^2} \frac{m_Q^2}{\Delta_H^2} \big(M_H - M_R\big) \log \frac{m_Q^2}{M_R^2} \nonumber\\
&& \qquad \qquad \qquad + \,(\tilde\delta_1 - \tilde \gamma_1 ) \frac{\Delta_Q^2}{\Delta_H} - \tilde\delta_1 \frac{\Delta_Q^2}{\Delta_H^2} (M_R - M_H + \Delta_H ) \Bigg) 
	   + \frac{\tilde \beta_{11}}{M_H^2}  \bigg( (M_H - M_R)^3 \log \frac{m_Q^2}{M_R^2} 
\nonumber\\
&& \qquad \qquad \qquad + \,\Delta_Q^3 \Big[ \log \big(M_R - M_H - \Delta_Q\big)-\log \big(M_R - M_H + \Delta_Q\big)\Big] \bigg) 
\nonumber\\
&& \qquad \qquad \quad \; + \frac{m_Q^2 \Delta_Q^2}{\Delta_H^3} \Bigg( (\tilde \alpha_2 - \tilde \alpha_1) \Big(\tilde\delta_2 + \tilde\delta_3 \log \frac{m_Q^2}{M_R^2} \Big) - \tilde \alpha_1 \Big( \tilde\delta_4 + \tilde\delta_5 \log \frac{m_Q^2}{M_R^2} \Big) \Bigg)\,
	\Bigg] \nonumber\\
&& \qquad \qquad \quad + \frac{M_H}{24} \big(M_R - M_H + \Delta_H \big)^2 \Bigg[ \tilde \beta_9\, \frac{\Delta_Q^2}{m_Q^2 \Delta_H} - \tilde \beta_{10}\, \frac{\Delta_Q^2}{m_Q^2 \Delta_H^2} (M_H - M_R) 
\nonumber\\
&& \qquad \qquad \qquad +\frac{\tilde \beta_6}{m_Q^2\, \Delta_H^2} \bigg( (M_H - M_R)\Big(\Delta_Q^2 - \frac{m_Q^2}{2}\Big) \log \frac{m_Q^2}{M_R^2} 
\nonumber\\
&& \qquad \qquad \qquad \quad + \Delta_Q^3 \Big[ \log \big(M_R - M_H - \Delta_Q\big)-\log \big(M_R - M_H + \Delta_Q\big)\Big] \bigg)
\nonumber\\
&& \qquad \qquad \qquad + \bigg( -\tilde \beta_7 \,\frac{\Delta_Q^2}{\Delta_H^3} + \tilde \beta_8 \,\frac{m_Q^2}{\Delta_H^3} \log \frac{m_Q^2}{M_R^2} \bigg)\,
	\Bigg] \Bigg\}\, ,
\nonumber\\ \nonumber\\
&& \qquad \qquad \qquad {\rm with }\qquad m_{QR}^2 = m_Q^2 - (M_R - M_H)^2\,.
\label{loop-HB-5B}
\end{eqnarray}

\clearpage


\begin{table}[b]
\renewcommand{\arraystretch}{1.1}
\setlength{\arraycolsep}{12mm}
\begin{center}
\begin{tabular}{cc|llllllllll}
 $a\,m_{\pi}$			     & $a\,m_{K}$			        &$a\,\mu_c$   & discr.  		&  $a\,m_{D}$       &  $a\,m_{D_s}$    &$a\,m_{D^*}$      &$a\,m_{D^*_s}$		\\ \hline
0.0703(4)	     &0.1697(3) &0.2230	 & $(\pm,\mp)$ & 0.6655(12)	  & 0.6981(4)	     & 0.7161(18)       & 0.7456(10)       	\\ 
                                 		     &					        &0.1919	 & $(\pm,\mp)$ & 0.6072(11)	  & 0.6402(3)	     & 0.6621(18)       & 0.6923(10)       	\\ 
                              		     &					        &0.2230	 & $(\pm,\pm)$ & 0.6706(15)	  & 0.7035(5)	     & 0.7078(24)       & 0.7430(10) 		\\ 
                              		     &					        &0.1919	 & $(\pm,\pm)$ & 0.6123(14) 	  & 0.6460(4)	     & 0.6536(23)       & 0.6898(10)  		\\ \hline
0.0806(3)	     &0.1738(5) &0.2227	 & $(\pm,\mp)$ & 0.6661(19) 	  & 0.6983(4)	     & 0.7209(26)       & 0.7452(12)  		\\ 
                          		     & 					        &0.1727	 & $(\pm,\mp)$ & 0.5712(14) 	  & 0.6041(4)	     & 0.6325(25)       & 0.6586(11)		\\ 
                              		     &					        &0.2227	 & $(\pm,\pm)$ & 0.6721(22) 	  & 0.7037(5)	     & 0.7209(20)       & 0.7452(10)    		\\ 
                              		     & 					        &0.1727	 & $(\pm,\pm)$ & 0.5775(17) 	  & 0.6102(4)	     & 0.6335(23)       & 0.6587(10)    		\\ \hline
0.0975(3)	     &0.1768(3) &0.2230	 & $(\pm,\mp)$ & 0.6666(16) 	  & 0.6980(5)	     & 0.7183(23)       & 0.7458(13)  		\\ 
                             		     & 					        &0.1727	 & $(\pm,\mp)$ & 0.5720(12) 	  & 0.6036(4)	     & 0.6308(24)       & 0.6587(13) 		\\ 
                        			     &					        &0.2230	 & $(\pm,\pm)$ & 0.6713(13) 	  & 0.7033(5)	     & 0.7169(19)       & 0.7451(8)    		\\ 
                                		     & 					        &0.1727	 & $(\pm,\pm)$ & 0.5770(12) 	  & 0.6098(4)	     & 0.6290(22)       & 0.6579(11) 		\\ \hline
0.1074(5)	     &0.2133(4) &0.2230	 & $(\pm,\mp)$ & 0.8473(10)  	  & 0.8780(5)	     & 0.9140(31)       & 0.9474(10)		\\ 
                                		     &					        &0.1727	 & $(\pm,\mp)$ & 0.7501(8) 	  & 0.7827(4)	     & 0.8262(29)       & 0.8601(9)		\\ 
                                		     &					        &0.2230	 & $(\pm,\pm)$ & 0.8588(16) 	  & 0.8922(7)	     & 0.9112(25)       & 0.9443(10)		\\ 
                                		     &			  		        &0.1727	 & $(\pm,\pm)$ & 0.7629(14) 	  & 0.7978(6)	     & 0.8224(24)       & 0.8566(10)		\\ \hline
0.1549(2)	     &0.2279(2) &0.2230	 & $(\pm,\mp)$ & 0.8543(5)	  & 0.8824(3)	     & 0.9268(11)       & 0.9536(7)		\\ 
                                		     &					        &0.1727	 & $(\pm,\mp)$ & 0.7549(5)	  & 0.7841(3)	     & 0.8362(11)       & 0.8637(8)		\\ 
                                		     &					        &0.2230	 & $(\pm,\pm)$ & 0.8666(8)	  & 0.8961(4)	     & 0.9218(11)       & 0.9500(6)		\\ 
                                		     &					        &0.1727	 & $(\pm,\pm)$ & 0.7683(7)	  & 0.7991(3)	     & 0.8322(17)       & 0.8597(7)		\\ \hline
0.1935(4)	    &0.2430(4)  &0.2230	 & $(\pm,\mp)$ & 0.8559(8)	  & 0.8784(5)	     & 0.9309(18)       & 0.9521(13)		\\ 
                                		     &					        &0.1727	 & $(\pm,\mp)$ & 0.7600(11)	  & 0.7850(4)	     & 0.8443(18)       & 0.8669(13)		\\ 
                                		     &					        &0.2230	 & $(\pm,\pm)$ & 0.8690(8)	  & 0.8928(5)	     & 0.9273(14)       & 0.9484(11)		\\ 
                                		     &					        &0.1727	 & $(\pm,\pm)$ & 0.7763(7)	  & 0.8007(5)	     & 0.8413(14)       & 0.8629(11)		\\ \hline
\end{tabular}
\caption{Masses for the $D$ mesons in units of the lattice scale $a$. The values in the table are provided to us by the authors 
of \cite{Kalinowski:2015bwa}.}
\label{Table_ETMC_3}
\end{center}
\end{table}

\clearpage

\begin{table}
\renewcommand{\arraystretch}{1.15}
\setlength{\arraycolsep}{12mm}
\begin{center}
 \begin{tabular}{cc|llllllllll}
 $a\,m_{\pi}$			    & $a\,m_{K}$			         &$a\,\mu_c$  & discr.            &$a\,m_{D}$  	  & $a\,m_{D_s}$	 &$a\,m_{D^*}$	 & $a\,m_{D^*_s}$   \\ \hline
0.1240(4)	    &0.2512(3)   &0.2772	 & $(\pm,\mp)$ & 0.8979(9)	  & 0.9412(2)	 & 0.9782(16)	 & 1.0225(7)	     \\ 
                          		    &  	 				         &0.2270	 & $(\pm,\mp)$ & 0.7994(8)	  & 0.8441(2)	 & 0.8880(16)	 & 0.9338(7)	     \\ 
                           		    &						 &0.2772	 & $(\pm,\pm)$ & 0.9154(14)	  & 0.9610(3)	 & 0.9759(15)	 & 1.0185(7)	     \\ 
	                       	            &     				         &0.2270	 & $(\pm,\pm)$ & 0.8181(12)	  & 0.8655(3)	 & 0.8859(15)	 & 0.9289(8)	     \\ \hline
0.1412(3)	    &0.2569(3)	 &0.2768	 & $(\pm,\mp)$ & 0.9002(10)	  & 0.9420(3)	 & 0.9776(20)	 & 1.0213(9)	     \\ 
                           		    &   					 &0.2389	 & $(\pm,\mp)$ & 0.8258(9)	  & 0.8692(3)	 & 0.9104(20)	 & 0.9545(9)	     \\ 
                         			    &						 &0.2768	 & $(\pm,\pm)$ & 0.9162(13)	  & 0.9623(4)	 & 0.9743(19)	 & 1.0169(9)	     \\ 
                         			    & 						 &0.2389	 & $(\pm,\pm)$ & 0.8433(12)	  & 0.8904(4)	 & 0.9067(18)	 & 0.9501(9)	     \\ \hline
0.1440(6)     &0.2589(4)	 &0.2768	 & $(\pm,\mp)$ & 0.9006(8)	  & 0.9425(3)	 & 0.9801(23)	 & 1.0252(8)	     \\ 
                         			    & 						 &0.2389	 & $(\pm,\mp)$ & 0.8268(12)	  & 0.8697(2)	 & 0.9153(19)	 & 0.9589(8)	     \\ 
                         			    &						 &0.2768	 & $(\pm,\pm)$ & 0.9160(13)	  & 0.9627(3)	 & 0.9813(16)	 & 1.0208(7)	     \\ 
                         			    &					  	 &0.2389	 & $(\pm,\pm)$ & 0.8432(11)	  & 0.8911(3)	 & 0.9145(15)	 & 0.9544(7)	     \\ \hline
0.1988(3)     &	0.2764(3)   &0.2929	 & $(\pm,\mp)$ & 0.9327(8)	  & 0.9668(5)	 & 1.0148(17)	 & 1.0496(12)	     \\ 
                         			    &  						 &0.2299	 & $(\pm,\mp)$ & 0.8164(13)	  & 0.8520(4)	 & 0.9092(16)	 & 0.9449(11)	     \\ 
                         			    &						 &0.2929	 & $(\pm,\pm)$ & 0.9500(12)	  & 0.9879(5)	 & 1.0098(44)	 & 1.0434(20)	     \\ 
                         			    &  					 	 &0.2299	 & $(\pm,\pm)$ & 0.8358(10)	  & 0.8746(5)	 & 0.9026(41)	 & 0.9381(18)	     \\ \hline
\end{tabular}
\caption{Masses for the $D$ mesons in units of the lattice scale $a$. The values in the table are provided to us by the authors 
of \cite{Kalinowski:2015bwa}.}
\label{Table_ETMC_4}
\end{center}
\end{table}


\clearpage

\bibliography{thesis}
\bibliographystyle{apsrev4-1}
\end{document}